%% file: paper.tex
\documentclass[twocolumn,showpacs,aps,floatfix,prd,nofootinbib]{revtex4}
\usepackage{graphicx}
\usepackage{dcolumn}
\usepackage{amsmath}
\usepackage{epsfig}
\usepackage{colordvi}
\usepackage{color}
\usepackage{hhline}
\usepackage{multirow}
\usepackage{subfigure}
\usepackage{rotating}
\usepackage{slashbox}
\usepackage{hyperref}
\RequirePackage{xspace}

\pdfoutput=1

\def\pep2{PEP-II}

\usepackage{relsize}
\def\babar{\mbox{\slshape B\kern-0.1em{\smaller A}\kern-0.1em
    B\kern-0.1em{\smaller A\kern-0.2em R}}}

\newcommand{\gev}{\ensuremath{\mathrm{\,Ge\kern -0.1em V}}\xspace}
\newcommand{\mev}{\ensuremath{\mathrm{\,Me\kern -0.1em V}}\xspace}
\newcommand{\gevc}{\ensuremath{\mathrm{\,Ge\kern -0.1em V}}\xspace}
\newcommand{\mevc}{\ensuremath{{\mathrm{\,Me\kern -0.1em V}}}\xspace}
\newcommand{\gevcc}{\ensuremath{{\mathrm{\,Ge\kern -0.1em V}}}\xspace}
\newcommand{\gevccs}{\ensuremath{{\mathrm{\,Ge\kern -0.1em V^2\!/}c^4}}\xspace}
\newcommand{\mevcc}{\ensuremath{{\mathrm{\,Me\kern -0.1em V}}}\xspace}

\def\epem       {\ensuremath{e^+e^-}\xspace}
\def\taup       {\ensuremath{\tau^+}\xspace}
\def\taum       {\ensuremath{\tau^-}\xspace}
\def\tautau     {\ensuremath{\tau^+\tau^-}\xspace}
\def\mtau       {\ensuremath{\tau}\xspace}
\def\q     {\ensuremath{q}\xspace}
\def\qbar  {\ensuremath{\overline q}\xspace}
\def\qqbar {\ensuremath{q\overline q}\xspace}

\def\uubar {\ensuremath{u\overline u}\xspace}

\def\ccbar {\ensuremath{c\overline c}\xspace}

\def\bbbar {\ensuremath{b\overline b}\xspace}

\def\piz   {\ensuremath{\pi^0}\xspace}
\def\pip   {\ensuremath{\pi^+}\xspace}
\def\pim   {\ensuremath{\pi^-}\xspace}
\def\Km    {\ensuremath{K^-}\xspace}
\def\KS    {\ensuremath{K^0_{\scriptscriptstyle S}}\xspace}
\def\pipi      {\ensuremath{\pi^+\pi^-}\xspace}
\def\pipm  {\ensuremath{\pi^\pm}\xspace}

\def\invfb     {\ensuremath{\mbox{\,fb}^{-1}}\xspace}
\newcommand{\qt}{\mathbf{q}_{T}}
\newcommand{\boldk}{\mathbf{k}}
\newcommand{\boldp}{\mathbf{p}}

\def\aulTh{\ensuremath{{ A_{12}^{UL}}}}
\def\aucTh{\ensuremath{{ A_{12}^{UC}}}}
\def\aul{\ensuremath{{ A_{0}^{UL}}}}
\def\auc{\ensuremath{{ A_{0}^{UC}}}}

\def\zbin{\ensuremath{(z_1,z_2)}}
\def\ptbin{\ensuremath{(p_{t1},p_{t2})}}

\def\pt         {\mbox{$p_t$}\xspace}
\def\Pperp        {\mbox{$P_h^\perp$}\xspace}
\def\Sq        {\mbox{$\mathbf{S}_q$}\xspace}
\def\PperpB        {\mbox{$\mathbf{P}^\perp_h$}\xspace}
\def\kbold   {\ensuremath{\mathbf{k}}}
\def\FF{\ensuremath{D(z)}\xspace}
\def\collins{\ensuremath{H_1^\perp}\xspace}
\def\upsbb   {\ensuremath{\FourS \to \BB}\xspace}

\mathchardef\Upsilon="7107
\def\Y#1S{\ensuremath{\Upsilon{(#1S)}}\xspace}
\def\FourS {\Y4S}
\newcommand{\dedx}{\ensuremath{\mathrm{d}\hspace{-0.1em}E/\mathrm{d}x}\xspace}
\def\B       {\ensuremath{B}\xspace}
\def\Bbar    {\kern 0.18em\overline{\kern -0.18em B}{}\xspace}

\def\BB      {\ensuremath{B\Bbar}\xspace}
\def\mumu       {\ensuremath{\mu^+\mu^-}\xspace}

 \def\eetott   {\ensuremath{\epem \to \tautau}\xspace}
 
 \def\uds   {\ensuremath{uds}\xspace}

\def\duephiz        {\ensuremath{2\phi_{0}}\xspace}

\def\Fcharm  {\ensuremath{F_c}\xspace}
\def\Fb          {\ensuremath{F_B}\xspace}
\def\Ftau       {\ensuremath{F_\tau}\xspace}
\def\fcharm   {\ensuremath{f_c}\xspace}
\def\fb           {\ensuremath{f_B}\xspace}

\def\Auds      {\ensuremath{A_\alpha}\xspace}
\def\Acharm  {\ensuremath{A_\alpha^{c}}\xspace}
\def\Ab          {\ensuremath{A_\alpha^\B}\xspace}  
\def\Atau       {\ensuremath{A_\alpha^\tau}\xspace}  
\def\Ameas    {\ensuremath{A_\alpha^{\rm{meas}}}\xspace}
\def\Adstar    {\ensuremath{A_\alpha^{\Dstar}}\xspace}
\def\Dstar   {\ensuremath{D^*}\xspace}
\def\Dstarpm {\ensuremath{D^{*\pm}}\xspace}
\def\Dz      {\ensuremath{D^0}\xspace}

\def\de{\mathrm{d}}
\def\jetset     {\mbox{\tt Jetset}\xspace}
\def\kktof     {\mbox{\tt KK2F}\xspace}

\def\afkqed    {\mbox{\tt AfkQed}\xspace}
\def\evtgen     {\mbox{\tt EvtGen}\xspace}
\def\geant      {\mbox{\sc Geant4}\xspace}

\newcommand{\BABARPubYear}    {13}
\newcommand{\BABARPubNumber}  {012}
\newcommand{\SLACPubNumber}   {15700}

\usepackage{relsize}
\def\babar{\mbox{\slshape B\kern-0.1em{\smaller A}\kern-0.1em
    B\kern-0.1em{\smaller A\kern-0.2em R}}}

\long\def\inst#1{\par\nobreak\kern 4pt\nobreak
    {\it #1}\par\vskip 10pt plus 3pt minus 3pt}

\begin{document}

\begin{flushleft}
\vspace{-2cm}
SLAC-PUB-\SLACPubNumber \\
\babar-PUB-\BABARPubYear/\BABARPubNumber
\end{flushleft}

\hyphenation{pro-duct}
\hyphenation{con-si-de-red}
\hyphenation{re-pre-sen-ted}
\hyphenation{de-fi-ni-tion}
\hyphenation{other-wise}

\title{
\Large \bf {\boldmath Measurement of Collins asymmetries in inclusive production 
of charged pion pairs in \epem annihilation at \babar}
} 
\input authors_may2013

\begin{abstract}
\noindent
We present measurements of Collins asymmetries in the inclusive process
$\epem\to\pi \pi X$,
where $\pi$ stands for charged pions,
 at a center-of-mass energy of 10.6 \gev.
We use a data sample of 468 \invfb collected by the \babar\ experiment at the PEP-II
$B$ factory at SLAC, and consider pairs of charged pions
produced in opposite hemispheres of hadronic events.
We observe clear asymmetries in the distributions of the azimuthal angles
in two distinct reference frames.
We study the dependence of the asymmetry on several kinematic variables,
finding that it increases with increasing pion momentum
and momentum transverse to the analysis axis,
and with increasing angle between the thrust and beam axis.
\end{abstract}


\maketitle
\noindent

\section{Introduction}\label{sec:Introduction}
Parton fragmentation functions  describe the probability for a parton to fragment 
into a hadron carrying a certain fraction $z$ of the parton momentum.
These functions  are denoted $D_h^i(z)$, where $i$ represents the fragmenting parton 
($g$, $u$, $\bar{u}$, $d$, $\bar{d}$,...), and $h$ is the produced hadron. 
Since the $D_h^i(z)$ incorporate the long distance, non-perturbative physics of the 
  hadronization processes, they cannot be calculated in perturbative QCD, but 
  can be evolved from a starting distribution at a defined energy scale.
Fragmentation processes have been studied in 
 lepton-hadron and hadron-hadron scattering, as well as in \epem 
annihilation, which provides the cleanest
 environment since no hadrons are present in the initial state.
Due to the large amount of experimental data 
collected at several \epem facilities, 
mainly LEP\cite{Buskulic:1996tt,Abreu:1997ir,Abbiendi:2004pr} 
and SLC~\cite{Abrams:1989rz,Petersen:1987bq,Abe:2003iy} at high
energies, and, recently,  \pep2~\cite{dave} and KEKB~\cite{Leitgab:2013rw} at
the center-of-mass energy $\sqrt{s}\sim10\gev$,
the unpolarized functions are presently well known.

 Transverse spin-dependent effects in fragmentation processes 
 were first proposed by Collins~\cite{Collins:1992kk,collins-2002-536}, who 
introduced the chiral-odd polarized Collins fragmentation function $H_1^\perp$.
 It describes the relation between the transverse spin of the fragmenting quark and the azimuthal distribution
 of  final-state hadrons around the quark momentum (spin-orbit correlation). 
In the transverse-momentum approach, $H_1^\perp$ depends on the hadron fractional energy $z=2E_h/\sqrt{s}$,
where $E_h$ and $\sqrt{s}/2$ are, respectively, 
the hadron and beam energy in the center-of-mass system,
and on the magnitude of the hadron transverse momentum
$\PperpB$ with respect to the three-momentum of the fragmenting quark.

The number density for finding a spinless hadron $h$,  with mass $M_h$, produced from a
transversely polarized quark ($q^\uparrow$) is defined 
in terms of the leading-twist unpolarized $D_1^q$, 
and Collins $H_1^{\perp q}$ fragmentation functions, as~\cite{PhysRevD.70.117504} 
\begin{equation}\label{eqn:Ndensity}
D_{h}^{q^\uparrow}(z,\PperpB) = D_1^q (z, \Pperp^2) \\
+ H_1^{\perp q}(z,\Pperp^2) \frac{(\hat{{\boldk}} \times \PperpB)\cdot \Sq}{zM_h}.
\end{equation}

\begin{figure}[!htb]
\centering
 \includegraphics[width=0.42\textwidth]{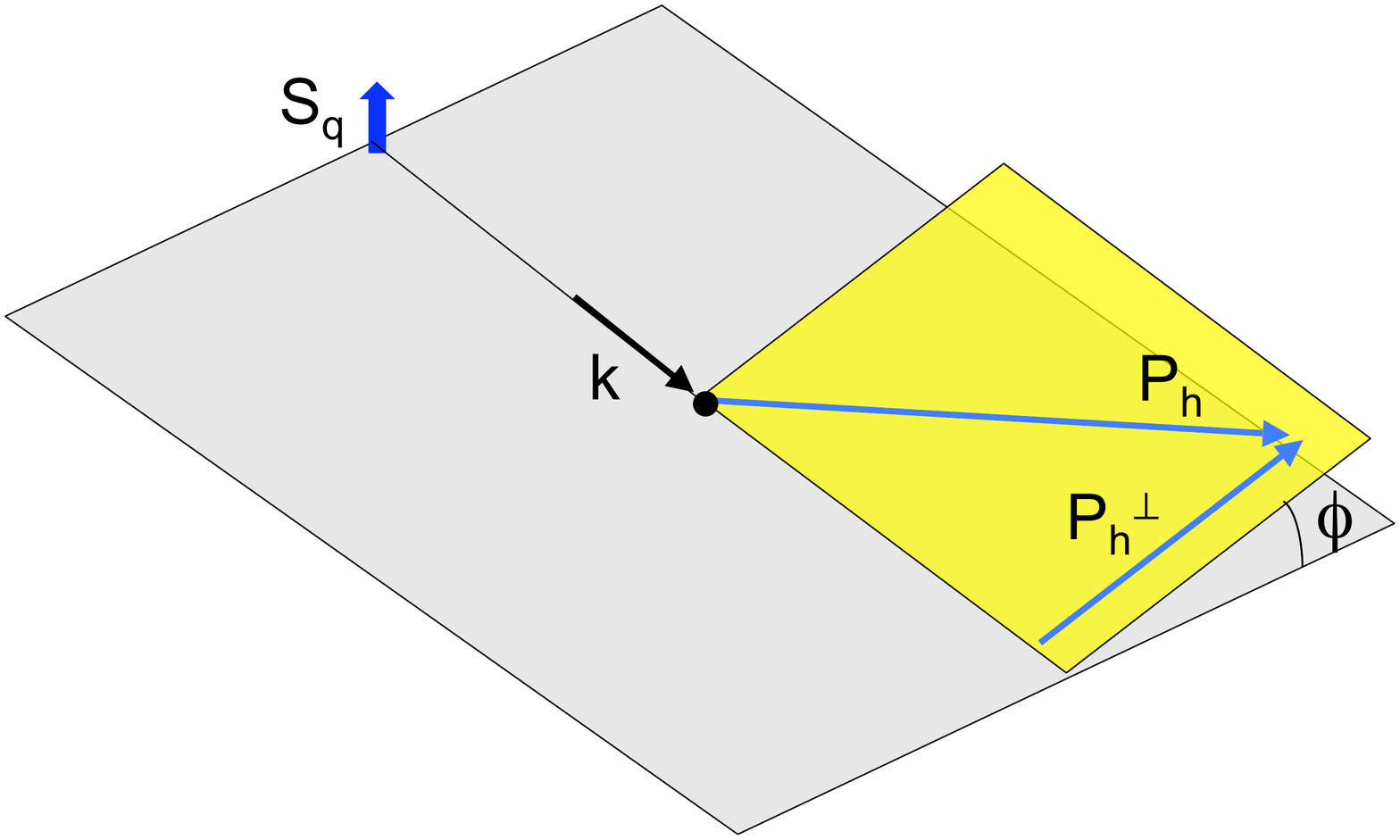}
\vspace{-0.5cm}
\caption{ (color online).
Definition of the azimuthal angle $\phi$
for a  quark with transverse spin $\Sq$ 
 which fragments into a spinless hadron of momentum \textbf{P}$_\mathrm{h}$ with a
component \PperpB\  transverse to the quark momentum
\kbold.
}
\label{fig:PhiAngle}
\end{figure}

The term  containing the Collins function depends on the spin vector of the quark
\Sq, and introduces an azimuthal asymmetry in the distribution of  hadrons around the quark
three-momentum direction $\hat{{\boldk}}$.
The triple product of Eq.~(\ref{eqn:Ndensity}), in fact,
produces a $\cos\phi$ modulation, where $\phi$ is
the azimuthal angle between the plane perpendicular to the 
quark spin, and the plane determined by  
\PperpB and $\hat{{\boldk}}$, as shown in Fig.~\ref{fig:PhiAngle}. 
In the literature, the amplitude of this modulation is called 
the Collins asymmetry or the Collins effect. 

Experimental evidence for a non-zero Collins function
was found by the HERMES~\cite{Airapetian201011,PhysRevLett.94.012002}
 and COMPASS~\cite{Adolph2012376,Alekseev2010240} collaborations,
from the analysis of semi-inclusive deep inelastic scattering (SIDIS)
of leptons off transversely polarized hydrogen and/or deuteron targets.
The cross sections for these processes are proportional to the
convolution of the Collins function and the chiral-odd 
transversity function~\cite{Ralston:1979ys}.
The latter is the least well known among the twist-two
parton distribution functions of the nucleon, and plays a crucial role
in understanding the spin properties.
In order to extract the transversity and the Collins functions
from SIDIS data, independent measurement of at least one of them is needed.

In \epem annihilation to a quark-antiquark pair, each quark spin is unknown:
the Collins asymmetry in a single jet ($ q\to hX$) will be zero.
However, the Collins effect can be observed when the
products of the quark and antiquark fragmentation are simultaneously considered. 
Experimentally, this is achieved by studying the  process $\epem\to\qqbar\to h_1 h_2 X$,
where $h_{1}$ ($h_2$) is a spinless hadron ($\pi$ or $K$) 
coming from the fragmenting quark $q$ ($\bar{q}$).
Events with a two-jet topology are selected, and
the correlation between the azimuthal angles of 
pairs of hadrons detected in opposite hemispheres
with respect to the plane perpendicular to the jet axis is analyzed.
The resulting azimuthal asymmetry is proportional to the product of two Collins functions. \\
\indent The first efforts to measure  Collins asymmetries 
in \epem annihilation, by studying $Z^0\to2 \,\mbox{jets}$ events,
were performed by the DELPHI Collaboration~\cite{efremov-1999-74},
while the first observation was obtained by the Belle
Collaboration~\cite{PhysRevLett.96.232002,PhysRevD.78.032011,PhysRevD.86.039905},
from a study of inclusive production of charged pion pairs at a
center-of-mass (c.m.) energy of approximately 10.6 \gev.
Assuming the universality of the Collins function~\cite{Metz2002139,PhysRevLett.93.252001},
a global analysis of SIDIS and \epem annihilation data has been carried out
by the authors of Refs.~\cite{PhysRevD.75.054032,Anselmino200998}, 
allowing the simultaneous extraction of the transversity  and Collins functions
for the pion system.

We report the measurements of the azimuthal modulation 
due to the Collins effect (Collins asymmetries) in the process
$\epem\to\qqbar\to\pi\pi X$, where $\pi$ stands for charged pion and  \q for a light quark: $u,\, d,\, s$.
We reproduce the Belle analysis~\cite{PhysRevD.78.032011,PhysRevD.86.039905}
of the azimuthal asymmetries as a function of the pions fractional energy in two reference frames.
We also perform a new measurement of the asymmetries 
as a function of the transverse momentum $p_t$ of pions with respect to the analysis axis.

\section{Analysis overview}\label{sec:analysis}
Charged pions are selected in opposite jets of hadronic events according  to the 
thrust axis of the event~\cite{Brandt196457,PhysRevLett.39.1587},
which permits the identification of  two hemispheres
(called 1 and 2, respectively, along and opposite to the thrust axis direction)  and
 to label the two pions as $\pi_1$ and $\pi_2$.
The analysis is performed in two convenient reference frames:
the thrust reference frame, defined in Sec.~\ref{subsec:rf12},
and  the second hadron momentum frame, defined in
Sec.~\ref{subsec:rf0}. This choice follows the scheme outlined 
by authors of Refs.~\cite{boer-1997-504, Daniel200923}, 
and allows a direct comparison of our results with the Belle measurements.
Section~\ref{sec:detector} summarizes the data sets used,
 while Sec.~\ref{sec:selection} describes in detail  the event and track selection.
The analysis method is discussed in Secs.~\ref{sec:rawasy} and~\ref{sec:dr}.
Dilutions of the asymmetries induced by background sources and by detector effects
not related to the Collins function are discussed in Secs.~\ref{sec:bkgd} and~\ref{sec:weight}, respectively.
Studies of possible systematic effects  are  summarized in Sec.~\ref{sec:syst}, while the 
final results on Collins asymmetry for light quark fragmentation are reported in Sec.~\ref{sec:results}. 

\subsection{Thrust reference frame: RF12}\label{subsec:rf12}
\begin{figure}[!htb]
 \begin{center}
  \includegraphics[width=0.48\textwidth]{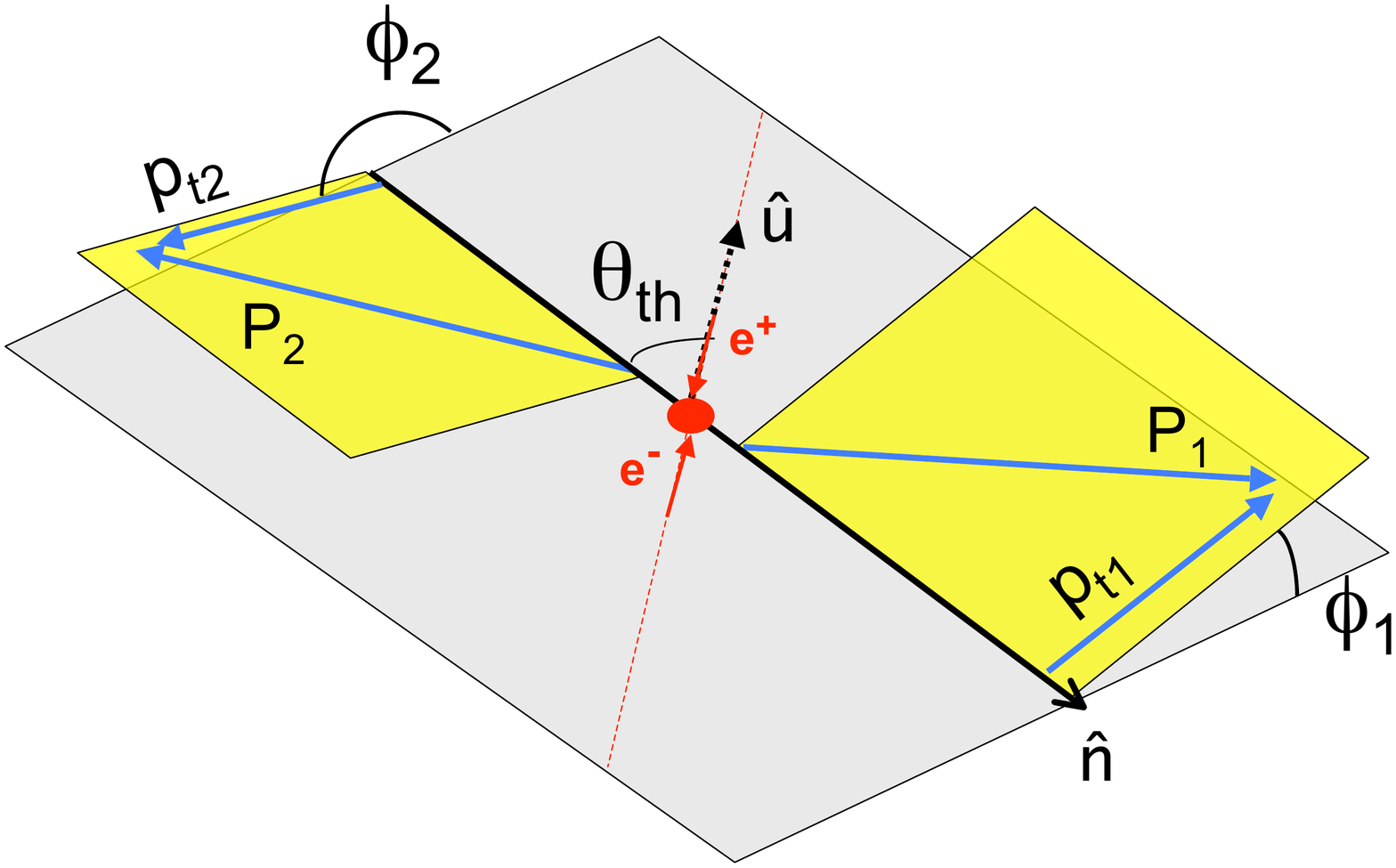}
\caption{ (color online).
Thrust reference frame (RF12).
The azimuthal angles $\phi_1$ and $\phi_2$ are the 
angles between the scattering plane and the
transverse pion momenta $\mathbf{p}_{ti}$ around the
thrust axis. 
The polar angle $\theta_{th}$ is defined as the angle between
the beam axis and the thrust axis $\bf{\hat{n}}$.
The pion transverse momenta $\mathbf{p}_{ti}$ used in the analysis 
differ from the corresponding  $\PperpB$, which refer to the true \qqbar direction. }
\label{fig:rf12}
\end{center}
\end{figure}

As mentioned in Sec.~\ref{sec:Introduction}, the Collins asymmetry ma\-ni\-fests itself as an azimuthal modulation of two
final state pions around the fragmenting quark-antiquark momentum.
The \qqbar direction is not accessible to a direct 
measurement and is approximated by the thrust axis of the event~\cite{PhysRevLett.39.1587}.
The kinematics in the \epem c.m. system corresponding to $\epem\to\pi_1\pi_2 X$, together 
with the definition of the two azimuthal angles, are schematically represented in Fig.~\ref{fig:rf12}.
 We refer to this frame as the thrust reference frame or RF12,
 since the thrust axis serves as reference axis for the azimuthal angles.
The correlation of the quark and antiquark Collins functions in opposite 
hemispheres gives a product of two modulations for the azimuthal angles $\phi_1$ and $\phi_2$, resulting in a 
$\cos(\phi_1+\phi_2)$ modulation.
The azimuthal angles are defined as
\begin{equation} \label{cross12}
\begin{split}
\phi_{1,2}  = \mathrm{sign}&[\hat{\mathbf{n}} \cdot \{ (\hat{\mathbf{u}} \times \hat{\mathbf{n}}) \times (\hat{\mathbf{n}} \times \hat{\mathbf{P}}_{1,2}) \} ]  \\
& \times \; \arccos \left( \frac{\hat{\mathbf{u}} \times \hat{\mathbf{n}}}{|\hat{\mathbf{u}} \times \hat{\mathbf{n}}|} \cdot \frac{\hat{\mathbf{n}} \times \mathbf{P}_{1,2}}{|\hat{\mathbf{n}} \times \mathbf{P}_{1,2}|} \right) , 
\end{split}
\end{equation}
where $\hat{\mathbf{u}}$ is a unit vector defined along the direction of the electron beam,  
$\hat{\mathbf{n}}$ is the thrust axis, and $\mathbf{P}_{1,2}$ is 
the three-momentum vector of the pion detected in the first ($\pi_1$) or in the second ($\pi_2$) hemisphere.\\
The differential  cross section depends on the fractional energies $z_1$ and $z_2$ of the two pions,
and on the sum of the azimuthal angles $\phi_1$ and $\phi_2$.
It can be written as~\cite{Daniel200923}:
\begin{widetext}
 \begin{align} \label{eqn:cross12} \nonumber
 \frac{\de\sigma(\epem\to\pi_1\pi_2 X) }{\de z_1 \de z_2 \de\phi_1 \de\phi_2
   \de\cos\theta_{th}} = &  \sum_{q,\overline{q}} \;\frac{3\alpha^2}{s} \;
 \frac{e^2_q}{4} \; z_1^2 z_2^2 \; 
\times \\ & \{  (1+\cos^2\theta_{th})\, D_1^{q,[0]} (z_1) \overline{D}_1^{q,[0]} (z_2)  
  + \sin^2\theta_{th} \; \cos(\phi_1+\phi_2)\, H_1^{\perp q,[1]} (z_1)
 \overline{H}_1^{\perp q,[1]} (z_2)  \},
 \end{align}
 \end{widetext}
where the summation runs over all quark flavors accessible at the c.m. energy $\sqrt{s}$,
$e_q$ is the charge of the quark \q in units of $e$, and the antiquark fragmentation function is denoted by a bar.
The so-called transverse moments of the fragmentation functions are defined as~\cite{Daniel200923}:
\begin{equation}
  F^{[n]} (z)= \int d | \mathbf{k}_T^2  | \left( \frac{|\mathbf{k}_T|}{M_\pi} \right)^n F(z, \mathbf{k}_T^2) \, ,
\end{equation}
with $F\equiv D_1^{q}$, $\overline{D}_1^{q}$, $H_1^{\perp q}$, and $\overline{H}_1^{\perp q}$.
In this equation, the pion transverse momentum has been rewritten in terms
of the quark intrinsic transverse momentum $\boldk_T$\footnote{Throughout the paper
we use the subscript $T$ to denote the transverse component of a vector
to the dipion axis in the frame where they are collinear, while the superscript $\perp$ indicates
the component transverse to the \qqbar axis.}:
$\PperpB=z\boldk_T$, and $M_\pi$ is the pion mass.
The Collins asymmetry can be extracted by measuring the
cosine modulation of the distribution of the quantity $(\phi_1+\phi_2)$
on top of the uniform distribution due to the unpolarized part
of the fragmentation function.
Dividing the full $(\phi_1+\phi_2)$ range into one hundred intervals,
we define the normalized azimuthal distribution as 

\begin{equation} \label{eqn:norm12}
R_{12}(\phi_1+\phi_2)=\frac{N(\phi_1+\phi_2)}{\langle N_{12}\rangle},
\end{equation}
with  $N(\phi_1+\phi_2)$ the di-pion yield in each $(\phi_1+\phi_2)$ 
subdivision, and  $\langle N_{12}\rangle$ the average bin contents.

\subsection{Second-pion reference frame: RF0}\label{subsec:rf0}
\begin{figure}[!htb]
 \centering
  \includegraphics[width=0.48\textwidth]{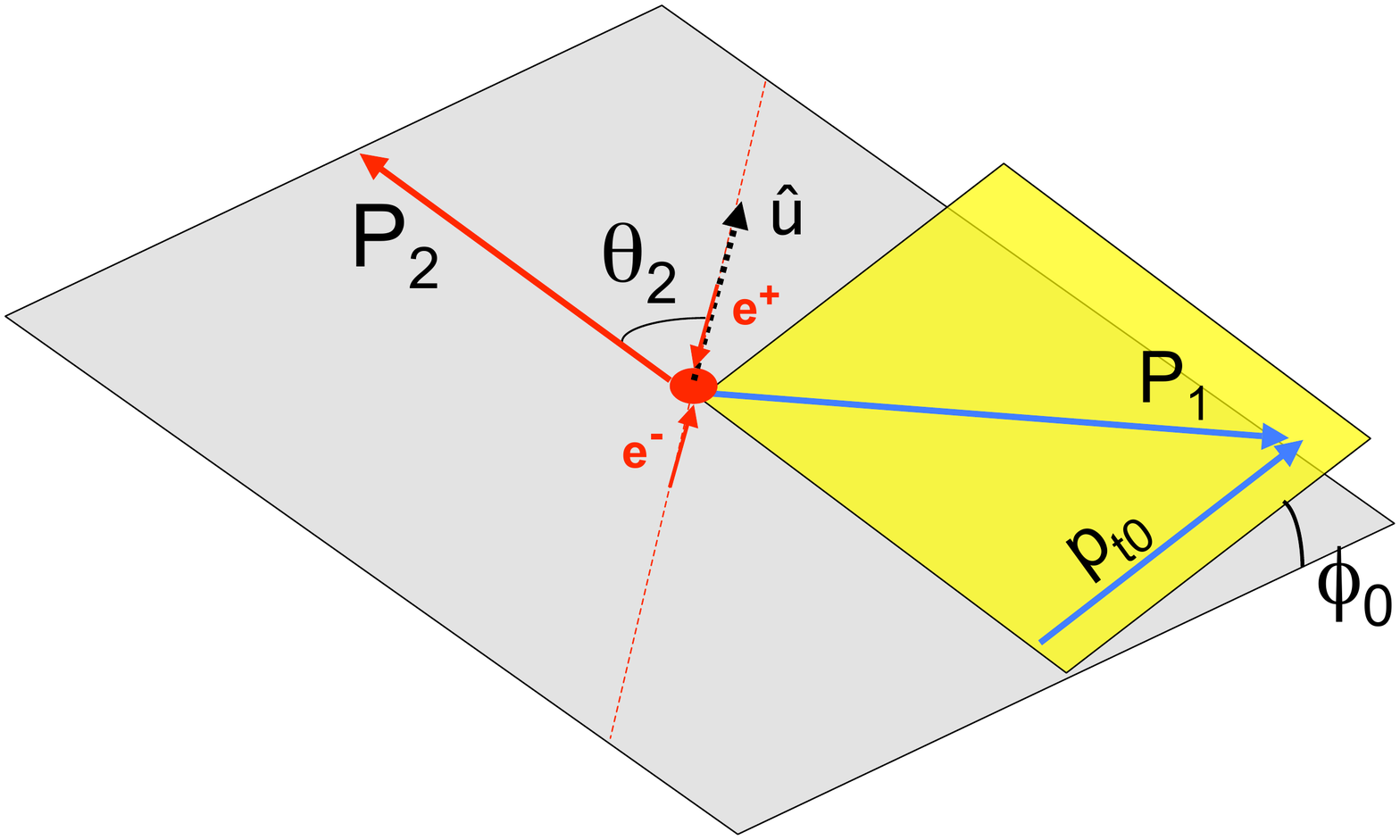}
\caption{ (color online).
Second-pion reference frame (RF0).
The azimuthal angle $\phi_0$ is defined as the angle between 
the plane spanned by the beam axis and the second pion momentum $\mathbf{P}_{2}$,
and the transverse momentum $\mathbf{p}_{t0}$ 
of the first pion around the second-pion direction.
The polar angle $\theta_2$ is defined as the angle between 
the beam axis and the momentum $\mathbf{P}_{2}$ of the second pion.}
\label{fig:rf0}
\end{figure}

The azimuthal asymmetries can also be measured in a different 
reference frame: following Ref.~\cite{boer-1997-504},
we use the direction of the observed pion $\pi_2$ as the reference axis,
and we define the scattering plane by the beam axis and the  momentum $\mathbf{P}_2$ of that pion,
as illustrated in Fig.~\ref{fig:rf0}.
Also in this frame, the kinematic variables are calculated in the \epem c.m. system, but
only one azimuthal angle, $\phi_0$, is defined:
\begin{equation} \label{form:0}
\begin{split}
\phi_0  = \mathrm{sign}&[ \mathbf{P}_{2} \cdot \{ (\hat{\mathbf{u}}
\times \mathbf{P}_{2} ) \times  ( \mathbf{P}_{2} \times
\mathbf{P}_{1}  )\} ] \\
 & \times \arccos \left( \frac{\hat{\mathbf{u}}
    \times \mathbf{P}_{2}}{|\hat{\mathbf{u}} \times \mathbf{P}_{2}|}
  \cdot \frac{ \mathbf{P}_{2} \times \mathbf{P}_{1}}{|
    \mathbf{P}_{2} \times \mathbf{P}_{1}|}  \right).  
\end{split}
\end{equation}
We refer to this frame as the second-pion reference frame, or RF0.
At leading order in the strong coupling $\alpha_s$,
the differential cross section is given by~\cite{boer-1997-504}
\begin{widetext}
\begin{align} \label{cross0} \nonumber
 \frac{\de\sigma (\epem\to\pi_1\pi_2 X)}{\de z_1 \de z_2
   \de^2 \qt \de \cos(\theta_2)d\phi_0} = \frac{3\alpha^2}{s}\; 
\frac{z_1^2 z_2^2}{4}  \times \bigg\{ & 
(1+\cos^2\theta_2) \;  \mathcal{F}(D_1 (z_1)  \overline{D}_1 (z_2)) + 
    \sin^2\theta_2\; \cos(2\phi_0) \\ 
&  \left. \times \mathcal{F}  \left[ (2\hat{\mathbf{h}} \cdot  \mathbf{k}_T \; 
\hat{\mathbf{h}} \cdot \mathbf{p}_T -\mathbf{k}_T \cdot \mathbf{p}_T ) 
 \frac{H_1^\perp (z_1)  \overline{H}_1^\perp (z_2)}{M^2_\pi}   \right] \right\},
\end{align}
\end{widetext}
where $|\qt|=Q_t$ is the transverse momentum of the virtual photon 
from \epem annihilation in the frame where $P_1$ and $P_2$ are collinear~\cite{Daniel200923}. 
$\mathcal{F}$ is a convolution integral over the transverse momenta
$\mathbf{P}_{1}^\perp=z_1\boldk_T$ and $\mathbf{P}_{2}^\perp=z_2\boldp_T$,
with $\boldk_T$ and $\boldp_T$ the transverse momenta
of the two fragmenting quarks:
\begin{align}\nonumber
\mathcal{F}(X\overline{X}) = \sum_{a,\bar{a}} e^2_a  \int & d^2 \mathbf{k}_T d^2 \mathbf{p}_T \delta^2 ( \mathbf{p}_T + \mathbf{k}_T - \qt   )\\ 
& X(z_1,z_1^2\mathbf{k}_T^2) \overline{X}(z_2,z_2^2\mathbf{p}_T^2),
\end{align}
and $\mathbf{\hat{h}}$ is the unit vector in the direction 
of the transverse momentum of the first hadron relative to the axis 
defined by the second hadron.

In this frame, the modulation due to the Collins effect 
is in the cosine of twice the azimuthal angle $\phi_0$, and the normalized distribution
is defined as
\begin{equation} \label{eqn:norm0}
R_{0}(2\phi_0)=\frac{N(2\phi_0)}{\langle N_{0}\rangle}.
\end{equation}

The  differential cross sections in Eq.~(\ref{eqn:cross12})
and Eq.~(\ref{cross0}) for the two reference frames are related to each other.
Integrating the first equation over 
$\phi_1$ and $\phi_2$, and the second over $\phi_0$ and $\qt$,
the same unpolarized cross section is obtained.
A similar relation exists for the Collins contributions.
However, due to the additional convolution over the transverse momentum $\qt$,  
the intrinsic transverse momentum dependence of the Collins function
should be known to exploit it.
The majority of authors assume that the Collins function is
a Gaussian in $k_T$, as is the unpolarized fragmentation function,
with a different width from the unpolarized case. 
More details can be found, for example, in
Refs.~\cite{PhysRevD.73.014020,PhysRevLett.85.712,PhysRevD.73.094025}.

\subsection{Unlike, Like, and Charged pion pairs}\label{subsec:ULC}

The possibility to select pion pairs with equal or opposite charges
allows sensitivity to ``favored" and ``disfavored" fragmentation processes.
A favored fragmentation function describes the fragmentation of a quark of flavor $q$ into
a hadron containing a valence quark of the same flavor:
for example, $u\rightarrow\pi^+$ and $d\rightarrow\pi^-$.
Vice versa, we refer to $u\rightarrow\pi^-$ and $d\rightarrow\pi^+$ as
disfavored fragmentation processes.
Therefore, the production of pion pairs with opposite charge
from the fragmentation of a $\uubar$ pair ($\epem\to\uubar\to\pi^\pm\pi^\mp X$)
can proceed through two favored fragmentation processes ($u\to\pi^+$ and $\overline{u}\to\pi^-$)
or through two disfavored fragmentation processes
($u\to\pi^-$ and $\overline{u}\to\pi^+$).
Neglecting heavy quark fragmentation and
introducing the favored functions 
$ D^{\rm{fav}}(z) = D_u^{\pi^+}(z) = D_d^{\pi^-}(z)$, and
$ \overline{D}^{\rm{fav}}(z) = D_{\bar{u}}^{\pi^-}(z) = D_{\bar{d}}^{\pi^+}(z)$,
as well as the disfavored functions
$D^{\rm{dis}}(z)=D_u^{\pi^-}(z)=D_d^{\pi^+}(z)=D_s^{\pi^\pm}(z)$ and
$\overline{D}^{\rm{dis}}(z)=D_{\bar{u}}^{\pi^+}(z)=D_{\bar{d}}^{\pi^-}(z)=D_{\bar{s}}^{\pi^\pm}(z)$,
 the cross section for charged pion pair production can be written 
 as~\cite{PhysRevD.73.094025}:

\begin{widetext}
\begin{align}\nonumber
 N^{U}(\phi)&=\frac{\de\sigma (e^+e^- \rightarrow \pi^{\pm} \pi^{\mp} X)}{\de\Omega \de z_1 \de z_2} 
\propto  \frac{5}{9}D^{\rm{fav}}(z_1) \overline{D}^{\rm{fav}}(z_2) + \frac{7}{9}D^{\rm{dis}}(z_1) \overline{D}^{\rm{dis}}(z_2) \\ \nonumber
 N^{L}(\phi)& =\frac{\de\sigma (e^+e^- \rightarrow \pi^{\pm} \pi^{\pm} X)}{\de\Omega \de z_1 \de z_2} \propto  \frac{5}{9}D^{\rm{fav}}(z_1) \overline{D}^{\rm{dis}}(z_2) + \frac{5}{9} D^{\rm{dis}}(z_1) \overline{D}^{\rm{fav}}(z_2) + \frac{2}{9}D^{\rm{dis}}(z_1) \overline{D}^{\rm{dis}}(z_2)  \\ 
N^{C}(\phi)& = \frac{\de\sigma (e^+e^- \rightarrow \pi \pi X)}{\de\Omega \de z_1 \de z_2}  = N^{U}(\phi) + N^{L}(\phi)  \propto 
 \frac{5}{9}[D^{\rm{fav}}(z_1) + D^{\rm{dis}}(z_1)] [\overline{D}^{\rm{fav}}(z_2) + \overline{D}^{\rm{dis}}(z_2)] + 
  \frac{4}{9}D^{\rm{dis}}(z_1) \overline{D}^{\rm{dis}}(z_2) 
\end{align}
\end{widetext}
where $\pi$ stands for a generic charged pion, 
$\phi$ is the azimuthal angle $\phi_1+\phi_2$ in RF12 or $\phi_0$ in RF0,
$\de\Omega=\de\phi \,\de\cos\theta$ with $\theta$ the polar angle
of the analysis axis, and the upper index indicates Unlike (U), Like (L) and Charged (C) sign pion pairs.

\section{ \babar\ experiment and data sample}\label{sec:detector}

The results presented here are based on a sample of data
collected with the \babar\ detector at the PEP-II
asymmetric-energy \epem collider, at the  SLAC National Accelerator Laboratory.
A total integrated luminosity of about 468~\invfb~\cite{Lees:2013rw}
is used, consisting of 424~\invfb collected at 
the peak of the \FourS resonance, and about 44~\invfb collected 40 \mev below the peak. 

The \babar\ detector is described in detail in references~\cite{Aubert:2001tu, newNIM}.
Charged particle momenta are measured by a combination of a 5-layer,
double sided silicon vertex tracker (SVT), and a 40-layer
drift chamber (DCH) that covers $92\%$ of the solid angle
in the c.m. frame, both located inside a 1.5 T superconducting solenoidal magnet.
Discrimination between charged pions, kaons, and protons
is obtained from measurements of the specific ionization (\dedx) 
in the tracking system, and from the Cherenkov light collected by an
internally reflecting ring-imaging Cherenkov detector (DIRC).
The DIRC covers $84\%$ of the c.m. solid angle in the central 
region of the \babar\ detector and has a $95\%$ ($91\%$) identification efficiency for pions  (kaons)
with momenta above $1.5$ \gev\footnote{Natural units are used throughout this article}.
Photons and electrons are identified and their energies measured with a high resolution CsI(Tl) crystal 
electromagnetic calorimeter (EMC).
Muons are identified in the instrumented flux return (IFR), which consists of 18 layers of steel interleaved 
with single-gap resistive plate chambers or limited-streamer tubes.

Detailed Monte Carlo (MC) simulation is used
to test and optimize the selection criteria, to study the 
detector acceptance, and to estimate the contribution of various  
background sources.
The simulation package \jetset~\cite{Sjostrand:1995iq} is used to
generate hadronic events in non-resonant \epem annihilation.
Separate MC samples are generated for light quarks,
$\epem\to\qqbar$ ($q=u,\,d,\,s$), called generic $uds$ MC,
and charm quarks, $\epem\to\ccbar$.
Samples of \BB events with generic \B decays are generated with the 
\evtgen~\cite{Lange:2001uf}  package.
Finally, \tautau and \mumu event samples are produced
with the \kktof~\cite{Ward:2002qq} generator,
and $\mumu\gamma$ events with \afkqed~\cite{akfqued}.
The generated events undergo a full detector simulation based on
\geant~\cite{Agostinelli:2002hh} and  are analyzed
in the same way as the experimental data. 
No transverse spin effects are implemented in the
MC generation, so it can be used to evaluate detector biases.
In addition, the \uds MC samples are reweighted to simulate
Collins asymmetries and to study the analyzing power of the method.

\section{Event and track selection}\label{sec:selection}
We focus on the measurement of the Collins effect in light quark fragmentation,
as the helicity is conserved only in the approximation of massless
quarks, and  the correlation between the fragmenting quark and
antiquark  may be lost for heavy quarks.
In this section, we summarize the event and track selection requirements.

Multi-hadronic events are selected by requiring at least three  reconstructed charged particles and the
value of the 2nd divided by the 0th Fox-Wolfram moment~\cite{Fox1979413},
calculated from charged tracks only, $R'_2<0.98$.
To  suppress  backgrounds  from $\epem\to\tautau$, $\gamma\gamma$ processes,
 and events characterized by emission of a very energetic photon via initial state radiation,
 we require the visible energy of the event in the laboratory frame ($E_{vis}$),
 defined as the sum of the energies of all reconstructed charged tracks and neutral candidates, 
 to be higher than 7 \gev.

Only good-quality reconstructed tracks with  momenta transverse to the beam direction
of at least 0.1 \gevc are considered for the asymmetry measurements.
Every track is required to originate from the vicinity
of the interaction point (IP) by requiring the distance  of closest approach to the IP
 in the transverse plane $d_{XY}<0.2$ cm and along the electron beam  $|d_Z|<1.5$ cm,
and to fall within the detector acceptance region: $0.41<\theta_{lab}<2.54$ rad, where
$\theta_{lab}$ is the polar angle of the track with respect to the beamline axis.

The thrust of the event is calculated using  tracks with relaxed cuts  
$d_{XY}<1.5$ cm and $|d_{Z}|<10$ cm, as well as neutral candidates lying 
within the calorimeter fiducial region  with  an energy greater than 0.030 \gev.
To avoid possible biases originating from the different
forward/backward detector configuration, the sign of the thrust axis is chosen at random.

\begin{figure}[!htb]
\centering
   \includegraphics[width=0.48\textwidth] {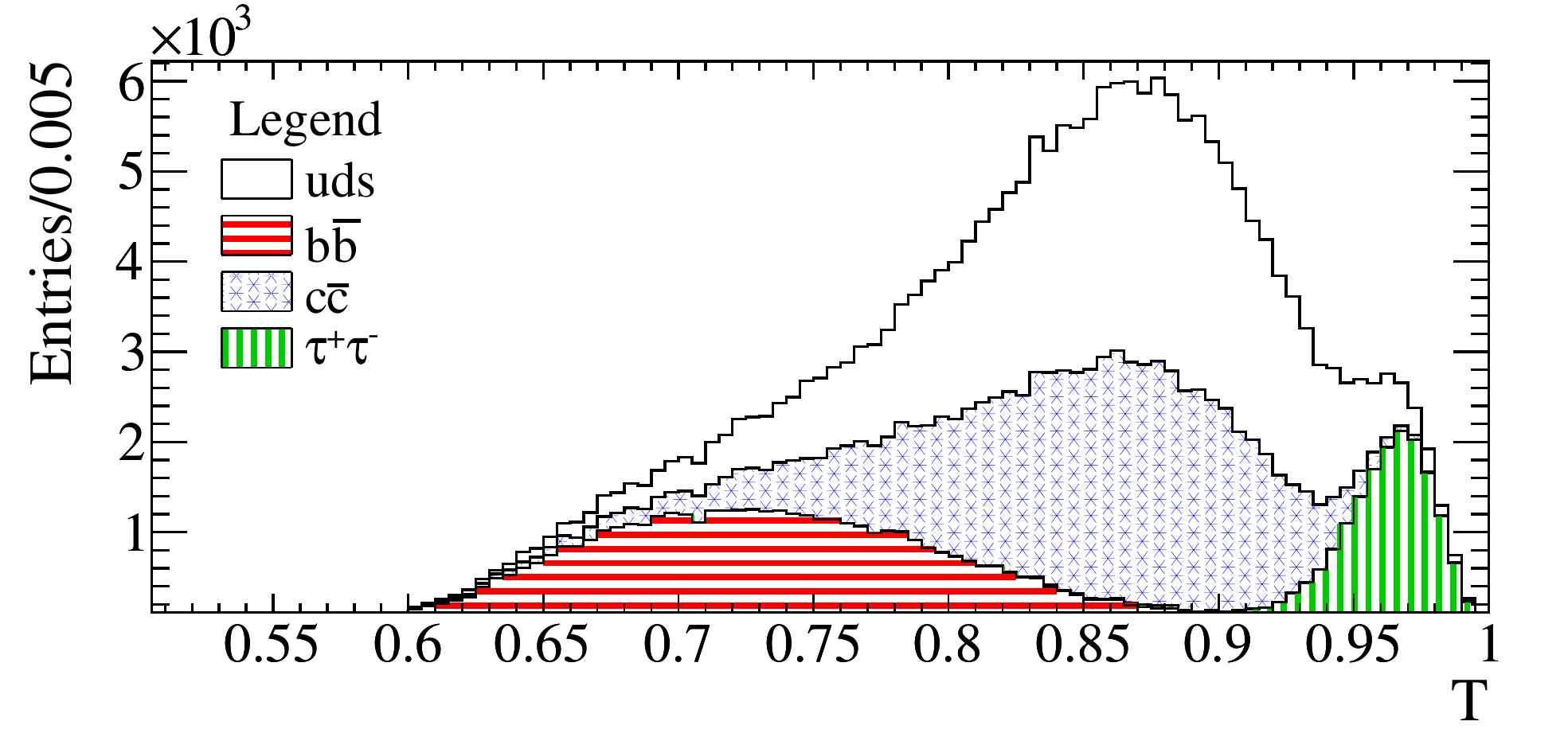} 
\caption{ (color online).
  Thrust distributions for simulated events under the
  \FourS resonance for multi-hadronic events
   with at least one pair of good quality tracks:
   $e^+e^- \rightarrow \BB$ (horizontal lines),
   $e^+e^- \rightarrow \ccbar$ (asterisks),  $e^+e^- \rightarrow \qqbar, \,q=uds$
    (white histogram)
   and $e^+e^- \rightarrow \tau\tau$ (vertical lines).
   The samples are normalized to an arbitrary luminosity.} 
\label{fig:thrust}
\end{figure}

Since the correlation between the \q and the \qbar spin is lost 
in the case of emission of energetic gluons, we  select the two-jet topology
and suppress $\epem\to\qqbar g$ events by requiring  a value of the event thrust $T>0.8$.
As shown in Fig.\,\ref{fig:thrust} the distribution of the thrust
for \uds events peaks at values higher than 0.85, but has a long tail
at lower values, which is mainly due to hard gluon radiation.
The requirement $T>0.8$ also removes the majority of the more spherical \BB events produced
in \FourS decays.
Events with charm quarks have a shape similar to the light
quarks; their contribution to the asymmetry is evaluated and
subtracted as described in Sec.~\ref{sec:bkgd}.

\begin{figure}[htb]
\begin{center}
     \includegraphics[width=0.46\textwidth]{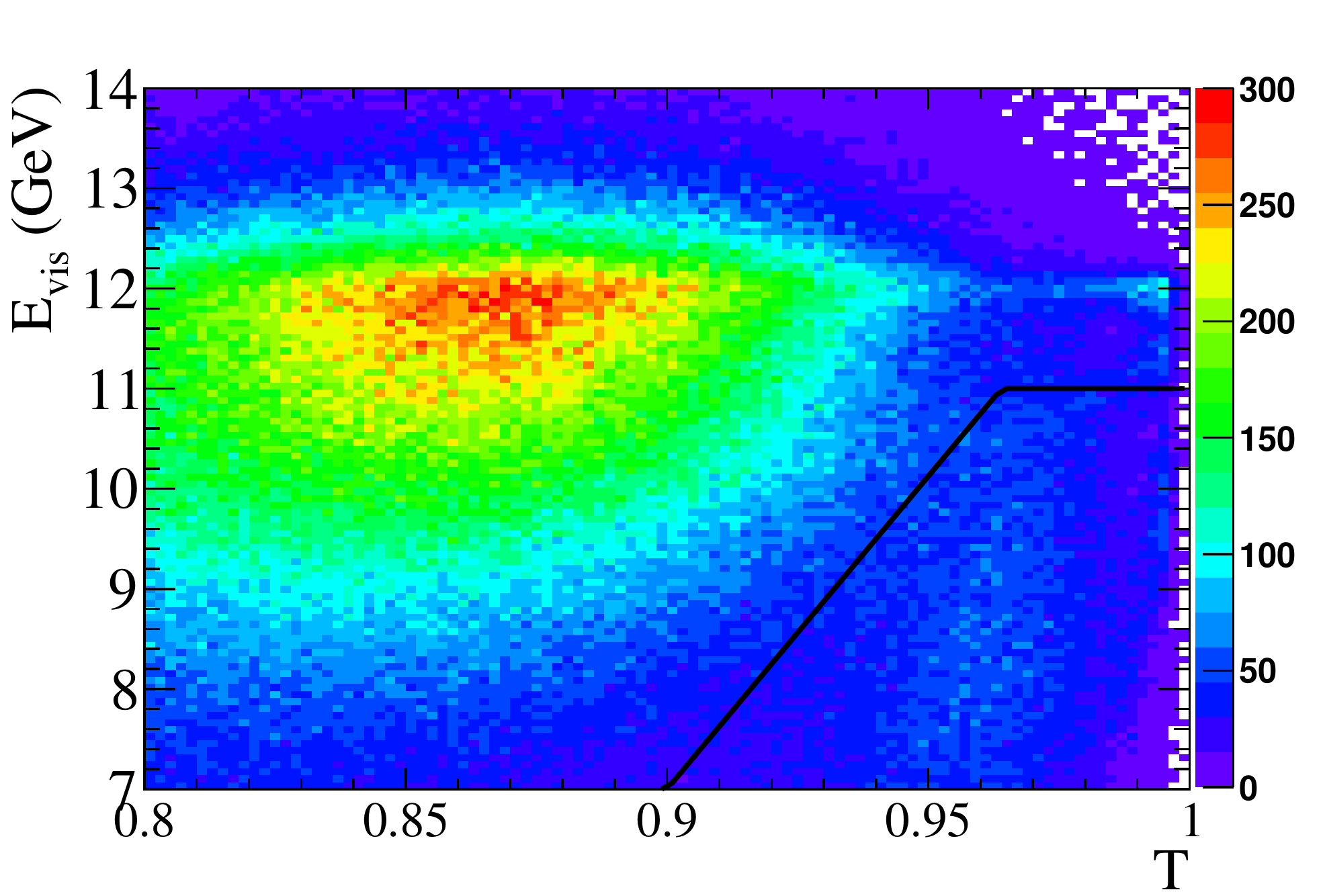}
 \caption{ (color online).
  Total visible energy of the event in the laboratory frame vs. the thrust
   value for the on-resonance data sample.
   The events at high thrust value and low total energy are due to the
   $\epem\to\taup\taum$ process. The black line is the cut applied in
   the analysis in order to remove this background. 
The peak at $E_{vis} \sim 12$~\gev and high thrust values,
   is due to radiative BhaBha and $\mumu(\gamma)$ events.}
 \label{fig:enthrust}
\end{center}
\end{figure}

Events from the $\epem\to\tautau$ reaction populate the region
at higher thrust values $T$ and lower $E_{vis}$, as is evident  
from Fig.~\ref{fig:enthrust}, which shows a scatter plot of $E_{vis}$ vs. $T$  
for events having at least one good hadron pair.
The small accumulation visible at lower energies and
$T>0.94$ is due to \tautau events, and it is removed 
by applying a cut around this region, as indicated by the line in Fig.~\ref{fig:enthrust}.

 Radiative $\epem\to\epem\gamma$ and $\epem\to\mumu\gamma$ events are the sources 
of the background peaking  at $E_{vis}=12$~\gev and concentrated in particular at high $T$.
 This kind of background is suppressed by requiring at least three charged hadrons in the event.
However, some $\mumu\gamma$ events, with the initial state photon
converting to a \epem pair, can pass this selection.
These events are characterized by small multiplicity and by two very energetic muons.
We reduce this contamination to a negligible level by  requiring that for 
events with multiplicity lower than five, the two most energetic tracks 
are not identified as muons, and no electrons are present.

 Reconstructed tracks in the selected events are used
 for the study of the Collins asymmetries if they are identified as pions
 and fail to pass specific muon and electron selectors.
 The efficiencies estimated for the latter are about 70\% and 98\%, and
 the pion mis-identification rate of about 2\% and 4\%,
 for muons and electrons, respectively.

Two-body decays of  $\bbbar$ bound states, mainly produced via initial state
radiation, generate a significant amount of unlike-sign pairs, 
with both tracks of c.m. momentum above 4.5 \gevc.
On the other hand, we expect to have very small signal from 
fragmentation processes with two such energetic tracks.
In particular no like-sign pairs are observed in the data sample
with $z_1$ and $z_2$  above 0.9.
We therefore limit the study to tracks with $z<0.9$.

The residual contributions of all other background sources ($\ccbar$,
$\BB$, and \tautau) are evaluated, and the measured 
asymmetry corrected as described in Sec.~\ref{sec:bkgd}.

 The fragmentation functions depend on the
 lightcone momentum fraction  $z$ of the produced hadron
 with respect to the fragmenting quark~\cite{boer-1997-504},
which is equivalent to the  
 fractional energy at large c.m. energy and not too small
 values of $z$~\cite{PhysRevD.75.054032},
  \begin{equation} \nonumber
  \frac{2E_{h}}{\sqrt{s}}=z+\frac{P_\perp^2}{zs}\simeq z.
\end{equation}
 It may be of interest to extend the study also for
very low $z$ values, in order to assess when this approximation fails.
On the other hand, low momentum tracks pose severe experimental difficulties
due to the  association of the hadrons to the incorrect jet.
For these reasons, the measurement of Collins asymmetry is performed only for
candidate pions with $z>0.15$.

The selected pions are separated in opposite hemispheres
according to the thrust axis ($\bf{\hat{n}}$), and are combined 
if they satisfy the following condition
\begin{equation}
W_{\rm{hemi}}=(\mathbf{P_1} \cdot \bf{\hat{n}}) (\mathbf{P_2} \cdot \bf{\hat{n}})<0,
\end{equation}
where $\mathbf{P_{1,2}}$ are the pions momenta.
For pairs with values of $W_{\rm{hemi}}$ near to zero there is a higher 
probability that one of the two tracks has been assigned to the wrong 
hemisphere. This effect is particularly evident for  pions with
low fractional energies.
The requirement that the pions are emitted within a cone of 
$45^\circ$ around the thrust axis removes the ambiguous tracks.

One of the most important contributions to azimuthal asymmetries 
not connected to the Collins effect originates from low energy gluon radiation $\epem\to\qqbar g$,
which is not completely removed by the event selection.
As reported in Refs.~\cite{Daniel200923,PhysRevLett.42.291},
the angular distribution of the gluon radiation process 
$\epem\to\qqbar g\to h_1 h_2 X$ is given by
\begin{equation}
\frac{dN}{d\Omega}\propto \frac{Q_t^2}{s+Q_t^2} \sin^2\theta \cos(2\phi).
\end{equation}
In addition, all the formalism used so far
is valid in the region where the transverse momentum $Q_t$
is small compared to $\sqrt{s}$ ($Q_t^2 \ll s$)~\cite{Daniel200923},
and a safe compromise is to require $Q_t<3.5$ \gevc. 

The same selection is applied to same-charge and opposite-charge
pion pairs. 
About $10^8$ pion pairs are selected and used in the analysis.

\section{Normalized azimuthal distributions}\label{sec:rawasy}

\begin{figure}[!htb]
\centering  
 \includegraphics[width=0.45\textwidth] {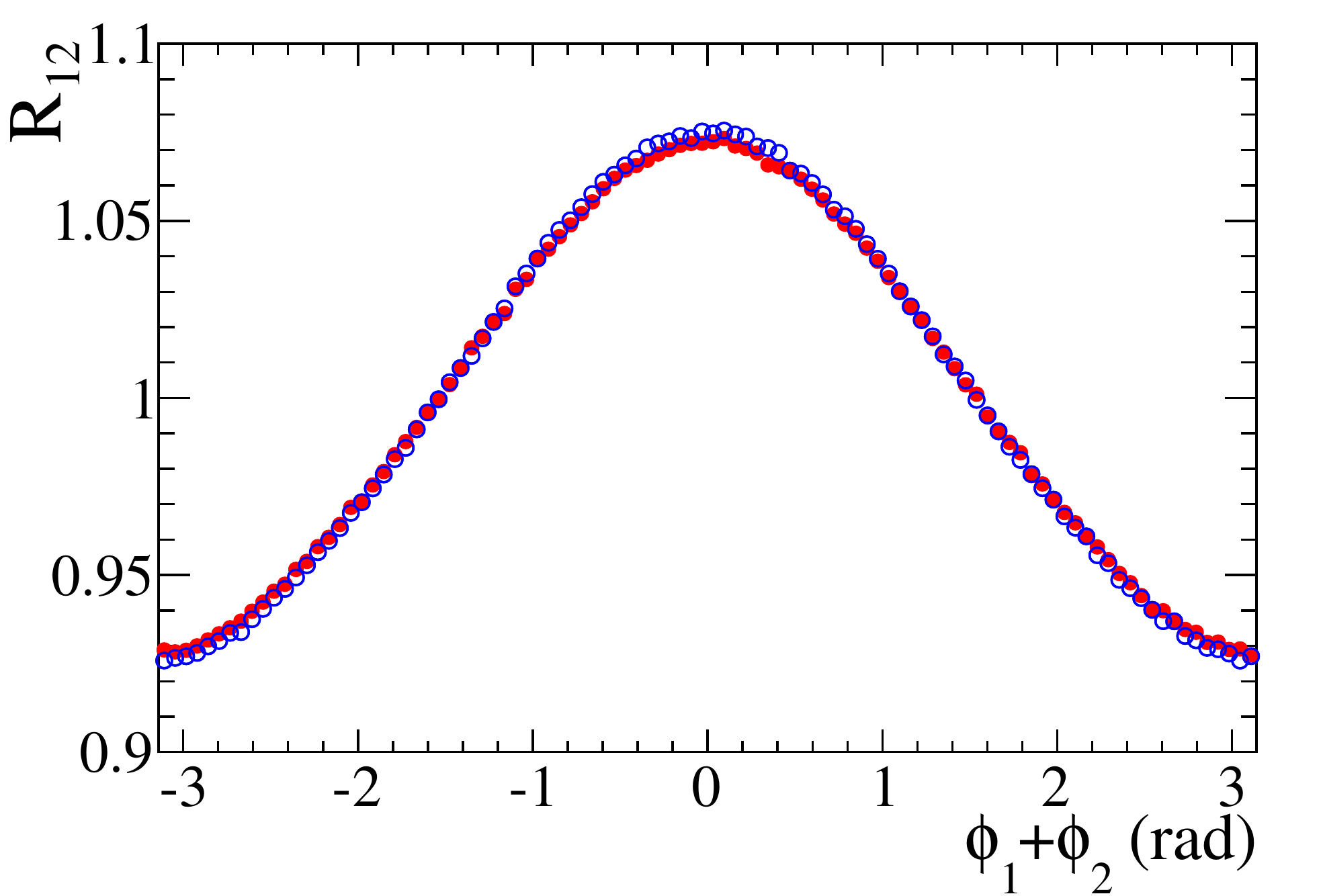} %
{\boldmath      \put(-50,120){(a)}    }   
\includegraphics[width=0.45\textwidth] {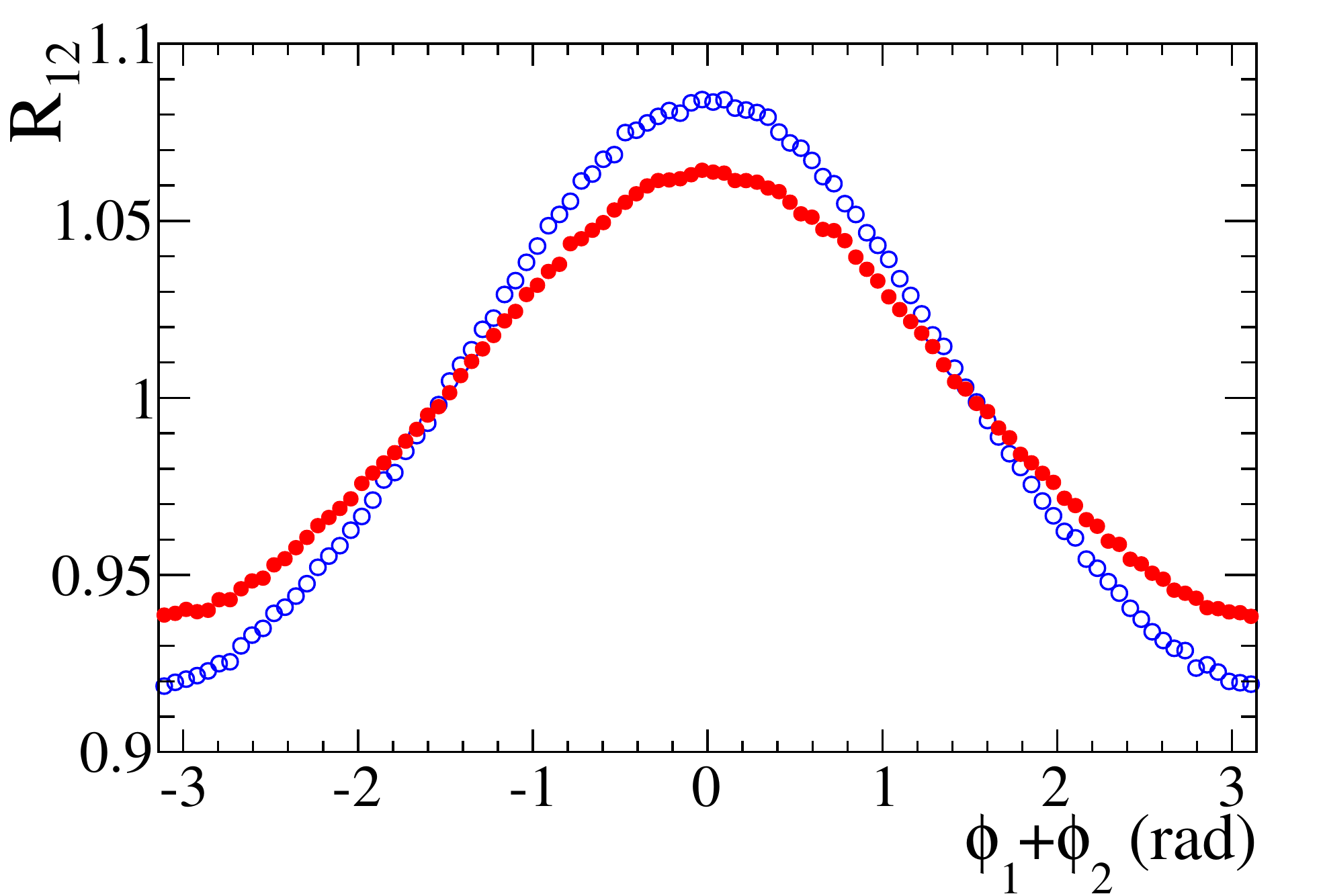} 
  {\boldmath      \put(-50,120){(b)}    }
  \caption{ (color online).
  Normalized azimuthal distributions for like-sign ($R^L$, full circles)
  and unlike-sign  ($R^U$, open circles) pion pairs,
  for  (a) MC simulation and (b) data, in RF12.} 
  \label{fig:rawasy}
  \end{figure}

Following Eqs.~(\ref{eqn:cross12}) and~(\ref{cross0}),
the azimuthal distributions of the normalized yields $R_\alpha$,
defined in Sec.~\ref{sec:analysis}, can be parametrized as
\begin{equation}
R^i_{\alpha}=b_\alpha+a_\alpha\cos(\beta_\alpha),
\label{eq:rawasy}
\end{equation}
where $\alpha=0,\, 12$ indicates the reference frame,
$i=U, \, L,\, C$ the charge combination of the pion pair,
and $\beta$ is the azimuthal angle combination $\phi_1+\phi_2$
or $2\phi_0$, according to the  frame used.
The parameter $b_\alpha$ should be consistent with unity,
while $a_\alpha$ gives the amplitude of the asymmetries.
The normalized azimuthal distributions, presented in Fig.~\ref{fig:rawasy} for MC
and data samples, are strongly affected by detector acceptances
and show apparent modulations.
This is clearly visible in the simulated sample, for which a flat
distribution is expected since the polarized \FF are not implemented
in the MC generator.
However, the $R^L$ and $R^U$ distributions
are almost coincident in the MC sample (Fig.~\ref{fig:rawasy}(a)), while a clear difference
is observed in data (Fig.~\ref{fig:rawasy}(b)).
This difference is the observable effect of the azimuthal asymmetry
produced by the Collins effect.

\begin{figure}[!htb]
\centering 
\includegraphics[width=0.5\textwidth] {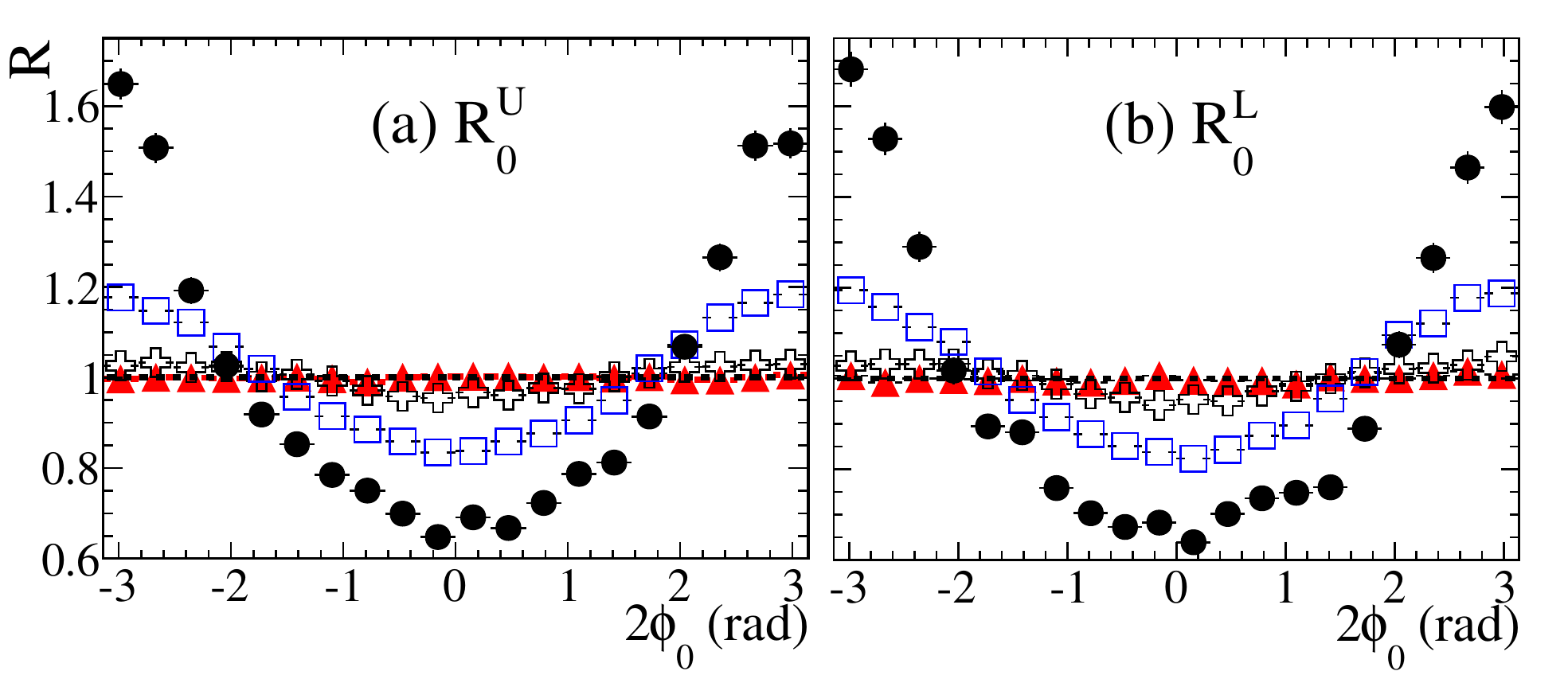}         
 \caption { (color online).
 Normalized azimuthal distributions for different intervals of
   $\cos\theta_{th}$   measured in the RF0 frame for unlike-sign (a) 
  and like-sign (b) pion pairs.
  The $\cos\theta_{th}$ intervals are as follows: 
  $0.8<\cos\theta_{th}<0.9$ for circles,
  $0.5<\cos\theta_{th}<0.7$ for  squares,
  $0.3<\cos\theta_{th}<0.5$ for  crosses,
  $0<\cos\theta_{th}<0.3$ for triangles.}
\label{fig:rawTheta}
\end{figure}

Detector effects depend on the jet direction.
When the \qqbar pair is created at low polar angle with respect to the beam axis,
there is a higher probability that part of the jet falls outside
the detector coverage, and the thrust can be badly reconstructed.
The result is a distortion of the distribution, as 
visible in Fig.~\ref{fig:rawTheta}, which shows $R^{U}$ and $R^{L}$ 
in the RF0 frame  for different intervals of $\cos(\theta_{th})$.
The same effect is also visible in the RF12 frame.
The triangles in Fig.~\ref{fig:rawTheta} also show the residual effects
of gluon radiation to be small.
We can parameterize the acceptance effects on the normalized distribution
as an additional contribution to the $\cos(\beta_\alpha)$ modulation,
whose amplitude varies with $\theta$:  
$  a_\alpha^{\epsilon}(\theta)$.
Therefore, Eq.\,(\ref{eq:rawasy}) becomes:
\begin{equation}
\begin{split} 
R^i_{\alpha} & =\, (1+a_\alpha^{\epsilon}(\theta)\cos(\beta_\alpha)) \cdot
(b_\alpha+a_\alpha\cos(\beta_\alpha)) \\
 & =\,  b_\alpha   + \left[ a_\alpha +  a_\alpha^{\epsilon}(\theta) b_\alpha \right]\cos(\beta_\alpha) + 
  a_\alpha  a_\alpha^{\epsilon}(\theta)  \cos^2(\beta_\alpha) \, ,
\end{split}
\label{eq:rawasy2}
\end{equation}
and shows a coupling between the Collins and detector acceptance
effects proportional to $\cos^2(\beta_\alpha)$.

In principle, it would be possible to estimate detector acceptance effects with
simulated events, and correct the asymmetries measured 
in the data sample, but this procedure would introduce large uncertainties.
All these considerations suggest the possibility to form a suitable
double ratio of azimuthal distributions,  
in order to reduce the effect of detector acceptance and 
perform a measurement almost independent from simulation.

\section{Double ratios}\label{sec:dr}
Given the difficulties in separating the Collins effect from asymmetries produced by
detector acceptances and radiative effects, we exploit 
the fact that most of the instrumental effects should largely cancel in ratios
of asymmetries, as for example, the ratio of unlike-sign over like-sign asymmetries:
\begin{equation}
\begin{split}
\frac{R_{12}^{U}}{R_{12}^L} & \simeq 
\frac{1+\left\langle\frac{\sin^2\theta_{th}}{1+\cos^2\theta_{th}} \right\rangle
 \,G^{U} \cos(\phi_1+\phi_2)}{1+\left\langle\frac{\sin^2\theta_{th}}{1+\cos^2\theta_{th}}\right\rangle
  \,G^{L}\cos(\phi_1+\phi_2)}  \\ 
& \simeq 1+\left\langle \frac{\sin^2\theta_{th}}{1+\cos^2\theta_{th}}\right\rangle 
\left\{ G^{U}-G^{L} \right\} \cos(\phi_1 + \phi_2).
\end{split}
\label{DR:UL}
\end{equation}
Here, $G^L$ and $G^U$ are, respectively:
\begin{equation}\label{eq:GUGL} 
\begin{split}
G^{U} & \propto \frac{5 H_1^{\rm{fav}} \overline{H}_1^{\rm{fav}}+ 7 H_1^{\rm{dis}}
  \overline{H}_1^{\rm{dis}}}{5 D_1^{\rm{fav}}\overline{D}_1^{\rm{fav}} +
  7 D_1^{\rm{dis}}\overline{D}_1^{\rm{dis}}}\,, \\  
G^{L} & \propto \frac{ 5 H_1^{\rm{fav}} \overline{H}_1^{\rm{dis}}+ 5 H_1^{\rm{dis}}
  \overline{H}_1^{\rm{fav}} + 2 H_1^{\rm{dis}} \overline{H}_1^{\rm{dis}} }{ 5 D_1^{\rm{fav}}\overline{D}_1^{\rm{dis}} +
  5 D_1^{\rm{dis}}\overline{D}_1^{\rm{fav}} + 2 D_1^{\rm{dis}} \overline{D}_1^{\rm{dis}} }\,.  
\end{split}
\end{equation}
where we omitted the $z$ and \pt dependence in order to simplify the notation.
The double ratio (DR) is performed after the integration
over the polar angle $\theta_{th}$, so that the average values
 of the quantity $\sin^2\theta_{th}/(1+\cos^2\theta_{th})$ appear. 
These average values do not differ for like-, unlike- and all
charged-pion pairs.
In Eq.\,(\ref{DR:UL}) we assume that the detector acceptance effects do
not depend on the charge combination of the pion pairs, that is 
$a^{\epsilon,L}(\theta_{th}) = a^{\epsilon,U}(\theta_{th})$. We also neglect
the extra term proportional to $\cos^2(\phi_1 + \phi_2)$, 
which couple the detector acceptance to the true Collins asymmetries,
and stop the series expansion at the first order in
$\cos(\phi_1 + \phi_2)$.
We have checked for the presence
of these and other terms in addition to the $\cos(\phi_1 + \phi_2)$
modulation and found them negligible.
Also the assumption of acceptance effects independent on the charge
combination of the pion pairs seems to hold, and  noting that also the
asymmetries produced by gluon radiation do not depend on the charge
combination,   the asymmetry amplitudes resulting from the double ratio
should mainly depend on a different combination  of favored and
disfavored fragmentation functions (see also discussion in
Sec.~\ref{sec:syst}).

Similarly,  the DR of the normalized distributions of
unlike-sign and  charged pion pairs is given by:
\begin{equation} \label{eqn:uc1}
\frac{R_{12}^{U}}{R_{12}^C} \simeq  1 +
\left\langle\frac{\sin^2\theta_{th}}{1+\cos^2\theta_{th}}\right\rangle
 \left\{ G^{U} - G^{C} \right\}    \cos(\phi_1+\phi_2), 
\end{equation}
with 
\begin{equation} \label{eqn:uc2}
G^{C}\propto \frac{5 (H_1^{\rm{fav}} + H_1^{\rm{dis}}) (\overline{H}_1^{\rm{fav}} +  \overline{H}_1^{\rm{dis}}) + 4 H_1^{\rm{dis}} \overline{H}_1^{\rm{dis}} }
{ 5 (D_1^{\rm{fav}} + D_1^{\rm{dis}}) (\overline{D}_1^{\rm{fav}} +  \overline{D}_1^{\rm{dis}}) + 4 D_1^{\rm{dis}} \overline{D}_1^{\rm{dis}}}. 
\end{equation}

The measured  U/L and U/C double ratios can be used to derive
information about the relative sign and magnitude of 
favored and disfavored fragmentation functions~\cite{PhysRevD.73.094025}.
Analogous expressions can be obtained in the RF0 reference frame,
with modulations in $\cos(2\phi_0)$ instead of $\cos(\phi_1+\phi_2)$.

The DRs are still parametrized by a cosine function
\begin{eqnarray} \label{eq:drfit}
  \frac{R_\alpha^i}{R_\alpha^j} = B_{\alpha}^{ij} + A_{\alpha}^{ij} \cdot \cos(\beta_\alpha)\,,
\end{eqnarray}
where $B$ and $A$ are free parameters.
 The constant term $B$ should be consistent with unity and 
 the parameter $A$, which depends on $z$, pt, and the average value
of $\sin^2\theta/(1+\cos^2\theta)$, should mainly contain the Collins effect.

\begin{figure}[!htb]
\centering
      \includegraphics[width=0.45\textwidth]{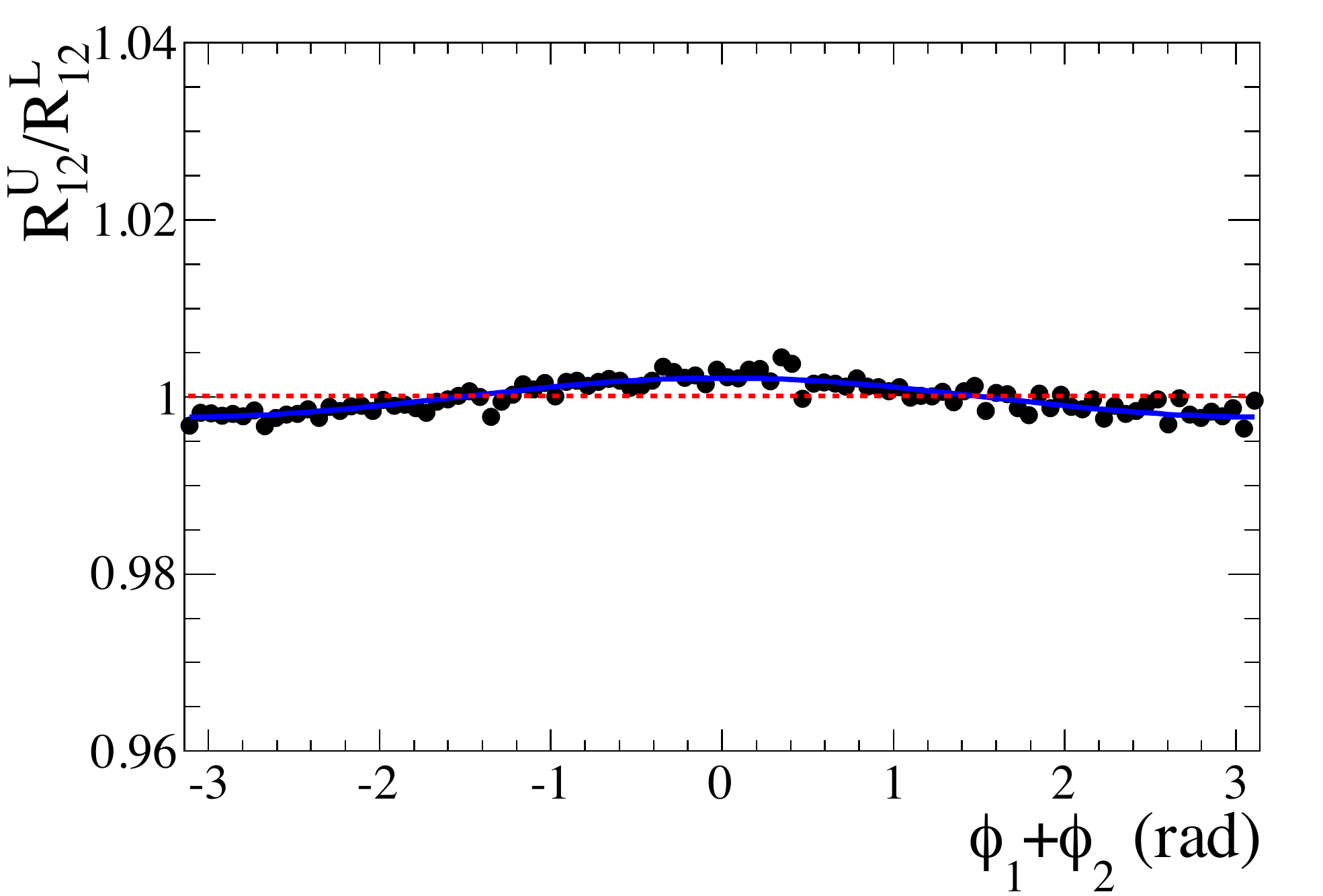} 
      {\boldmath      \put(-50,130){(a)}    }
  \includegraphics[width=0.45\textwidth]{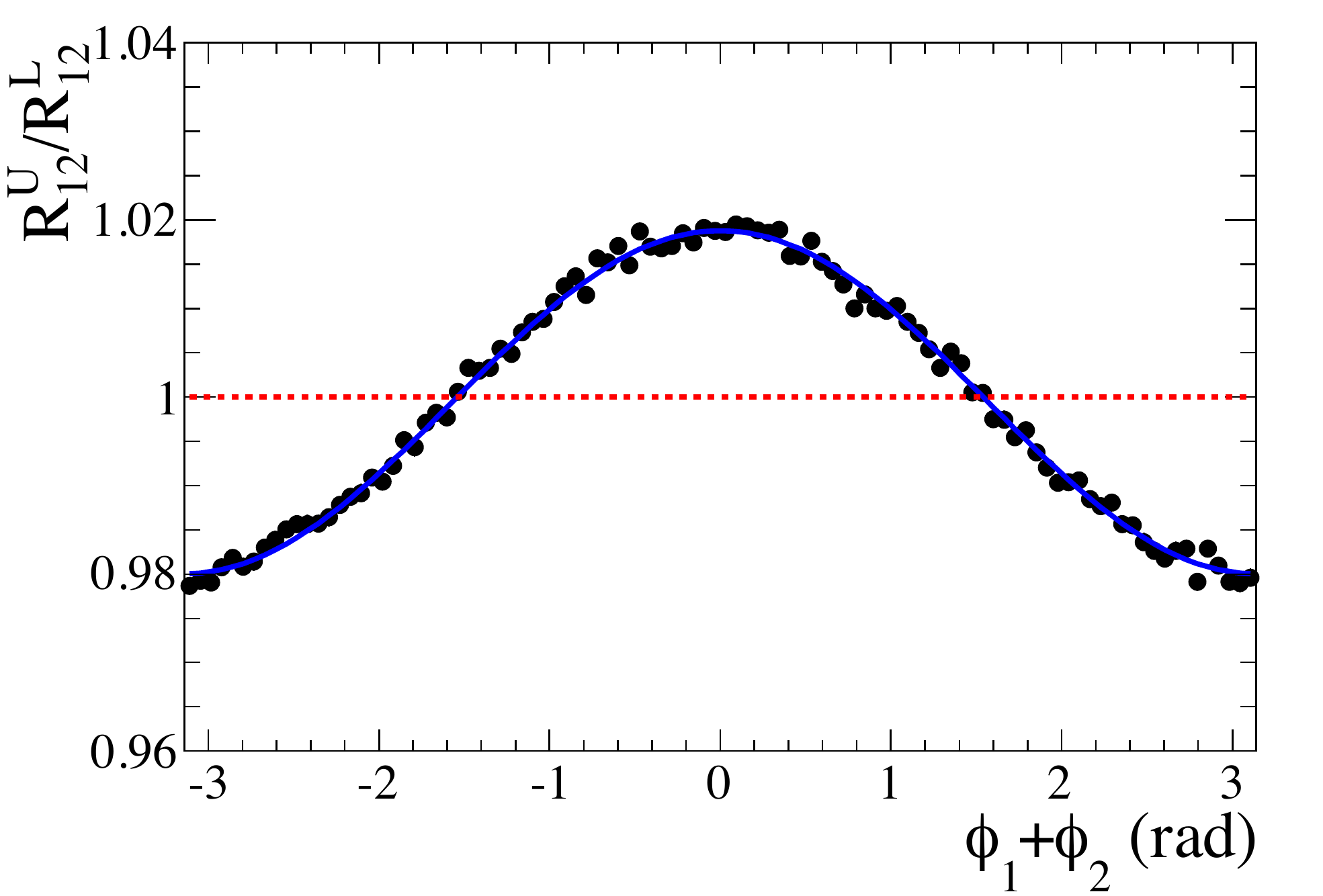}
   {\boldmath      \put(-50,130){(b)}    }
  \caption{
Double ratio of azimuthal distributions of unlike over like sign pion pairs 
for Monte Carlo (a) and data (b) samples, in the RF12 system.    
The solid lines are the result of the fits with the function reported in Eq.~\ref{eq:drfit}. 
    }
  \label{fig:DRtot}
\end{figure}

Figure~\ref{fig:DRtot} shows the DR of unlike to like sign pion pairs
for samples of simulated and data events.
The distribution for the MC sample is now essentially flat as expected;
however, a slight deviation from zero asymmetry, of the order of 0.2\%, is measured.
The origin and the effect of this  bias will be discussed in Sec.~\ref{sec:syst}a. 
A clear cosine modulation is instead visible in the data sample (Fig.~\ref{fig:DRtot}(b)),
which can be attributed to the Collins effect.

Thanks to the large amount of data, we can study the dependence of the asymmetry as a 
function of fractional energies ($z_1$ and $ z_2$)
and transverse momenta ($p_{t1}$ and $p_{t2}$, and $p_{t0}$) of the selected pions.
We choose $6\times6$ ($z_1,z_2$)-bins, with the following $z$ intervals:
 $[0.15-0.2]$, $[0.2-0.3]$, $[0.3-0.4]$, $[0.4-0.5]$, $[0.5-0.7]$, $[0.7-0.9]$;
we use   $4\times4$  ($p_{t1},p_{t2}$) bins in the RF12 frame,
and 9 $p_{t0}$ bins in the RF0 frame.
The $p_t$ intervals are defined in Tab.~\ref{tab:totPt}.

\begin{figure}[!htb]
\centering   
\includegraphics[width=0.48\textwidth]{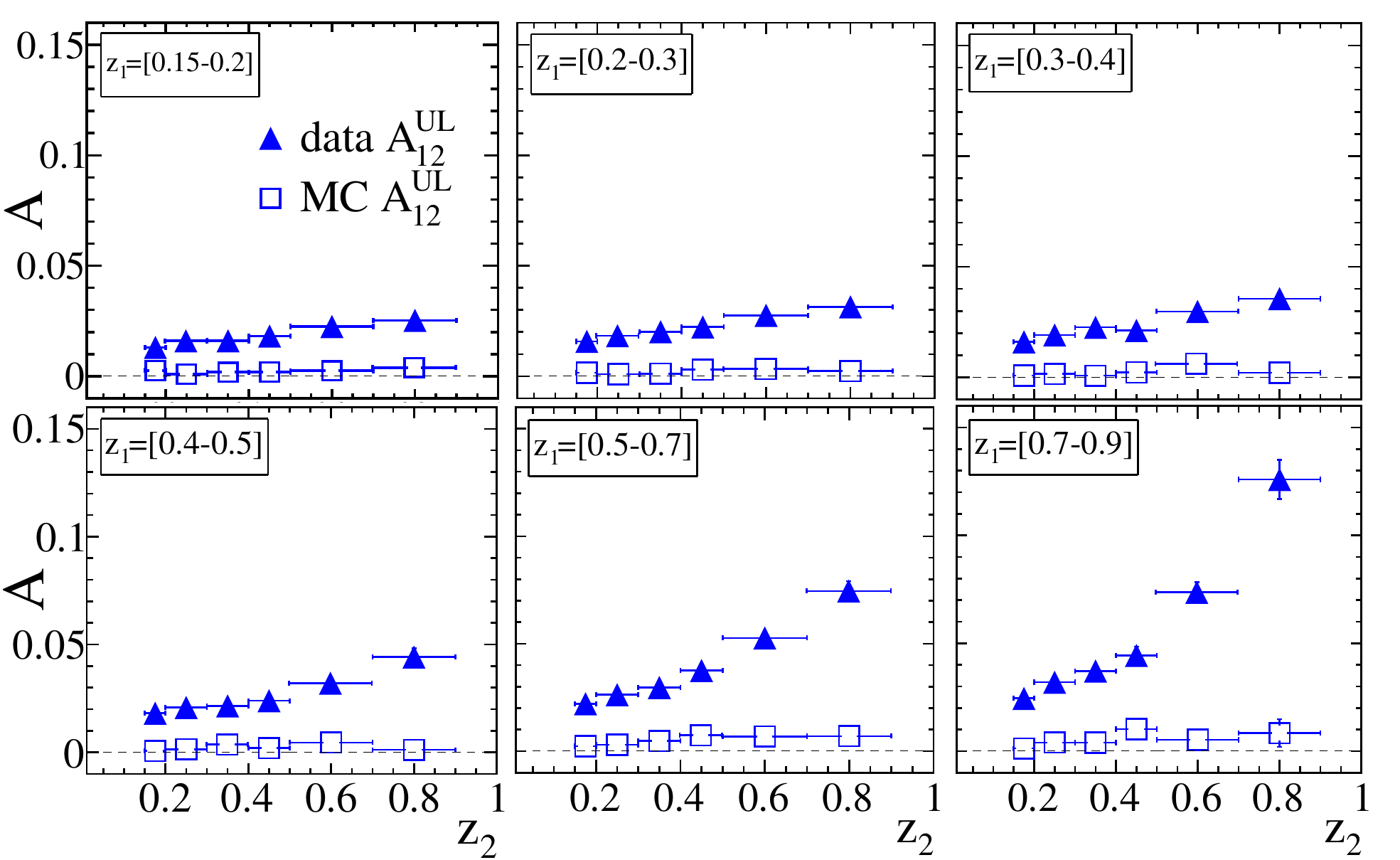} 
\caption{Comparison of raw DR asymmetries as a function of $6\times 6$
 ($z_1,z_2$)-bin subdivision calculated in data (triangles) 
 and MC samples (squares).
 In each plot, $z_1$ is fixed following the interval subdivisions
 described in the text, and $z_2$ ranges between 0.15 to 0.9.}
  \label{fig:rawdataZ}
 \end{figure}

Figure~\ref{fig:rawdataZ} shows the asymmetries obtained
from fits to the UL double ratio ($A^{UL}$)  in the RF12 frame
for data and MC  samples  in every $(z_1,z_2)$ bin.  
The asymmetries are  not corrected for the effects described in the next
three sections, and report only the statistical errors.
 Similar results are obtained for the asymmetries measured in the RF0 frame,
 and as a function of pion transverse momenta.

\section{Contribution of background events to the asymmetries.}\label{sec:bkgd}
The presence of background processes modifies the measured asymmetry $A_\alpha^{\rm{meas}}$. 
This is obtained by fitting the double ratio of the selected sample, and can be written as:
\begin{equation}
A_\alpha^{\rm{meas}}=\left( 1-\sum_i F_i \right) \cdot A_\alpha+ \sum_i F_i \cdot A_\alpha^i\,.
\label{eq:asy-bkg}
\end{equation}
Here,  $A_\alpha$, is the true Collins asymmetry produced by the fragmentation of light quarks, 
while $A_\alpha^i$ and $F_i$ are, respectively, the asymmetry and the
fraction of  pion pairs in the selected sample due to the $i^{th}$ background component.

The background processes giving a significant contribution
after the selection procedure are $\epem \rightarrow \tau^+\tau^-$,
$\epem \rightarrow \ccbar $, and  $\epem\to\FourS\to\BB$. 
We refer to them as the $\tau$, charm and bottom backgrounds, respectively.
In the former process, azimuthal asymmetries can arise from parity violation
in the weak decay of the heavy leptons.
For  charm processes the Collins effect is expected to be suppressed by the 
heavy mass of the fragmenting quarks. 
The study of the azimuthal asymmetries for \ccbar processes 
would be interesting on its own, but  larger data samples and
an optimized analysis would be necessary to perform precise measurements. 
No asymmetries arising from the Collins effect are expected from
$\FourS\to\BB$ decays.

The fractions $F_i$ and the asymmetries $A^i_\alpha$ of the background components
are determined using both MC and data control samples specific to each process,
and evaluated for each bin of $z$ and $p_t$.

\subsubsection*{The $\epem\to\tau^+\tau^-$ background}\label{sec:tau}
In order to assess whether a significant asymmetry is produced by 
\eetott processes we study a $\tau$-enhanced data sample, consisting
of the events in the lower-right side of the $E_{vis}$-vs.-$T$
distribution of Fig.~\ref{fig:enthrust}, and rejected by the cut
shown in the same picture. The purity of this control sample is estimated 
to be about $75\%$;  the fitted asymmetries are very small and 
consistent with zero within about two standard deviations.
We also perform the analysis  on a sample of simulated \tautau events,
applying the same event selection as for the data, and obtain asymmetries consistent with 
the small bias observed in the \uds MC sample.

The contribution of the \eetott background appears in Eq.~(\ref{eq:asy-bkg})
as the product of the asymmetry $A_\alpha^\mtau$ multiplied by the pion pairs fraction \Ftau.
We estimate \Ftau from the number of pion pairs selected in a MC sample of \tautau events
scaled by the data/MC luminosity ratio, independently for every $z$ and $p_t$ bin.
The values of \Ftau range from about $1\%$ at low $z_i$, to more than $18\%$ 
at high $z_i$, and are around $2\%$ independently of \pt.

Considering that the asymmetries measured in the \mtau-enhanced
samples are consistent with zero or give only very small deviations from zero, 
and that the contamination from \tautau events is significant only at large
$z_i$, where the Collins effect from $uds$ is  large (see Fig.~\ref{fig:rawdataZ}),
we set $A_\alpha^\mtau=0$ everywhere.

\subsubsection*{The $\epem\to\ccbar$ and $\epem\to\BB$ backgrounds} \label{sec:charm}
The fraction of pion pairs due to $\epem\to\ccbar$ events is much larger than
the $\taup\taum$ component, because of the higher production
 cross section and of event shapes similar to those for light quark production.
The fraction \Fcharm, estimated  with a generic \ccbar-MC sample, 
amounts to about 25\% for the whole data sample, 
roughly independent of \pt, but ranging from about 
30\% for pairs with low fractional energies down to less than one percent at the highest $z_i$ values.  

The \BB events are strongly suppressed by the event selection,
mainly because of the cut on the event thrust, and the 
fractions \Fb are estimated to be at most 2\% for low fractional energies, with no tracks with $z>0.5$.
As a consequence even a sizable asymmetry in the \BB sample would have negligible effect
on the measured asymmetry, and we set $\Ab=0$.

On the contrary, given the large fractions, the azimuthal asymmetries of the selected charm sample,
even if small, can have a significant impact on the total asymmetries,  
and therefore have to be independently measured. 
For this purpose, we  select a charm-enhanced data sample requiring at least one \Dstarpm
 candidate from the decay  $\Dstarpm\rightarrow\Dz\pipm$, with the  \Dz candidate
reconstructed  in the following Cabibbo-favored decay channels:
 $\Km\pip$, $\Km\pipi\pip$, $\KS\pipi$, and $\Km\pip\piz$. 
Note that a control sample built in this way will contain also a fraction of \BB events.

The reconstructed \Dz mass is required to be within $30\,\mevcc$
of the world average value~\cite{PhysRevD.86.010001}.
A low momentum pion is then combined with the \Dz candidate in order
to obtain the \Dstar candidate.
We retain events with at least one \Dstar candidate for which
$0.1425<\Delta M<0.1485$ \gevcc, where $\Delta M$ is the mass
difference between the reconstructed \Dstar and \Dz candidates.

As for the full data sample, we measure
the azimuthal asymmetry $A_\alpha^{\Dstar}$ by fitting the double ratio  
of pion pair distributions in the \Dstar-enhanced control sample.
The estimated fractions   \fcharm of pion pairs in this sample that are from \ccbar events
 average about $90\%$, with values for the individual $(z_1,z_2)$-bins ranging from more 
than $90\%$ to  about $60\%$ with decreasing fractional energies, and almost constant with \pt.
Fractions \fb from \BB events amount to a few percent at low energies, 
and vanish for $z_i>0.5$.

\subsubsection*{Corrections to the measured asymmetries}
Using the asymmetries  $A_\alpha^{\rm{meas}}$  and $A_\alpha^{\Dstar}$
fitted respectively in the full and \Dstar-enhanced data samples,
together with the fractions $F_i$ and $f_i$, and assuming 
$\Atau=\Ab=0$ and that the charm asymmetry is the same in both samples, we can write
\begin{equation}\label{eq:Acharm1}
\begin{split}
\Ameas  &= (1-\Fcharm-\Fb-\Ftau)\cdot\Auds+ \Fcharm\cdot\Acharm\,,\\
\Adstar   &= (1-\fcharm-\fb)\cdot\Auds \,+\, \fcharm\cdot\Acharm\,.
\end{split}
\end{equation}
The unknown background-corrected Collins asymmetries $A_\alpha$ 
and the charm contribution \Acharm are obtained solving these equations
in each bin of $z$ and $p_t$.

\begin{figure}[!htb]
\centering
 \includegraphics[width=0.48\textwidth] {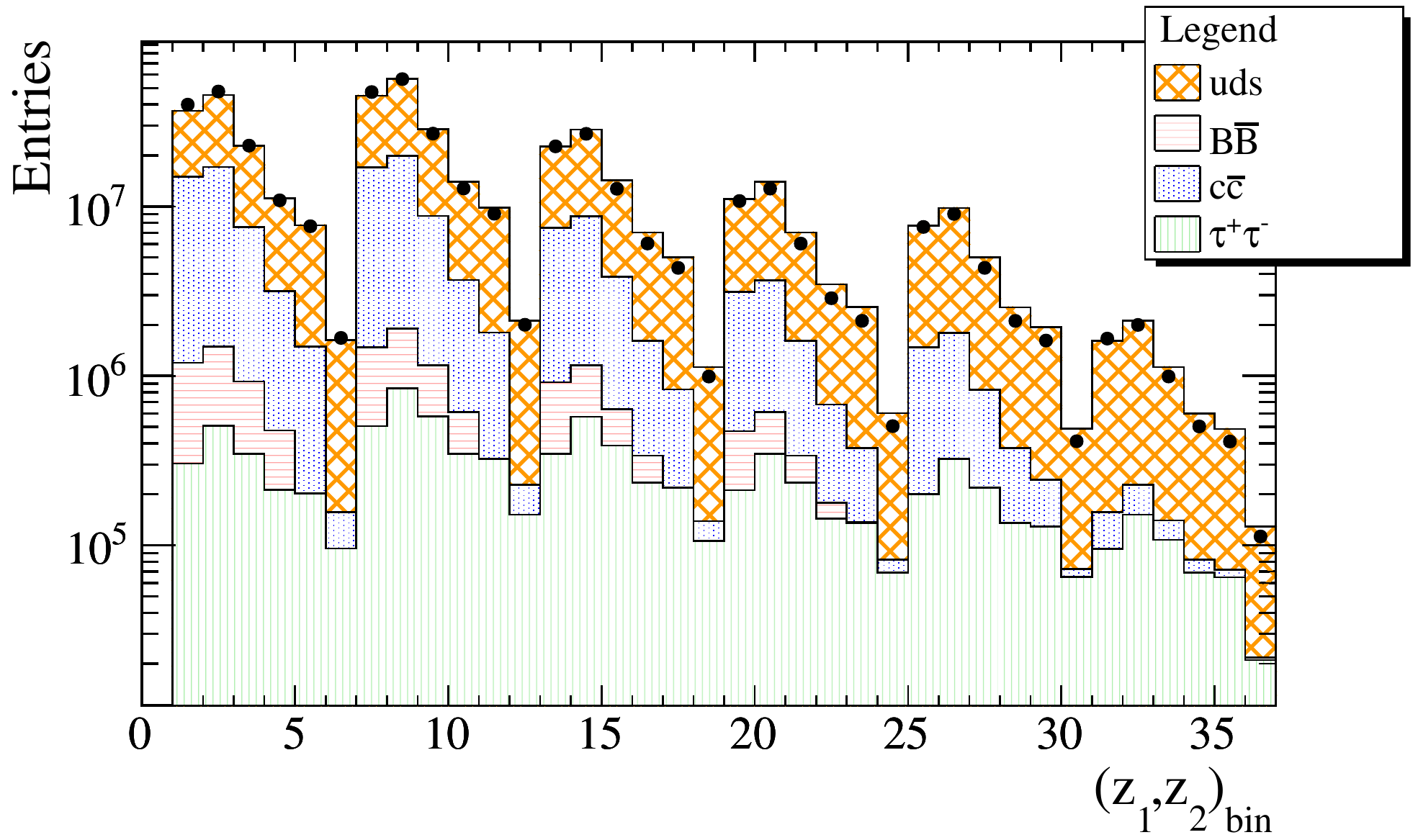}
 {\boldmath      \put(-70,120){(a)}    }
    \vskip+2ex 
  \includegraphics[width=0.48\textwidth] {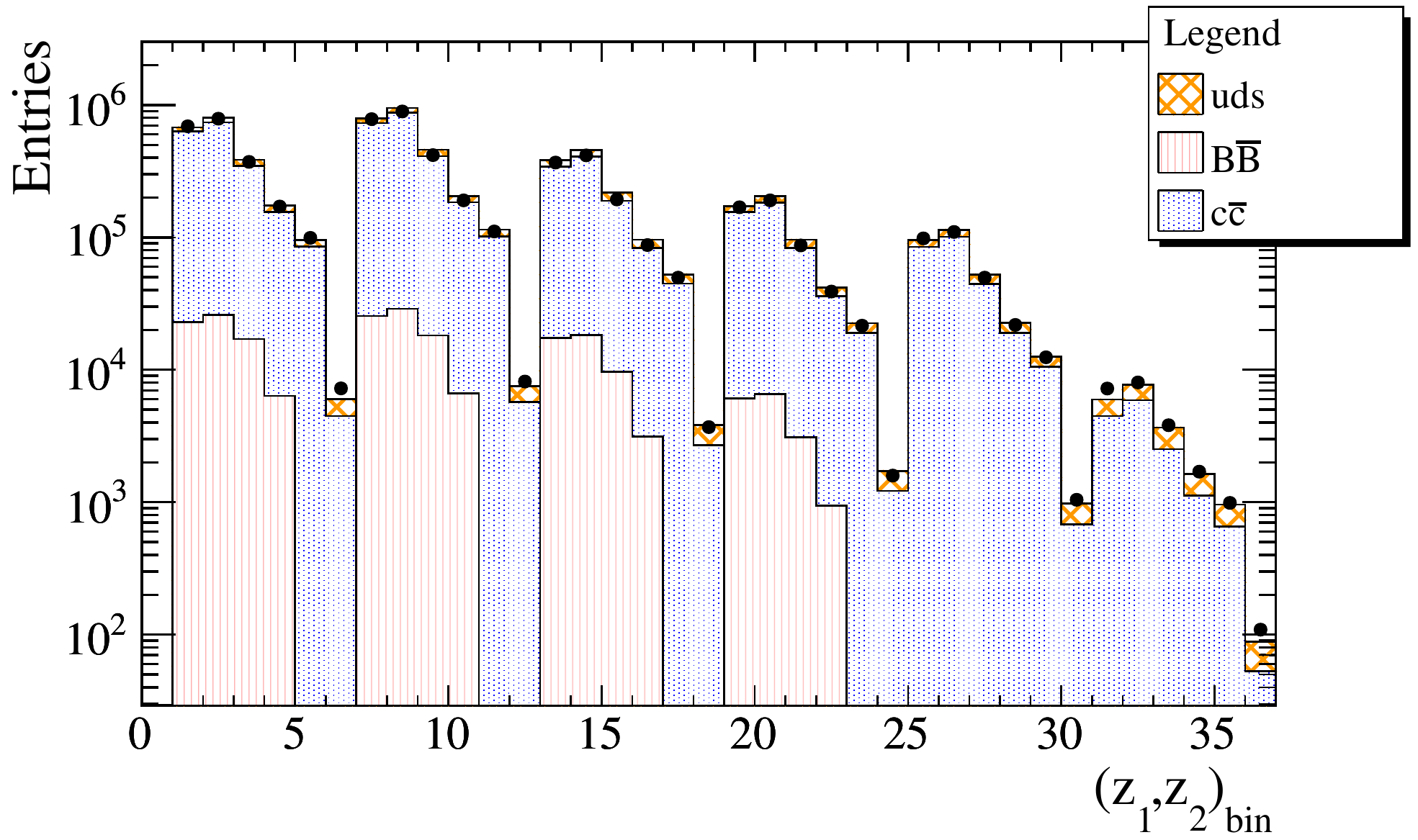} 
   {\boldmath      \put(-70,120){(b)}    }
 \caption{ (color online).
    Comparison between the number of pion pairs in the data (a)
    and $D^*$-enhanced data (b) samples  
   (points)  and the sum of the  contribution due to 
   \tautau, \BB,  \ccbar, and $uds$ components estimated with MC simulation as a function of
    \zbin\ bins. 
  } \label{fig:frac}
\end{figure}
A significant source of systematic error in this procedure can arise from
 the fractions $F_i$ and $f_i$, which are estimated with MC simulation.
 The $\pi^\pm$ cross sections in $\epem\to\qqbar$ processes are known 
at no better than the few percent level; 
furthermore, only a fraction of all charmed-hadrons and $\B$-meson decays 
 have been measured and included in the EvtGen generator.
 Also \mtau decays with many hadrons in the final state are known with significant uncertainties only.
In order to evaluate the effect of these uncertainties on the
measured fractions, we compare bin-by-bin the number of pion pairs selected in the data with those
selected in the \uds, \mtau, charm, and bottom MC samples, summed according to the nominal production cross sections. 
The observed data-MC differences are at most at the few percent level as can be deduced by Fig.~\ref{fig:frac}.
Conservatively, we assign these differences as additional uncertainties
on the charm (\Fcharm, and \fcharm) and tau (\Ftau) fractions,
which are the most significant contributions in the extraction of
$A_\alpha$ and $A_\alpha^c$ from Eq.~(\ref{eq:Acharm1}).
This choice has a very little effect on the final result, given that the uncertainties on the background subtraction procedure 
are dominated by the statistical errors of the fit to the  \Dstar control sample,
in particular for bins of high fractional energies. 

We check the consistency of the $D^*$-enhanced sample by performing the correction of the measured asymmetries, and the
estimation of the charm contributions independently for the four $D^0$ decay modes,
finding no significant differences.

\section{Asymmetry dilution due to detector effects}\label{sec:weight}

\begin{figure}[htb]
 \begin{center}
  \includegraphics[scale=0.35]{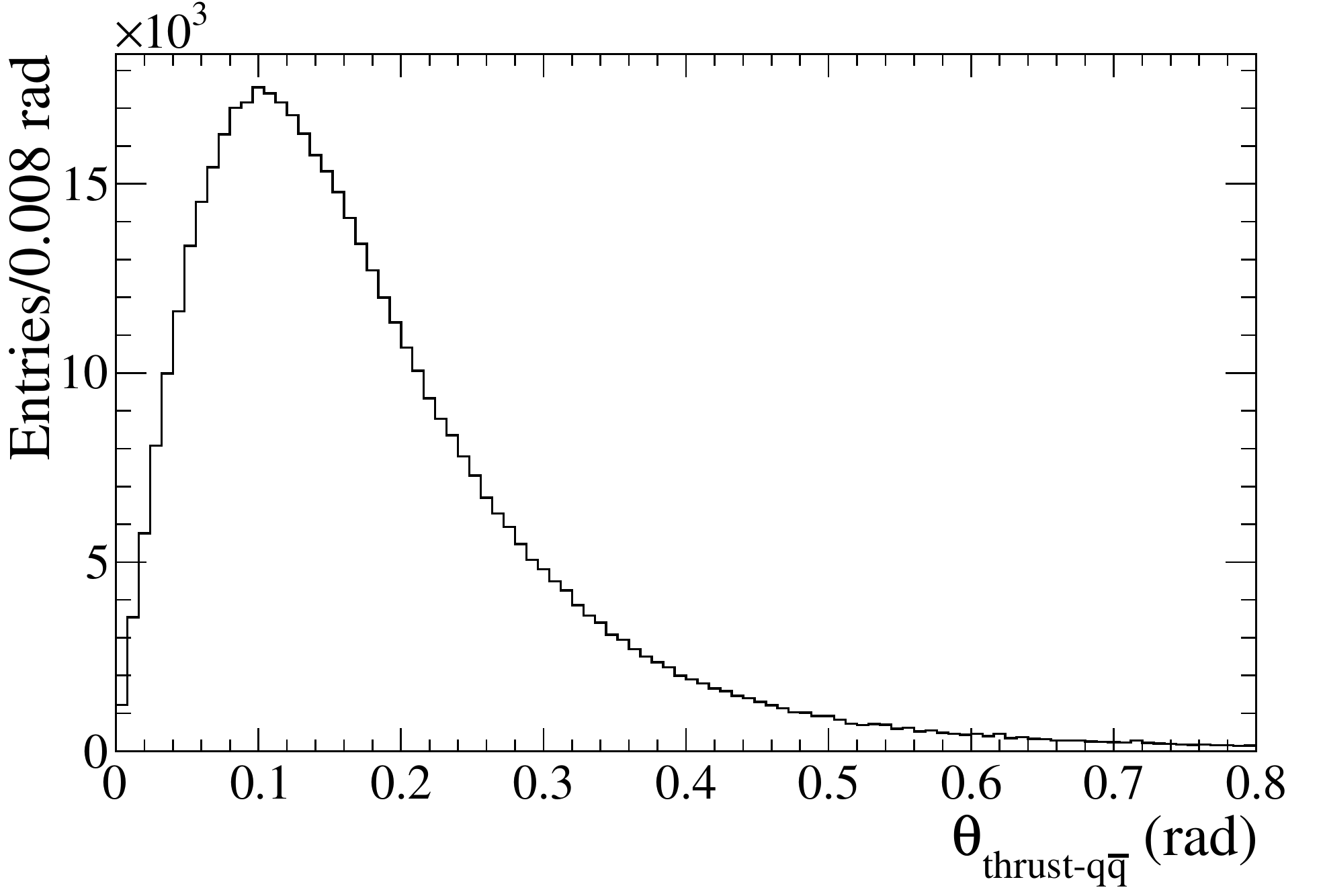}
  \caption{Opening angle in radians between the thrust axis 
  calculated from reconstructed particles in simulated events
 and the  generated \qqbar axis.  } 
  \label{fig:thrustdiff}
 \end{center}
  \end{figure}

The experimental method uses  the thrust axis
to estimate the \qqbar axis.
The distribution of the opening angle between the two axes
for simulated events, 
shown in Fig.~\ref{fig:thrustdiff}, 
peaks at about 100 mrad, and has a long tail at higher values.
This produces a dilution of the asymmetries, in
particular  in the thrust reference frame, where the azimuthal angles
$\phi_1$ and $\phi_2$ are calculated with respect to the thrust axis
(see Fig.~\ref{fig:rf12}). 
The impact on the measurement of the azimuthal 
angle $\phi_0$ in RF0 is small.
Particle identification and tracking resolution have small effects in both frames. 

An additional smearing of the azimuthal angles can occur when the 
track is very close to the analysis axis,
where a small mis-measurement of the track results in a large shift 
in the azimuthal angle.
A larger effect is expected in RF12 where,
by construction, energetic tracks are close to  the thrust axis.

These and other detector effects can be studied via MC
simulation, comparing the fitted asymmetries with 
those introduced in the simulation. 
The MC generator does not include the Collins effect,    
so we remodulate the generated azimuthal distributions applying 
 to every selected pion pair a weight
defined as $w^i=1+a^i\cdot \cos(\beta_{\alpha, \rm{gen}}),$
with $i= U, \,L, \, C$ and $\alpha=12$ or $0$.
In RF12 the angle $\beta_{12,\rm{gen}}$ is the sum of the azimuthal angles 
$\phi_{1,\rm{gen}}$ and $\phi_{2,\rm{gen}}$ for generated particles 
calculated with respect to the true quark-antiquark axis,
while in  RF0  the angle $\beta_{0,\rm{gen}}=2\phi_{0,\rm{gen}}$
is calculated with respect to the 3-momentum of one of the generated pions
which  makes the pair.
The ratio of the fitted to the simulated asymmetry should be unity for
perfect reconstruction, i.e. if all $\beta_{\alpha}=\beta_{\alpha,\rm{gen}}$,
where $\beta_\alpha$ are the proper combinations of azimuthal angles 
in the reconstructed MC sample.
We verify that for $a_i=0$ the fitted asymmetry is consistent with the biases
observed in Sec.~\ref{sec:dr}. 
We consider several $a_i$ values between zero to 0.1, independent of $z$,
\pt and $\theta_{th(2)}$. Subtracting the bias and taking the ratio for each,
we obtain dilution values that vary by less than their statistical error in each
$z$, $p_t$, and $\sin^2\theta_{th(2)}/(1+\cos^2\theta_{th(2)})$ bin.
We therefore average these values in each bin and use them to correct the
background-corrected data.

\begin{figure}[!htb]
 \includegraphics[width=0.3\textwidth] {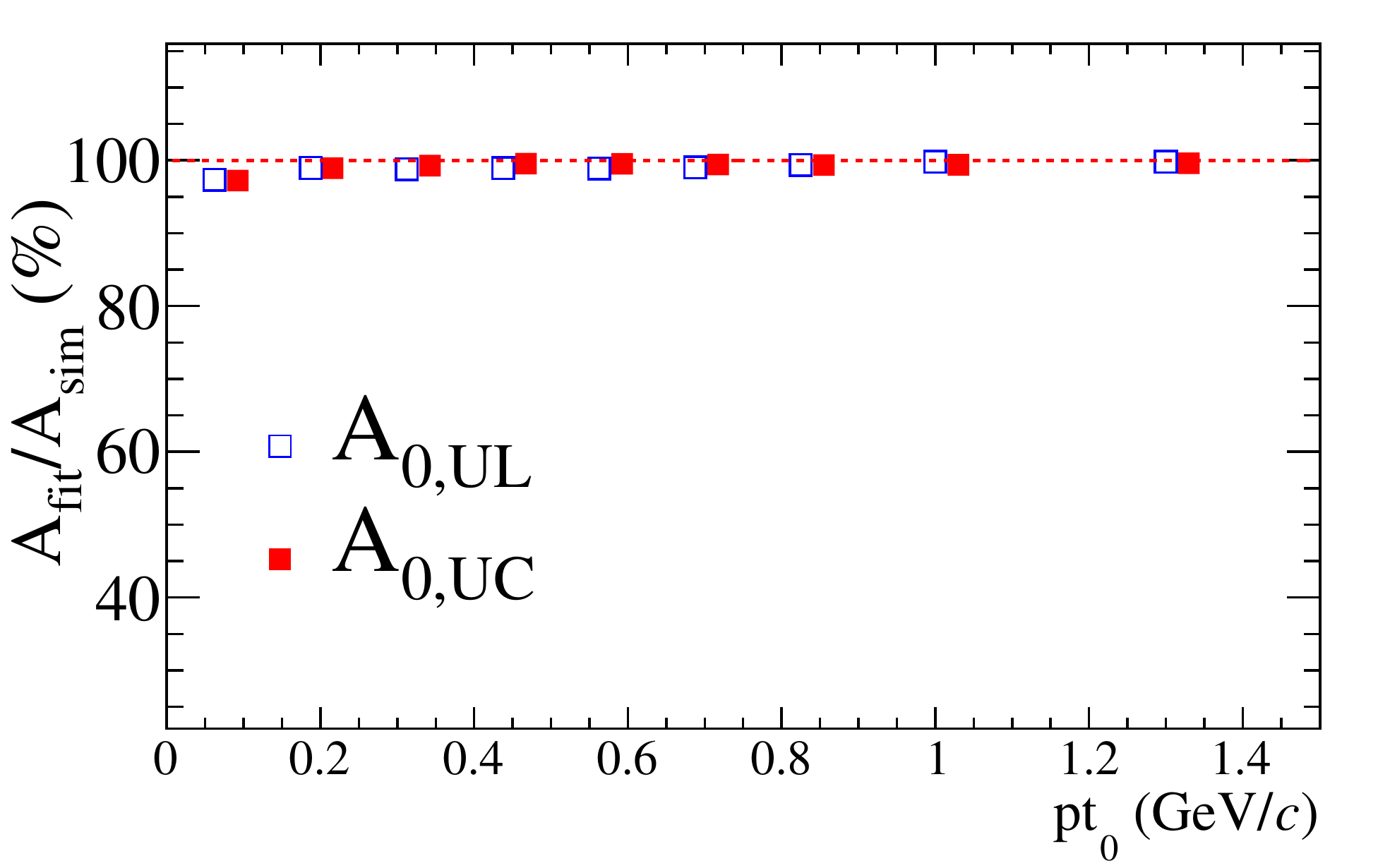} 
    {\boldmath      \put(-30,40){(a)}    }
    \vskip+2ex 
\includegraphics[width=0.45\textwidth] {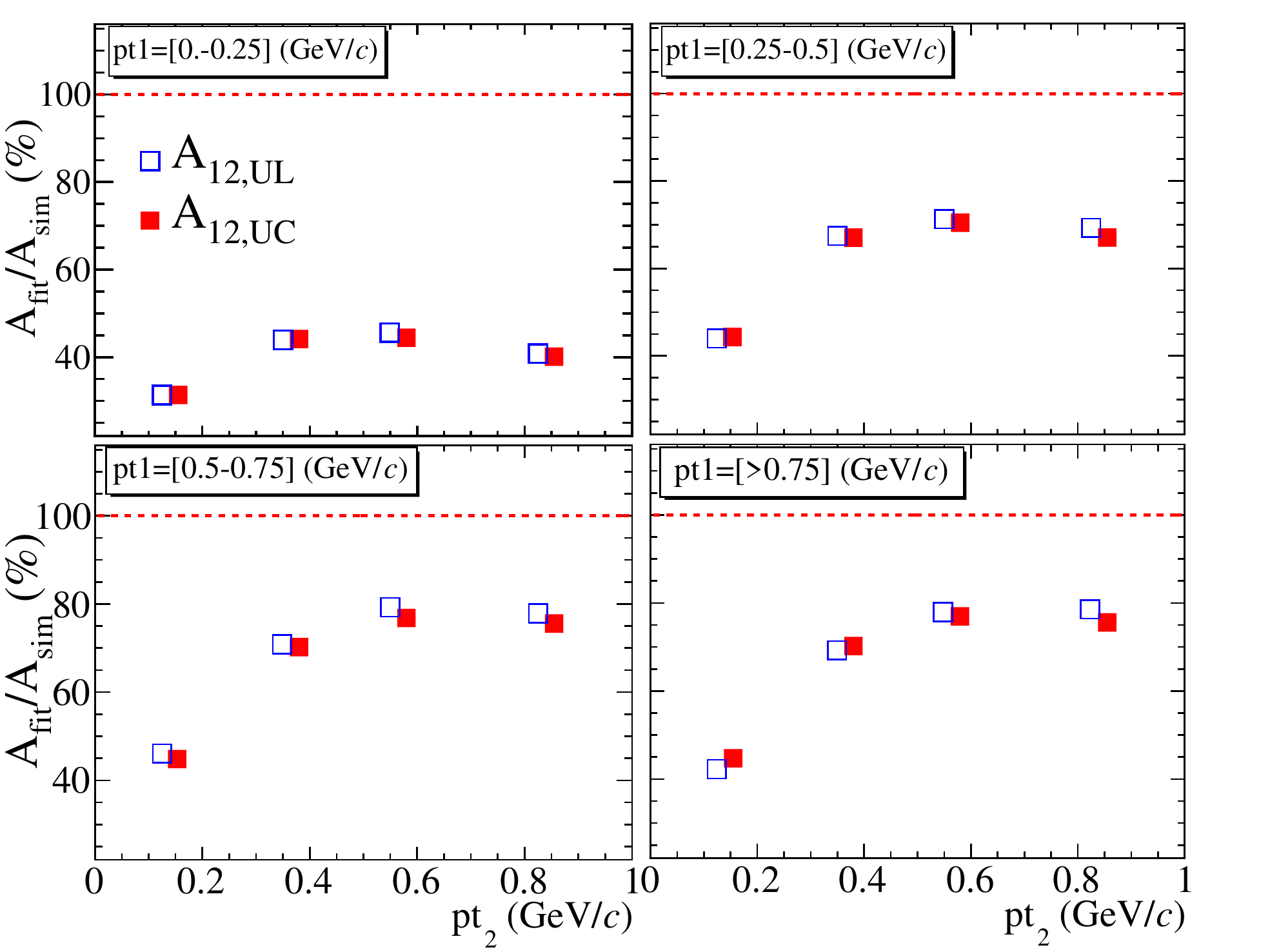}  
    {\boldmath      \put(-40,30){(b)}    }
 \caption{
 Ratio (in percent) of the fitted to the simulated asymmetries
as described in the text,  as a function of $p_t$ bin.
 Open and full squares  refer to the UL and UC double ratios, respectively.
 In the RF0 frame (a) the fitted asymmetry is consistent
 with the simulated one, while in the RF12 frame (b) it is systematically
 underestimated. }
\label{fig:wei}
\end{figure}

As shown in Fig.~\ref{fig:wei}(a), the results for RF0  are
essentially consistent with the simulated
asymmetries for every bin. 
On the contrary, the fitted asymmetries for RF12 (Fig.~\ref{fig:wei}(b))
systematically underestimate the generated values, 
with correction factors ranging from about 1.3 to 2.3 with
increasing values of $z$, and from about 3 to 1.3 with
increasing values of $\pt$.
The errors on weighted averages of the correction factors are 
assigned as systematic errors. 

Collins asymmetries are expected to depend on the polar angle 
of the analysis axis, as well as on the two pions' $z$ and $p_t$,
and any strongly dependence might affect the dilution factors.
We have evaluated these factors using weights in which $a^i$ has 
the expected linear dependence on the quantity $\sin^2\theta_{th(2)}/(1+\cos^2\theta_{th(2)})$,
and weights that contain a linear combination of $z$,
such as $a^i(z_1,z_2)=a^i\cdot z_1\cdot z_2$.
The measured dilutions are in good agreement each other
and the shifts in the va\-lues are within the uncertainties.
Since from the results of the tests performed there are no
indications of the dependence of the dilution on the true asymmetry,
we report the results using uniform weighting.

\section{Study of systematic effects} \label{sec:syst}

We have identified a number  of systematic effects that can potentially influence the measurement of the asymmetries.
The sizable effects are corrected as described above; here we discuss the origins
and uncertainties of these and other effects.
Unless otherwise stated, all the systematic uncertainties are evaluated in each bin.

\paragraph{Test of the DR method on Monte Carlo sample.}\label{subsec:mc}
The  MC generator describes the radiative gluon
effects, but does not contain asymmetries  based on the Collins effect.
Nevertheless, as discussed in Sec.\ref{sec:dr} and visible in
Fig.~\ref{fig:rawdataZ}, the fitted asymmetries for  the \uds-MC samples show deviations from zero
at all values of the pion transverse momenta and fractional energies.
The fitted asymmetries are  small with respect to the data everywhere,
but not negligible in several cases, being of the same order as 
other systematic uncertainties.
In order to understand the origin of this bias we compare three different MC samples:
the first sample is built at generator level, that is taking the momenta of pions as produced by the event generator;
the second sample also uses  generator level momenta, but  only pions with associated reconstructed and identified tracks;
the last sample is the standard fully reconstructed MC sample. 
We also compare the results using as reference axis the true \qqbar
axis, instead of the thrust axis, and varying the criteria  that define
the track detector acceptance,  as track polar angle and number of hits per track in the DCH. 
As a result of these investigations we conclude that the small bias observed in RF0 is due to a non-perfect
cancellation of the detector effects in the double ratio procedure,
while in RF12 the main effects come from the use of the thrust axis as a reference axis.
We subtract the estimated bias from the asymmetry, and take as systematic uncertainty the
combination in quadrature of the largest variation of the performed
tests with the statistical error of the bias measured with the standard fit. 
 
\paragraph{Uncertainties due to the \pt resolution.}
The Collins effect also depends on the pion momentum
transverse to the direction of the fragmenting quark ($P^\perp$, as in Fig.~\ref{fig:PhiAngle}).
In the RF12 frame, this quantity is not accessible because the
momenta of the fragmenting quarks is not known, and the asymmetries
are measured as a function of the momenta of the pions transverse to the thrust axis ($\pt_1$ and $\pt_2$).  
In order to convert the measured asymmetries to the 
asymmetries with respect to the true $P^\perp$ momenta, one must
account not only for the dilution effect discussed in the prevoius section,
but also for the migration of pion pairs  from one momentum-bin to another.
The  $P^\perp$  resolution function is obtained for different \pt ranges making use of the $uds$ MC sample. 
We fit the distribution of the difference between the true
$P^\perp$ and the reconstructed \pt transverse momentum
with a double gauss function, or a crystal-ball function for
$\pt<0.25$ \gev; the width of the dominant gaussian component results of the order of 100 \mev.
These resolution functions are then used to redistribute the pion
pairs in the various $(\pt_1,\pt_2)$  bins according to the probability that they were
generated  in that particular bin, and the fit to the 
azimuthal distributions of the new bin contents are performed.
The relative difference between the asymmetries obtained with this and the
standard procedure are of the order of 10\% for every bin, with the
exception of the lower $(\pt_1,\pt_2)$ bin, were it amounts to about 30\%;
these differences are assigned as systematic uncertainties.

\paragraph{Uncertainties due to particle identification.}
With the algorithm used, the probability of misidentifying kaons and protons
as pions has been measured to be a few percent per track. 
This results in a purity for the selected pion pair sample of about $96\%$,
with the remaining $4\%$ of pairs made of a true pion and a true kaon.
We repeat the study with both more stringent and more
loose  selection criteria, and compare the results with the standard selection. 
Good agreement is found among the different selections; the absolute values of the 
differences amount to at most a few percent of the measured asymmetries and
are assigned as systematic errors.

\paragraph{Uncertainties due to the fit procedure.}
The dependence of the measured asymmetries on the binning
is checked by comparing the results with three different
bin sizes of the azimuthal angles $\beta_{12}=\phi_1+\phi_2$ and $\beta_0=\duephiz$:
$18^\circ$, $4.5^\circ$, and $1.8^\circ$ (used for the standard analysis).
The largest deviations are less than 1\% and are taken as systematic errors.\\
\indent The DR distributions are fitted by Eq.~\ref{eq:drfit}, which approximates
the series expansion to the first order in $\cos(\beta_\alpha)$, 
with $\beta$ the azimuthal angles in the respective reference frame and $\alpha=12,\,0$,
and neglects possible $\cos^2(\beta_\alpha)$ contribution due to the 
detector acceptance.
In order to check for the sensitivity to these contributions,
we use different fitting functions with additional higher harmonic terms.
No significant changes in the values of the cosine moments
with respect to the standard fits are observed.\\
\indent A certain level of correlation among the entries of the double ratio
distributions is expected because the  same pion can be used to form
several pion pairs, so that the statistical error returned by the
fits could be underestimated.
We check for this effect, performing a set of 3000 pseudo-experiments.
For each pseudo-experiment we randomly generate according to 
the fit model a statistical sample of the same size of  that selected by 
the analysis procedure.
Gaussian fits to the pull distributions of the values of the fitted
asymmetries give results consistent with a vanishing mean 
and a unit width, as expected for an unbiased fit model.

\paragraph{Test of the double ratio with same sign pion pairs.}
The Collins effect does not depend on the electric charge,
but only on the combination of favored and disfavored 
fragmentation functions in the particular charge combination of the 
paired pions.
In particular, the same combination appears when a $\pip\pip$
or a $\pim\pim$ pair is considered. 
Gluonic radiative effects do not depend on the electric charge either.
Therefore, we can test the double ratio procedure and the possible charge dependence of the detector
response by probing the ratio of normalized azimuthal distributions
for positively charged  over negatively charged pion pairs.
Results consistent with unity are obtained.

\paragraph{Subtraction method and double ratio.}
As a cross check of the double ratio method we also extract the
asymmetry by using  a different procedure which consists of taking the
difference, instead of the ratio, of pion pair rates. 
In this case, gluon radiation effects cancel at all orders, while 
the cancellation of  the acceptance effects could be non optimal.  
The asymmetries measured with the two methods are consistent,
making us confident that possible radiative and detector effects not
canceling in the double ratio procedure do not significantly affect the results.

\paragraph{Study of beam polarization effects.}
Charged particles circulating in a magnetic field become polarized transverse to the beam
direction due to the emission of spin-flipping synchrotron radiation,
known as the Sokolov-Ternov effect~\cite{Sokolov:1963zn}. 
Beam polarization can affect the angular distribution of produced
hadrons in $\epem \rightarrow hX$ introducing an azimuthal asymmetry
with respect to the beam spin direction.
This asymmetry has in common with the Collins asymmetry that both are
transverse single spin asymmetries: the former concerning lepton
spins, the latter quark spins. 
The beam polarization is expected to be negligible at the PEP-II
interaction point. This can be verified by studying the reaction
$\epem \rightarrow \mu^+\mu^-$, whose cross section can be written as~\cite{PhysRevLett.35.1688}: 
\begin{equation}
\frac{d\sigma(\epem \rightarrow \mu^+\mu^-)}{d\Omega} \propto 1+\cos^2\theta +P^2\sin^2\theta \cos(2\phi)
\end{equation}
where $P$ is the degree of transverse polarization of the beams, and
$\theta$ and $\phi$ are the polar and azimuthal angles of the produced
muons in the $\epem$ center of mass system.

We analyze the $\cos\theta$ and $\phi$ distributions of 
selected muon pairs, emitted at a polar angle $|\cos\theta|<0.75$,
in order to ensure that they fall within the SVT coverage. 
We perform the fit to the whole sample and separately to
the samples corresponding to the different data taking periods.
In all cases the fits are consistent with expectations for 
unpolarized beams. 
We conclude that no significant systematic errors need to be assigned for possible
buildup of beam polarization.

\paragraph{Consistency of asymmetries in different data sets.}
The results reported in this paper are obtained combining the 
data taken at two different c.m. energies, at the peak of the \FourS resonance
and 40\,\mev below.  While the slightly different energy is not a problem, 
the two sets of data differ for the background due to \upsbb. 
We perform a consistency check of the results obtained 
fitting separately the two data sets ($a$ and $b$) as follows:
\begin{equation}
\chi^2=\sum_i \frac{(A_i^a-A_i^b)^2}{(\delta A_i^a)^2+(\delta A_i^b)^2},
\end{equation}
with $\sum_i$ the summation over $z$ or $p_t$ bins,
and $\delta A$ the statistical error on the measured asymmetry $A$.
We find the overall $\chi^2$ per degree of freedom ranging between 1.2 and 0.7.
Analogous checks performed on sub-samples of data collected
in different data-taking periods show a general consistency of the results.

\begin{figure}[!htb]
 \includegraphics[width=0.45\textwidth] {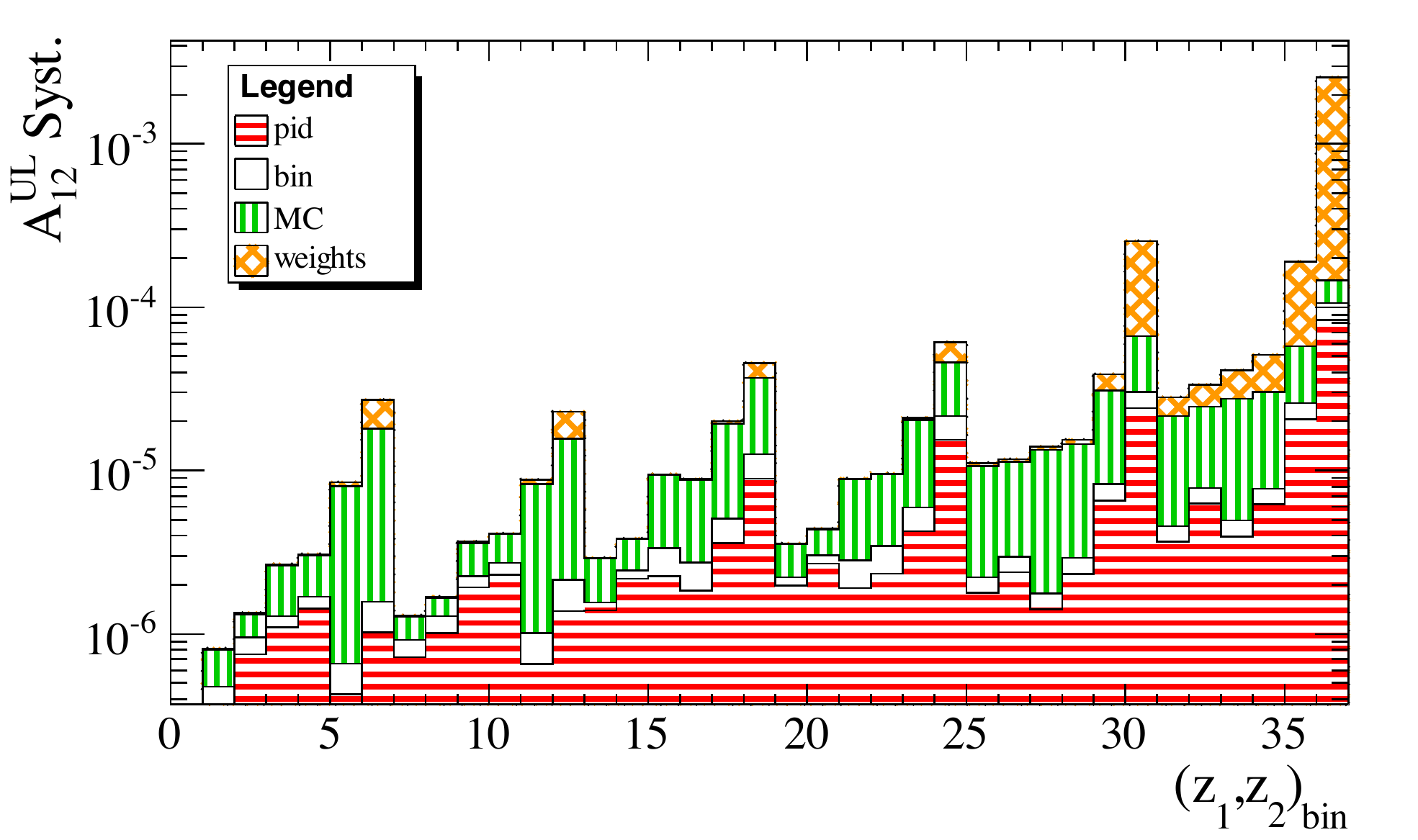} 
{\boldmath      \put(-60,115){(a)}    }
 \includegraphics[width=0.45\textwidth] {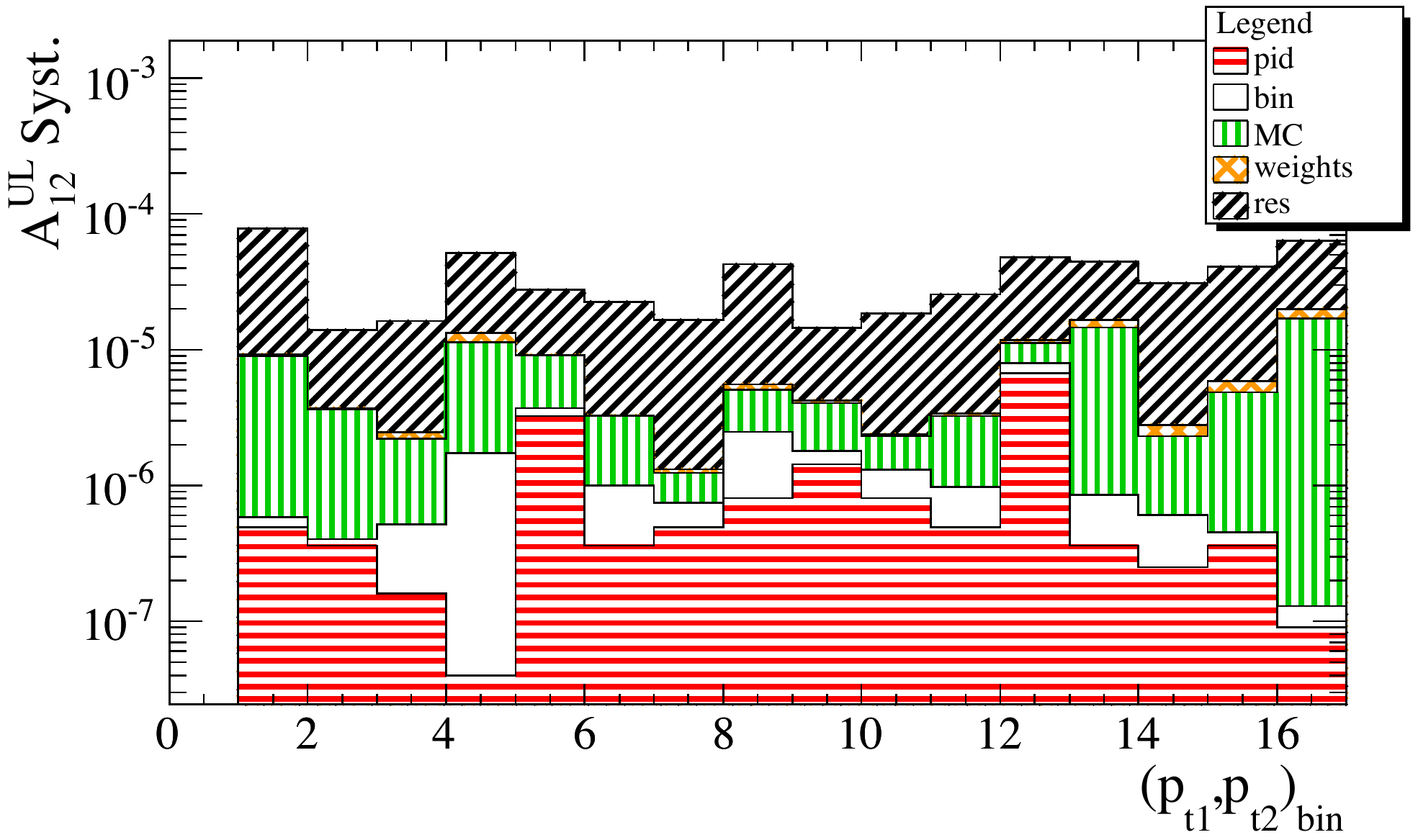} 
    {\boldmath      \put(-60,115){(b)}    }
 \caption{ (color online).
 Squared contributions to the systematic error on the
 asymmetry   (for the UL double ratio in the RF12 frame) 
as a function of \zbin\ bins (a) and \ptbin\ bins (b).
 The reported uncertainties are 
due to particle identification (pid),
 the binning in the azimuthal angle (bin), 
the bias observed in MC (MC), 
the correction for the asymmetry dilution (weights), and
the \pt resolution with respect to the \qqbar axis (res).
 }
\label{fig:systematic}
\end{figure}

\paragraph{Summary of systematic uncertainties.}
All  systematic effects are evaluated for each bin of fractional energy 
and pion transverse momentum.
As an example, the sizable contributions to the absolute systematic errors for
$A^{UL}_{12}$ are shown as a function of the 36 ($z_1,z_2$) bins in
Fig.~\ref{fig:systematic}(a), and as a function of the 16
($p_{t1},p_{t2}$) bins in Fig.~\ref{fig:systematic}(b). 
The histograms report the squared errors assigned
for uncertainties due to particle identification (pid), bin size of the
azimuthal distributions (bin), estimate of the bias observed in the MC
sample (MC), and estimate of the correction factors for the dilution
of the asymmetry (weights).  
The two latter sources dominate at high fractional energies, 
while at low-$z$ values all contributions are comparable.
In Fig.~\ref{fig:systematic}(b) is also reported the squared uncertainty
due to the \pt resolution (res).
This, and the uncertainty in the estimate of the bias
are the dominant sources at all values of transverse momenta in the RF12 frame.
The total systematic error is obtained by adding in quadrature the individual contributions.

The contribution of the various background sources (\ccbar, \bbbar, and \tautau events)
to the measured asymmetries is subtracted with the procedure described in Sec.~\ref{sec:bkgd}.
Through Eq.~(\ref{eq:Acharm1}), the statistical error of the Collins asymmetries
account for the statistical uncertainties of the asymmetries measured in the full 
and in the $D^*$-enhanced data sample,
and for the statistical and systematic uncertainties in the determination 
of the relative fractions $F_i$ and $f_i$.

\section{Results}\label{sec:results}
This section summarizes the measured asymmetries as a function of 
 fractional energies $z$, transverse momenta $p_t$, and 
 polar angle of the analysis axis, after all corrections and systematic 
uncertainties discussed in the previous sections are applied.
We also report the asymmetry measured in  RF12 
in a four-dimensional space, as a function of ($z_1,z_2,p_{t1},p_{t2}$).

\begin{figure*}[h]
  \centering
 \includegraphics[width=0.8\textwidth] {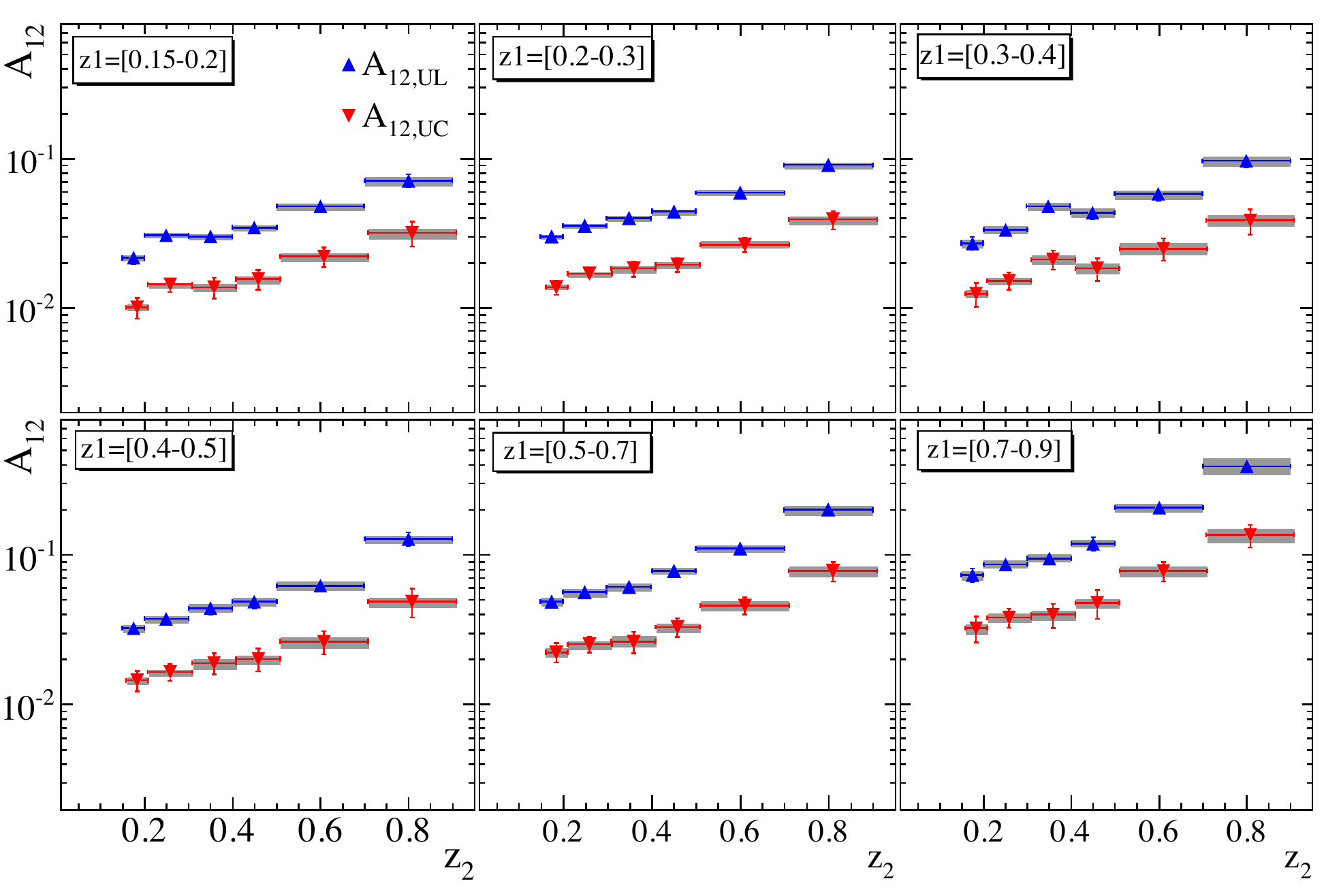} 
  \caption{ (color online).
    Collins asymmetries for light quarks measured in bins
    of fractional energies \zbin, in RF12.
    Asymmetries for the UL (up triangles) and UC  (down triangles) ratios are reported,
    with statistical  error bars and systematic uncertainties  represented by the 
     bands around the points.
  }
  \label{fig:finalZ12}
\end{figure*}
\begin{figure*}[!htb]
  \centering
  \includegraphics[width=0.8\textwidth] {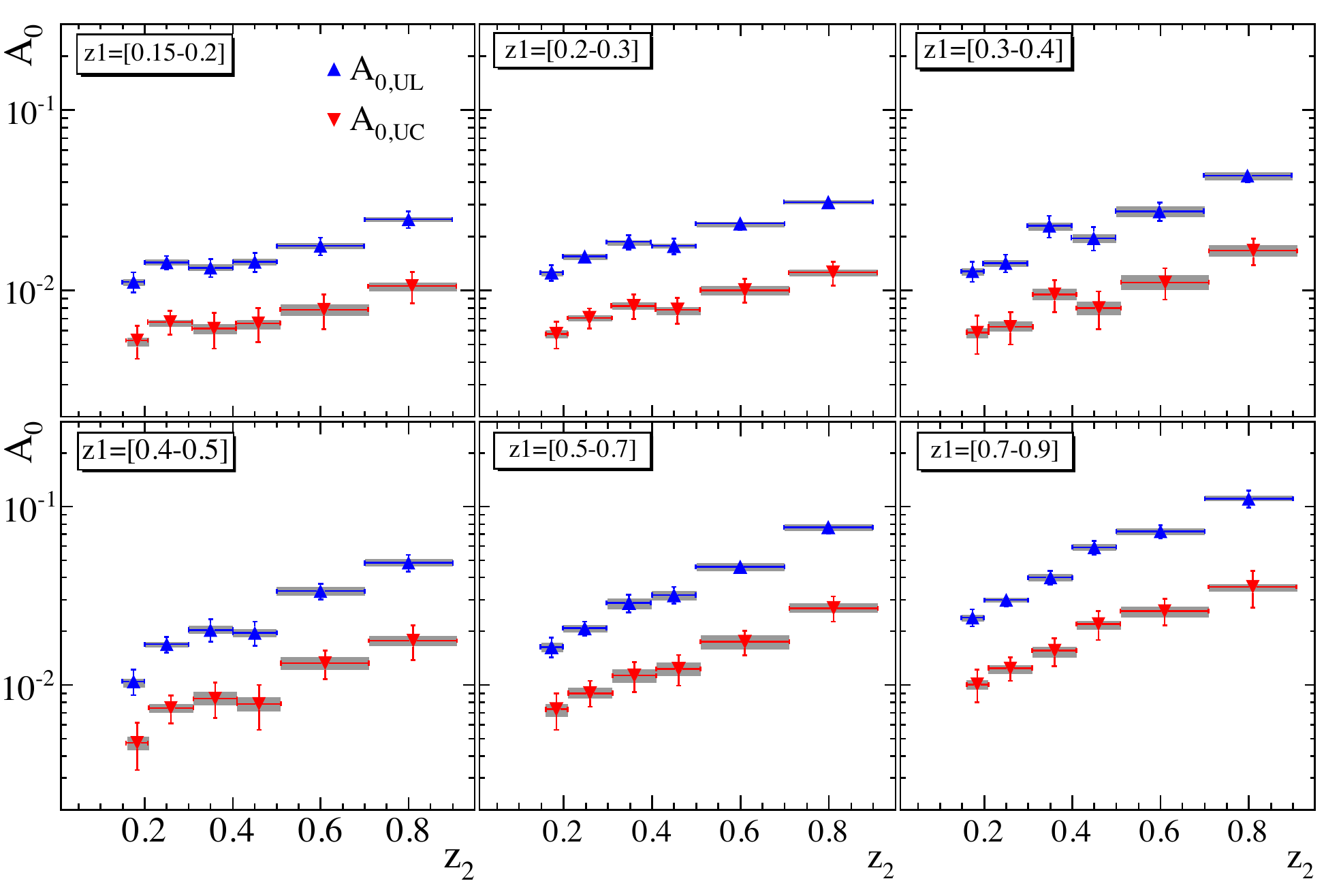} 
  \caption{ (color online).
    Collins asymmetries for light quarks measured in bins
    of fractional energies \zbin, in RF0.
    Asymmetries for the UL (up triangles) and UC  (down triangles) ratios are reported,
    with statistical  error bars and systematic uncertainties  represented by the 
     bands around the points.
  }
  \label{fig:finalZ0}
\end{figure*}

\subsection{Collins asymmetries {\textit vs.} fractional energies \label{sec:AvsZ}}
The Collins asymmetries measured for each \zbin\ bin are
summarized in Tabs.~\ref{tab:totZul} and~\ref{tab:totZuc},
and illustrated in Fig.~\ref{fig:finalZ12} and 
Fig.~\ref{fig:finalZ0},  for RF12 and RF0, respectively.
In each plot the asymmetries are reported for every $z_2$ bin in a
given interval of $z_1$.
We note a very good consistency among symmetric bins, with 
$z_1$ and $z_2$ exchanged, giving additional confidence on the 
correctness of the fitting procedure.
A rise of the asymmetries with increasing pion energies is clearly visible
in all plots, in agreement with theoretical 
predictions~\cite{Bacchetta2001155,Bacchetta2008234,PhysRevD.64.094021} and 
Belle results~\cite{PhysRevD.78.032011,PhysRevD.86.039905}.
The measured values  span more than an order of magnitude, being
about 1-2\% in the lower \zbin\ bins, and close to 40\% for $A_{12}^{UL}$
 at the highest energies.

The measured UC asymmetries are smaller than the UL asymmetries
by roughly a factor of 2. 
This behavior was already observed by Belle, and
should reflect the different contribution of 
favored and disfavored fragmentation functions to the UC and UL
ratios, as discussed in Sec.\,\ref{sec:dr}.
An analysis of Belle data, under the assumption 
\begin{equation}
  H_1^{\rm{fav}(\rm{dis})}(z) \,=\, C_{\rm{fav}(\rm{dis})} z D_1^{\rm{fav}(\rm{dis})}(z) \, , 
\end{equation}    
found values for the parameters $C_{\rm{fav}}$ and $C_{\rm{dis}}$ consistent with
a large disfavored Collins fragmentation function
 with sign opposite to the favored one~\cite{PhysRevD.73.094025},
as also suggested by the HERMES experiment~\cite{PhysRevLett.94.012002}.

\subsection{Collins asymmetries {\textit vs.} transverse momenta \label{sec:AvsPT}}
\begin{figure*}[htb]
  \begin{center}
    \includegraphics[width=0.5\textwidth] {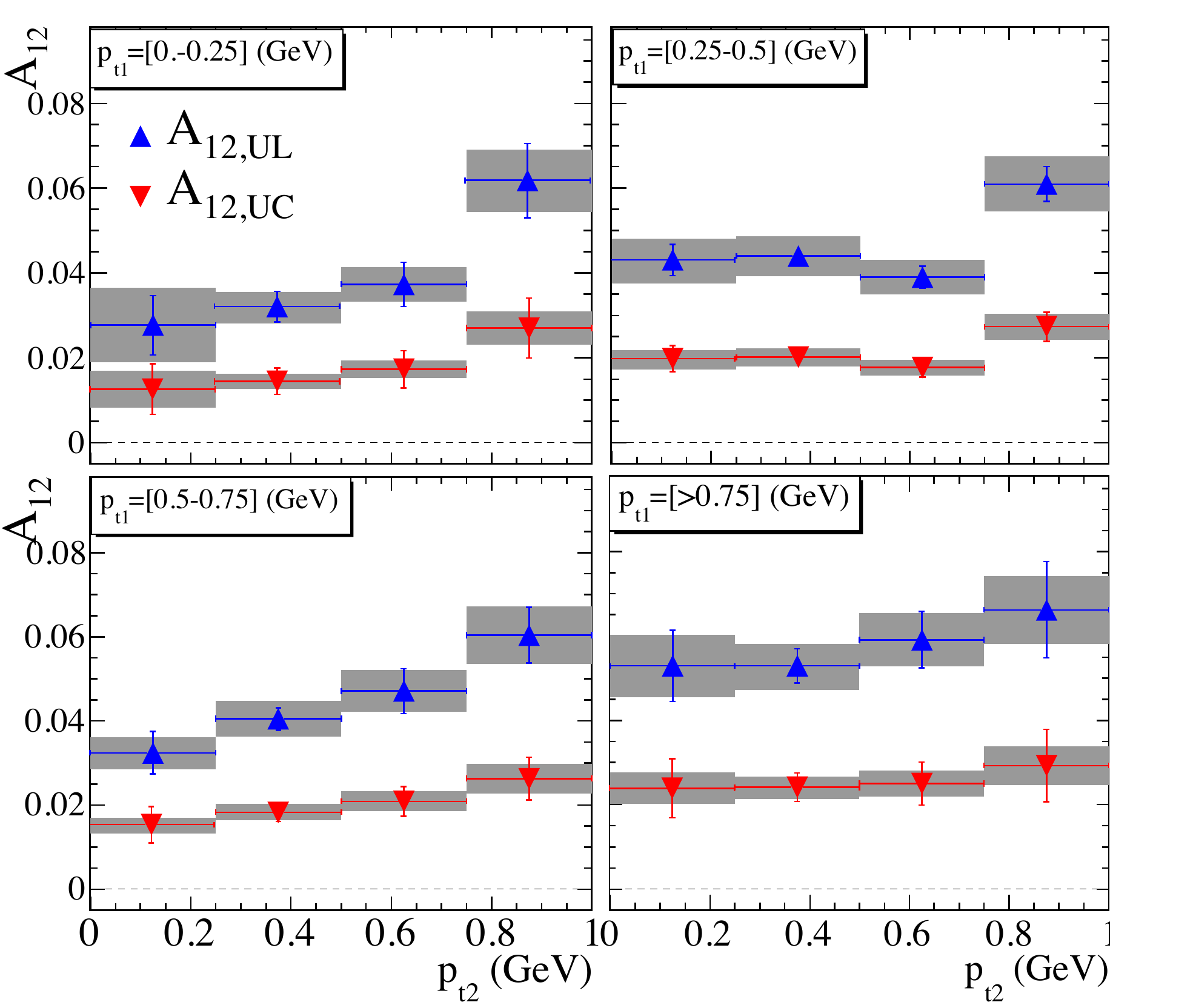}  
    \includegraphics[width=0.4\textwidth] {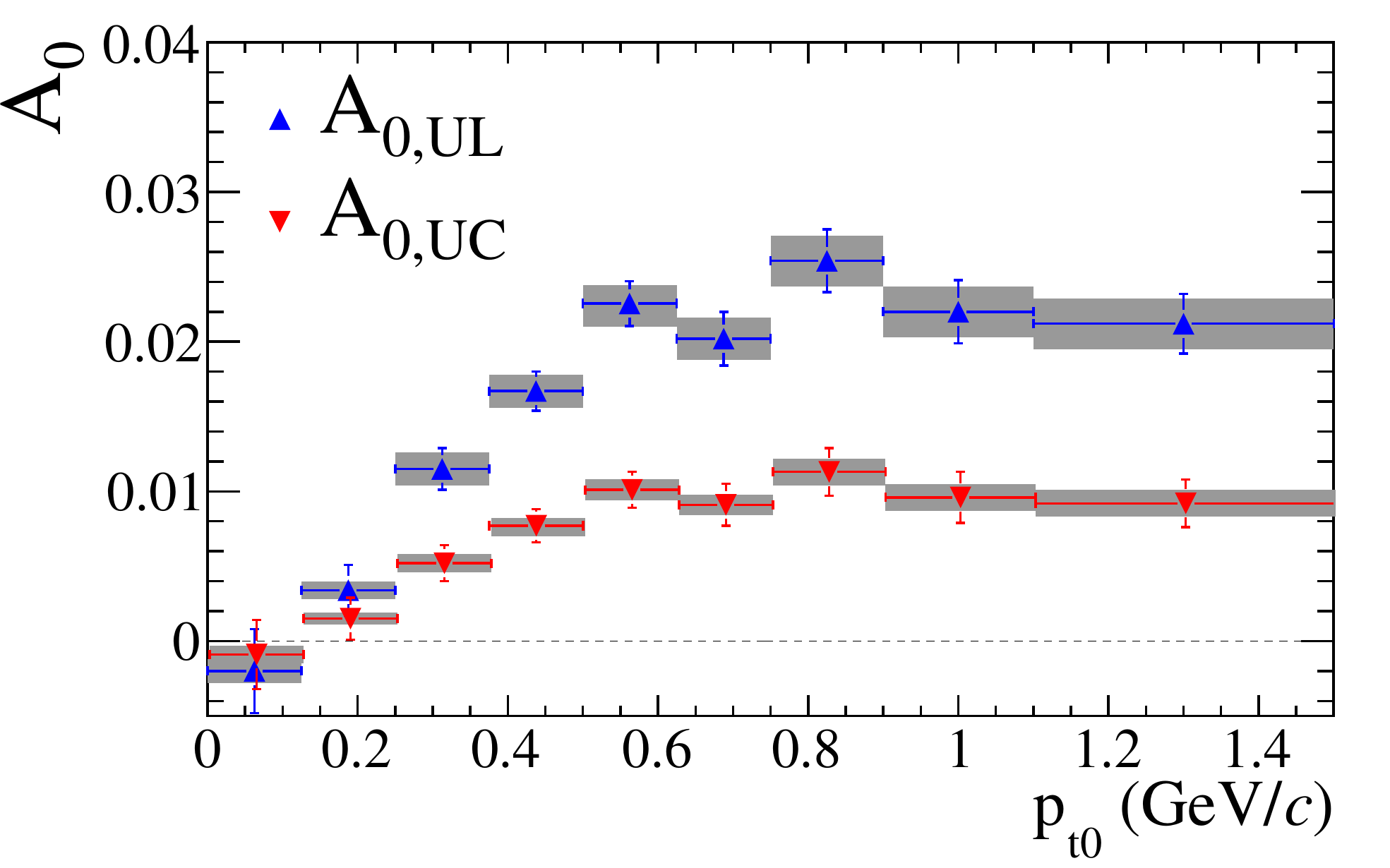} 
  \end{center}
  \caption{ (color online).
    Collins asymmetries for light quarks measured 
    in $(p_{t1},p_{t2}$ bins in RF12 (left plots), 
    and in nine bins of $p_{t0}$ (right plot) in RF0.
    Asymmetries for the UL (up triangles) and UC  (down triangles)
    ratios are reported,  with statistical  error bars and systematic
    uncertainties  represented by the bands around the points. 
  }
  \label{fig:finalPt}
\end{figure*}
The Collins asymmetries measured in the two reference frames, 
in bins of \ptbin\ and $p_{t0}$, are shown in Fig.~\ref{fig:finalPt}
 and Tab.~\ref{tab:totPt}.
The results from the two double ratios are reported.
This is the first measurement of the dependence on the pion transverse momenta in \epem annihilation, and
 is important for a theoretical understanding of the evolution of the
 Collins fragmentation function.
In RF0 the measured asymmetries are consistent with zero at very low
$p_{t0}$, rise almost linearly up to about 2\% for UL and 1\% for UC,
at 0.8 \gevc, and then flat.
In RF12 the asymmetries slightly differ from zero at low transverse
momenta, and exhibit also in this case a smooth rise of the
asymmetries with \ptbin\ up to a maximum of about 7\% and 3\% for UL and UC, respectively. 
Due to the limited resolution at very low transverse momenta it is not possible
to verify the expected vanishing of $A_{12}$ with $p_{t1,t2}$ going to zero.
The average $p_{t1,t2}$ value for the lowest bin is, in fact, $0.16$ \gev.

\subsection{Collins asymmetries  {\textit vs.} $z$ and $p_t$ }
\begin{figure*}[!htb]
\centering
 \includegraphics[width=0.85\textwidth] {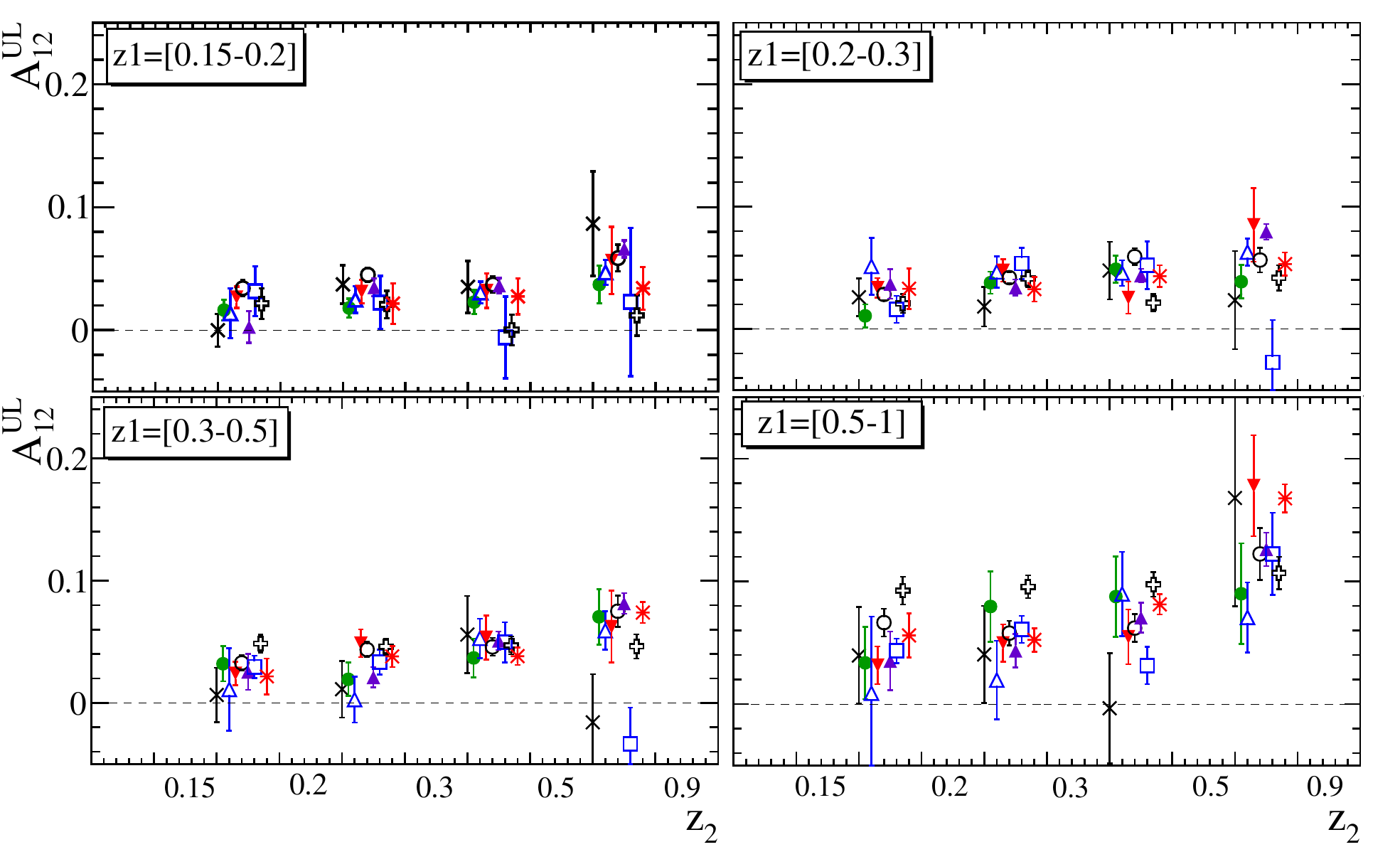} \\
 \includegraphics[width=0.85\textwidth] {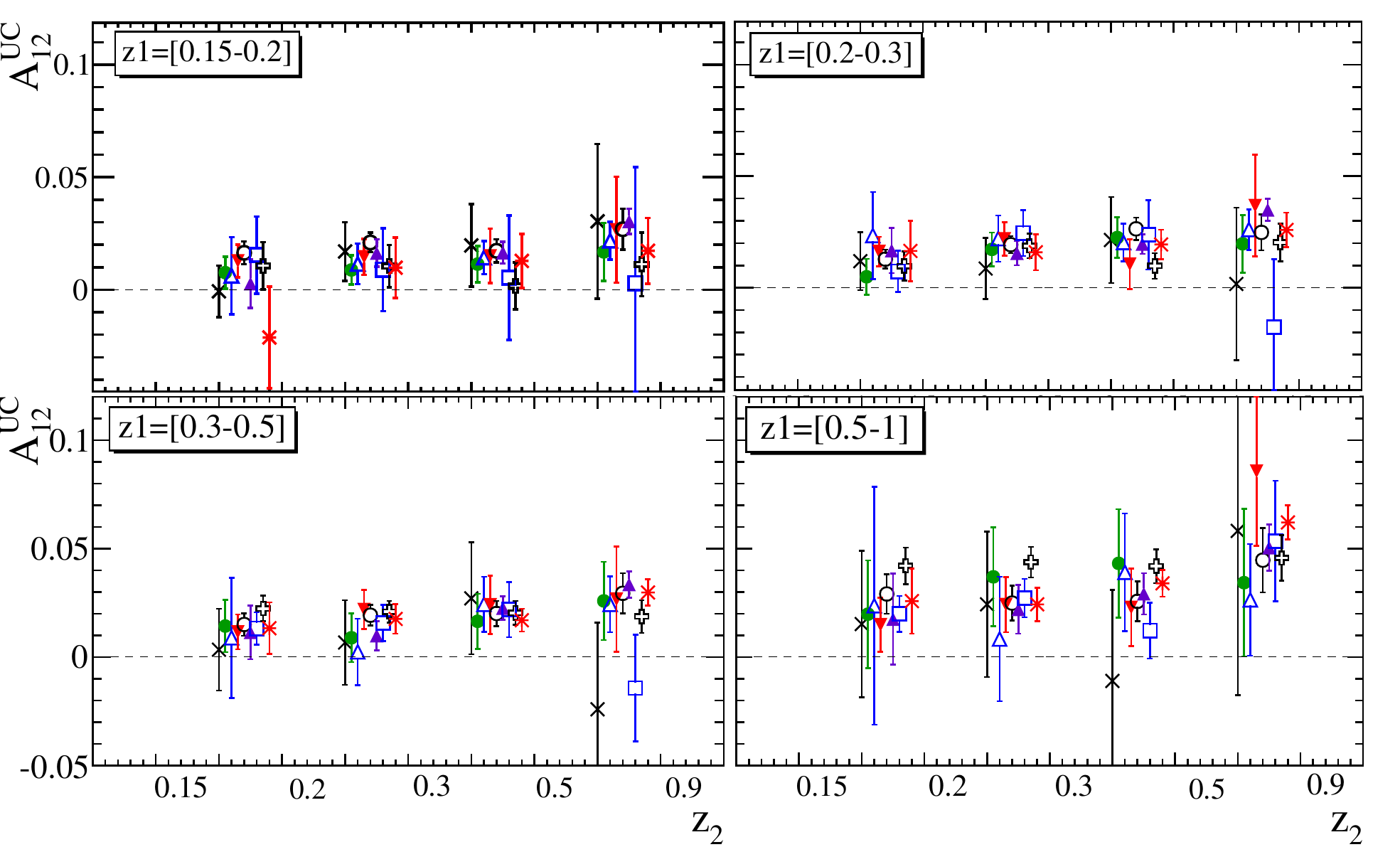} \\
  \includegraphics[width=0.8\textwidth] {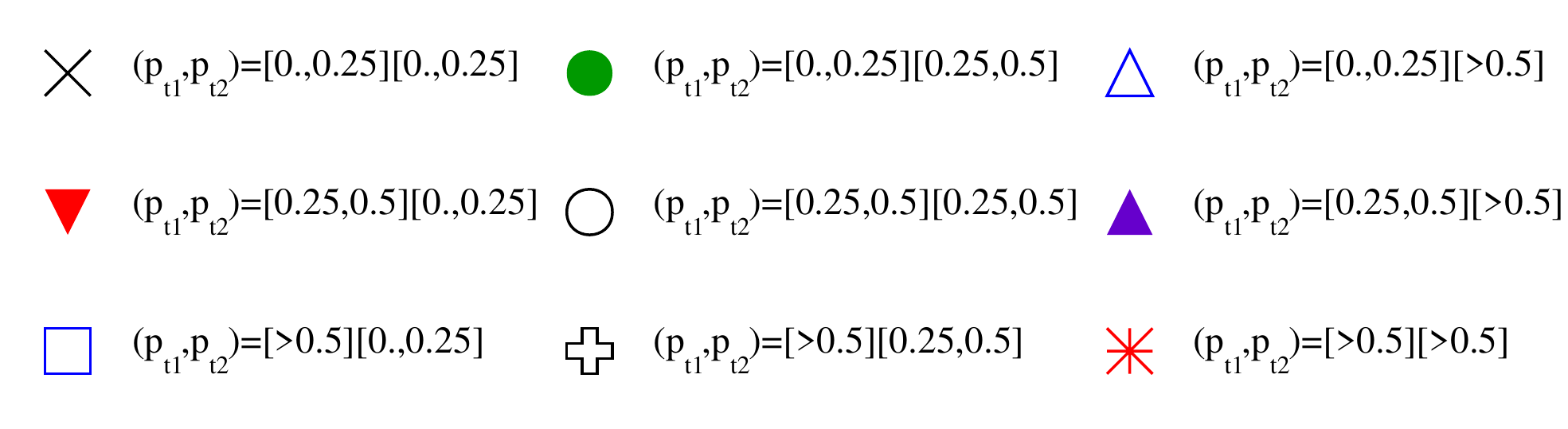} 
 \caption{ (color online).
 Light quark asymmetry $A^{UL}_{12}$ (up) and
 $A^{UC}_{12}$ (down) calculated in the RF12 frame 
  as a function of $(z_1,z_2,p_{t1},p_{t2})$.
The plots show the \zbin\ dependence for each 
\ptbin\ interval, identified by the different markers and colors as described in the legend.}
\label{fig:zpt}
\end{figure*}

The study of the asymmetry behavior  as a function of both
pion fractional energies $z_{1,2}$ and transverse momentum $p_{t1,t2}$ is an
important test to probe the factorization of the Collins fragmentation function
assumed in Ref.~\cite{PhysRevD.75.054032} and,
in general, provides a powerful tool to access $p_t-z$ correlations in the Collins asymmetries.
We perform this study in the thrust reference frame, using four $z_i$  and three $\pt_i$ ($i=1,2$) intervals.
The boundaries for the fractional energies
intervals are set to $z_i =0.15,\,0.2,\,0.3,\,0.5,$ and $\,0.9$, while
the  intervals for the transverse momenta are  $\pt_i<0.25$ \gev,
$0.25<\pt_i<0.5$ \gev, and $\pt_i>0.5$ \gev. 
We estimate background contributions, dilution effects, 
and systematic uncertainties independently for each 
$(z_1,z_2, \pt_1,\pt_2)$ bin, following the procedures described in the previous sections.
The results are summarized in Tabs.~\ref{tab:zpt},~\ref{tab:zptUC} and~\ref{tab:zptmean}, and in Fig.~\ref{fig:zpt}.

\subsection{Collins asymmetries {\textit vs.} polar angles \label{sec:AvsTheta}}
The transverse polarization of the original \qqbar
pair created in \epem annihilation should be proportional to
$\sin^2\theta$, where $\theta$ is the polar angle
of the \qqbar axis with respect to the beam axis.
The Collins asymmetry should manifest a similar dependence, as shown by
Eqs.\,(\ref{DR:UL}) and (\ref{eqn:uc1}) for the UL and UC ratios, respectively. 
We can test this prediction, and in particular that the asymmetries
vanish for $\theta=0$, by studying the asymmetries as a function of the
quantity $\sin^2\theta/(1+\cos^2\theta)$ after integration over $z$ and \pt, with the 
\qqbar polar angle estimated by  the polar angle of the thrust axis
($\theta_{th}$) or of the reference hadron ($\theta_2$).
\begin{figure*}[htb]
\begin{center}
 \includegraphics[width=0.48\textwidth] {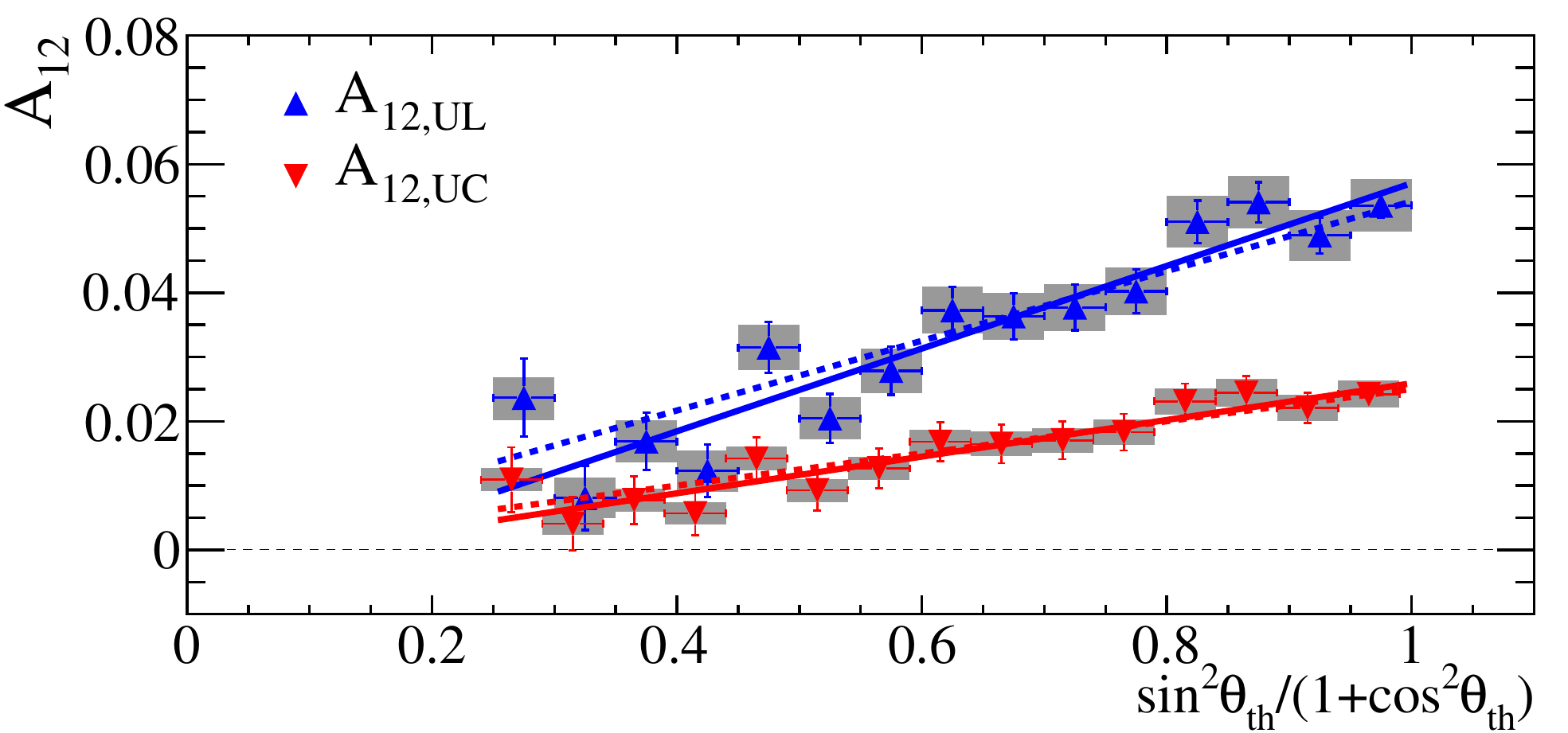}\label{fig:theta12}
{\boldmath      \put(-110,95){(a)}    }
  \includegraphics[width=0.48\textwidth] {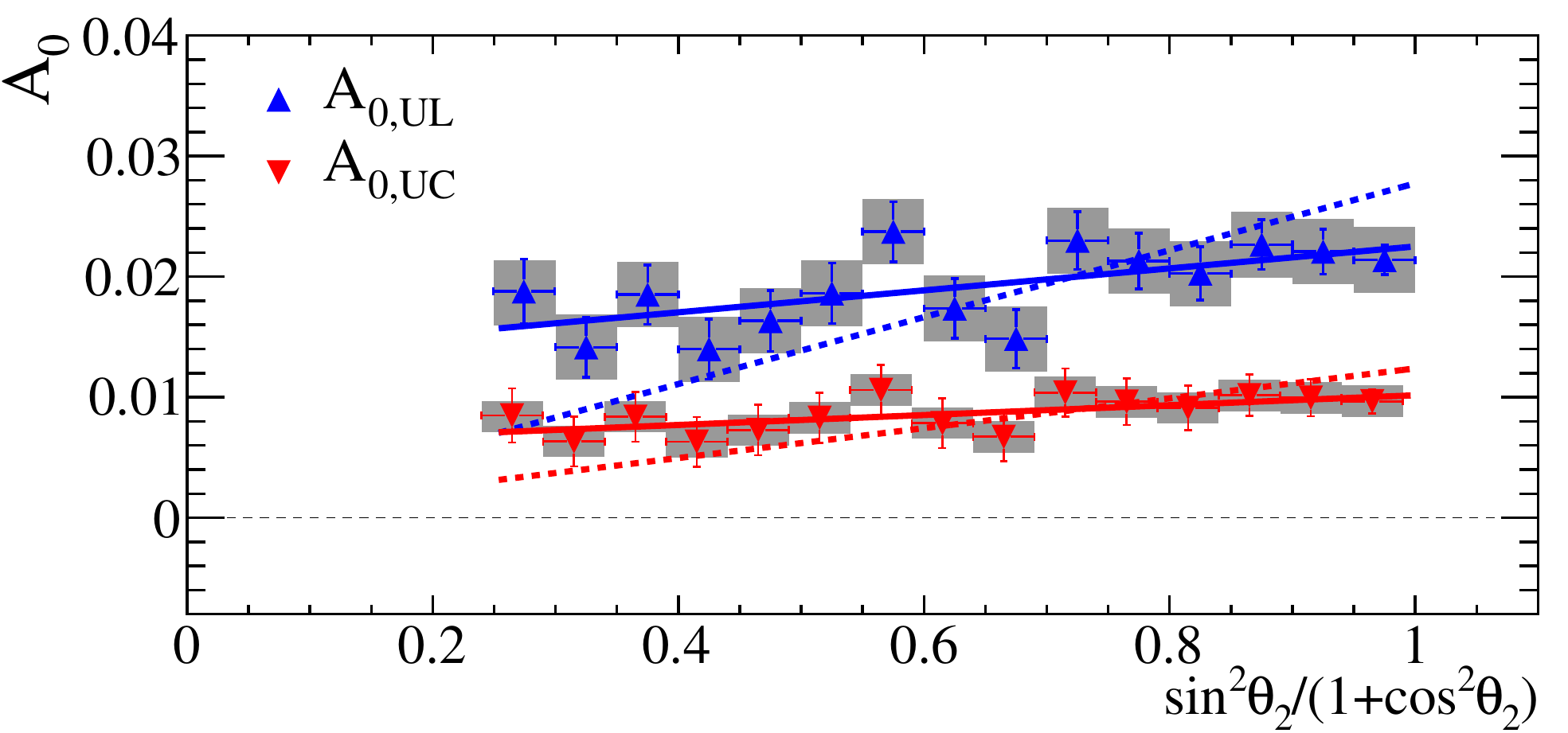}\label{fig:theta0}
  {\boldmath      \put(-110,95){(b)}    }
\end{center}
\caption{ (color online). Light quark asymmetry parameters calculated in RF12 (a)
 and in RF0 (b), as a function of $\sin^2\theta/(1+\cos^2\theta)$,
 where $\theta \equiv \theta_{th}$ in (a) and $\theta\equiv\theta_2$
 in (b). 
 The up-triangles (blue online) refer to the UL asymmetry, while the
 down-triangles (red online) to the UC asymmetry.
 Systematic contributions are shown by gray boxes.
 The result of the linear fits are shown  as solid lines of the
 corresponding colors, and  summarized in Tab.~\ref{tab:thetaTh}.
 The dashed lines represent fitted lines through the origin.
}
\label{fig:finalTheta}
\end{figure*}
The results are displayed  in Fig.~\ref{fig:finalTheta} and summarized in Tab.~\ref{tab:thetaasy}.
The asymmetries are corrected for background contributions and dilution
effects, following the same procedure described in
Secs.~\ref{sec:charm}  and \,\ref{sec:weight}.
As systematic uncertainties we assigned the average values of 
 the significant contributions studied in Sec.~\ref{sec:syst},
 which are added in quadrature in Fig.~\ref{fig:finalTheta} (gray error bands). 
 
We subject each set of data points to a linear fit  ($p_0+p_1\cdot
x$), both with the intercept parameter floating and constrained to the origin of the axes ($p_0=0$). 
The results of these fits are summarized in Tab.~\ref{tab:thetaTh}.
For $A_{12}$, all four fits have a good $\chi^2$, and
both fitted $p_0$ are consistent with zero (Fig.~\ref{fig:finalTheta}(a)).
Also the $A_0$   asymmetries are consistent with a linear dependence
on $\sin^2\theta/(1+\cos^2\theta)$, but the results 
 clearly favor a non zero constant term for both $A_0^{UL}$ and $A_0^{UC}$
 (Fig.~\ref{fig:finalTheta}(b)).
This behavior is in contradiction with the cross section formula reported in Eq.~\ref{cross0},
the origin of which is not understood.

\section{Summary}
We have presented a set of measurements of  azimuthal asymmetries 
in  inclusive production of charged pion pairs in $\epem\to\qqbar$ 
annihilation, where the two pions arise from the fragmenting quarks in opposite jets.
We consider two reference frames, and extract asymmetries from suitable ratios
of normalized azimuthal distributions: the ratio of 
opposite-sign to same-sign pion pairs (UL double ratio) and
the ratio of opposite-sign to any charge pion pairs (UC double ratio).
We observe clear, non-vanishing asymmetries that can be related
to the Collins fragmentation functions.

We measure the Collins asymmetries as a 
function of several variables, including  the transverse momenta \pt
($p_{t1}$ and $p_{t2}$, and $p_{t0}$) of the pions with respect to the analysis axis,
their fractional energies $z_{1,2}$, and polar angle $\theta$
of the analysis axis with respect to the beam axis.

The  asymmetries rise with $z_1$ and $z_2$,  as expected from
theoretical predictions, and span more than an order of magnitude.
Our data are also consistent with the previous measurements performed by the 
Belle Collaboration~\cite{PhysRevLett.96.232002,PhysRevD.78.032011},
with the exception of the bins where the highest fractional energies
are involved. In particular we measure an asymmetry about three
standard deviations higher in the highest ($z_1,z_2$) bin.
It must be noted, however, that the two data sets are not directly
comparable, because of the different width of the highest $z$
interval, which is  $0.7 < z < 1.0$ for Belle, 
while we limit our study to $0.7 < z < 0.9$ 
for the reasons explained in Sec.~\ref{sec:selection}.  

Theoretical calculations with different approaches have been proposed
to make predictions or to reproduce the available data (see for example
 Refs.~\cite{PhysRevD.73.094025,Bacchetta2008234} 
 and references therein). The new precise data presented here
can be used to improve the tuning of the various models and
possibly discriminate among the different assumptions.
These results can also be combined with Belle  and SIDIS data
 to improve the simultaneous extraction of favored and disfavored \collins, and the
transversity and other chiral-odd parton distributions, following, e.g.,
the approach of Ref.~\cite{PhysRevD.75.054032,Anselmino200998}.

There is an increase of the asymmetries with increasing \pt,  
more pronounced in the RF0 system, and there is
an indication that a maximum value of the asymmetry is reached at $p_{t} \simeq 0.8$ \gev. 
The available data sample is not sufficient to
effectively study the region above 1~\gevc, and determine if the 
asymmetries decrease after the peak is reached.
  No previous data from \epem annihilation  are available to compare with.
Assuming that the $Q^2$ evolution\footnote{In SIDIS processes $Q^2$ is the 4-momentum
transferred from the lepton to the struck hadron, while in $e^+e^-$ annihilation it is equal to s.}
of the Collins function is the same as
for the unpolarized function, the authors of Ref.~\cite{PhysRevD.75.054032} extracted the
\pt dependence at  $Q^2=2.4$ $\gev^2$.
However, such an assumption on the evolution with $Q^2$ 
is not established, and even questionable, given the chiral-odd nature of the Collins function.
Alternative choices have been proposed, including the extreme
case that the Collins function does not evolve with  the energy scale~\cite{Bacchetta2008234}.
The measurement of the asymmetry as a function of \pt  obtained by 
\babar\ at $Q^2\simeq110$ $\gev^2$, that is an energy scale much 
higher with respect to the data presently available from SIDIS, 
will be very valuable in shedding light on this important question. 

We also measured the asymmetries in bins of the quantity
$\sin^2\theta/(1+\cos^2\theta)$, where $\theta$ is
$\theta_{th}$ or $\theta_2$  according to the 
reference frame used, as defined in Fig.~\ref{fig:rf12}
and Fig.~\ref{fig:rf0}, respectively.
The expected linear dependence is observed in
both reference frames. However in RF0 the fit result is
inconsistent with a line crossing the origin, in disagreement with expectations.

\section{ACKNOWLEDGMENTS}
\label{sec:Acknowledgments}
The authors would like to  thank Ralf Seidl of the Belle 
 Collaboration for the fruitful discussions on the analysis method.
 We are also grateful to Mauro Anselmino  and  Marco Contalbrigo,
for the helpful discussions on the theoretical aspects of the measurement.\\
\indent We are grateful for the extraordinary contributions of our \pep2\ colleagues in
achieving the excellent luminosity and machine conditions
that have made this work possible.
The success of this project also relies critically on the 
expertise and dedication of the computing organizations that support \babar.
The collaborating institutions wish to thank 
SLAC for its support and the kind hospitality extended to them. 
This work is supported by the US Department of Energy
and National Science Foundation, the
Natural Sciences and Engineering Research Council (Canada),
the Commissariat \`a l'Energie Atomique and
Institut National de Physique Nucl\'eaire et de Physique des Particules
(France), the Bundesministerium f\"ur Bildung und Forschung and
Deutsche Forschungsgemeinschaft (Germany), the
Istituto Nazionale di Fisica Nucleare (Italy),
the Foundation for Fundamental Research on Matter (The Netherlands),
the Research Council of Norway, the Ministry of Education and Science of the Russian Federation, 
Ministerio de Econom\'{\i}a y Competitividad (Spain), and the
Science and Technology Facilities Council (United Kingdom).
Individuals have received support from the Marie-Curie IEF program (European Union) and the A. P. Sloan Foundation (USA).

\begin{table*}[!hbt]
\begin{center}
\caption{Azimuthal asymmetries, in percent, obtained by fitting the UL double ratio in bins
\zbin. The errors are statistical and systematic.
The table also reports the average values of $z_{1,2}$, $p_{t1,t2,t0}$,  
 and $s^2_{i}/(1+c^2_{i})$, with $s=\sin\theta,\, c=\cos\theta, \, i=th,2$.  
} 
\begin{tabular}{ c c c c c c c c c c c c}
\toprule 
\multirow{2}{*}{$z_1$} & \multirow{2}{*}{$\langle z_1 \rangle$} & $\langle p_{t1} \rangle$ & \multirow{2}{*}{$z_2$} & \multirow{2}{*}{$\langle z_2 \rangle$} & $\langle p_{t2} \rangle$ & $\langle p_{t0} \rangle$ & \multirow{2}{*}{$ \left\langle\frac{s_{th}}{1+c_{th}}\right\rangle$} & \multirow{2}{*}{$\left\langle\frac{s_{2}}{1+c_{2}}\right\rangle$} &  \multirow{2}{*}{\aulTh (\%)} & & \multirow{2}{*}{ \aul  (\%)} \\
& & \footnotesize{(GeV)}  & & & \footnotesize{(GeV)} & \footnotesize{(GeV)}    & & & & \\
 \toprule
 $[0.15,0.2]$ & 0.174 & 0.308 & $[0.15,0.2]$ & 0.174 & 0.307 & 0.372 & 0.716 & 0.687 & 2.16 $\pm$ 0.20 $\pm$ 0.09 & & 1.11 $\pm$ 0.14 $\pm$ 0.05\\
 $[0.15,0.2]$ & 0.174 & 0.310 & $[0.2,0.3]$ & 0.244 & 0.369 & 0.364 & 0.715 & 0.683 & 3.07 $\pm$ 0.18 $\pm$ 0.12 & & 1.43 $\pm$ 0.12 $\pm$ 0.05\\
 $[0.15,0.2]$ & 0.174 & 0.313 & $[0.3,0.4]$ & 0.344 & 0.431 & 0.356 & 0.711 & 0.676 & 2.99 $\pm$ 0.27 $\pm$ 0.16 & & 1.34 $\pm$ 0.16 $\pm$ 0.06\\
 $[0.15,0.2]$ & 0.174 & 0.315 & $[0.4,0.5]$ & 0.444 & 0.475 & 0.351 & 0.706 & 0.671 & 3.42 $\pm$ 0.28 $\pm$ 0.17 & & 1.44 $\pm$ 0.17 $\pm$ 0.06\\
 $[0.15,0.2]$ & 0.174 & 0.318 & $[0.5,0.7]$ & 0.577 & 0.506 & 0.346 & 0.697 & 0.663 & 4.77 $\pm$ 0.39 $\pm$ 0.29 & & 1.76 $\pm$ 0.20 $\pm$ 0.07\\
 $[0.15,0.2]$ & 0.174 & 0.322 & $[0.7,0.9]$ & 0.772 & 0.439 & 0.339 & 0.683 & 0.655 & 7.02 $\pm$ 0.71 $\pm$ 0.50 & & 2.48 $\pm$ 0.26 $\pm$ 0.08\\
 $[0.2,0.3]$ & 0.244 & 0.369 & $[0.15,0.2]$ & 0.174 & 0.309 & 0.511 & 0.715 & 0.685 & 2.99 $\pm$ 0.18 $\pm$ 0.11 & &1.24 $\pm$ 0.12 $\pm$ 0.05\\
 $[0.2,0.3]$ & 0.244 & 0.372 & $[0.2,0.3]$ & 0.244 & 0.372 & 0.490 & 0.713 & 0.682 & 3.55 $\pm$ 0.18 $\pm$ 0.13 & &1.53 $\pm$ 0.11 $\pm$ 0.06\\
 $[0.2,0.3]$ & 0.244 & 0.376 & $[0.3,0.4]$ & 0.344 & 0.435 & 0.467 & 0.710 & 0.676 & 3.99 $\pm$ 0.27 $\pm$ 0.19 & &1.83 $\pm$ 0.17 $\pm$ 0.06\\
 $[0.2,0.3]$ & 0.244 & 0.378 & $[0.4,0.5]$ & 0.444 & 0.479 & 0.453 & 0.706 & 0.672 & 4.38 $\pm$ 0.27 $\pm$ 0.20 & &1.76 $\pm$ 0.17 $\pm$ 0.06\\
 $[0.2,0.3]$ & 0.244 & 0.380 & $[0.5,0.7]$ & 0.577 & 0.511 & 0.438 & 0.698 & 0.665 & 5.93 $\pm$ 0.36 $\pm$ 0.30 & &2.32 $\pm$ 0.19 $\pm$ 0.08\\
 $[0.2,0.3]$ & 0.244 & 0.379 & $[0.7,0.9]$ & 0.772 & 0.442 & 0.412 & 0.685 & 0.658 & 8.98 $\pm$ 0.65 $\pm$ 0.49 & &3.07 $\pm$ 0.24 $\pm$ 0.08\\
 $[0.3,0.4]$ & 0.344 & 0.430 & $[0.15,0.2]$ & 0.174 & 0.311 & 0.707 & 0.710 & 0.682 & 2.69 $\pm$ 0.27 $\pm$ 0.17 & &1.27 $\pm$ 0.16 $\pm$ 0.06\\
 $[0.3,0.4]$ & 0.344 & 0.434 & $[0.2,0.3]$ & 0.244 & 0.375 & 0.661 & 0.710 & 0.680 & 3.34 $\pm$ 0.25 $\pm$ 0.20 & &1.41 $\pm$ 0.16 $\pm$ 0.06\\
 $[0.3,0.4]$ & 0.344 & 0.438 & $[0.3,0.4]$ & 0.344 & 0.438 & 0.613 & 0.706 & 0.674 & 4.80 $\pm$ 0.42 $\pm$ 0.31 & &2.26 $\pm$ 0.31 $\pm$ 0.12\\
 $[0.3,0.4]$ & 0.344 & 0.440 & $[0.4,0.5]$ & 0.444 & 0.482 & 0.580 & 0.702 & 0.671 & 4.33 $\pm$ 0.41 $\pm$ 0.30 & &1.95 $\pm$ 0.29 $\pm$ 0.11\\
 $[0.3,0.4]$ & 0.344 & 0.439 & $[0.5,0.7]$ & 0.577 & 0.515 & 0.548 & 0.696 & 0.665 & 5.71 $\pm$ 0.53 $\pm$ 0.45 & &2.74 $\pm$ 0.32 $\pm$ 0.18\\
 $[0.3,0.4]$ & 0.344 & 0.433 & $[0.7,0.9]$ & 0.772 & 0.442 & 0.493 & 0.684 & 0.659 & 9.62 $\pm$ 0.93 $\pm$ 0.71 & &4.31 $\pm$ 0.35 $\pm$ 0.22\\
 $[0.4,0.5]$ & 0.444 & 0.475 & $[0.15,0.2]$ & 0.174 & 0.313 & 0.901 & 0.706 & 0.680 & 3.21 $\pm$ 0.27 $\pm$ 0.19 & &1.03 $\pm$ 0.17 $\pm$ 0.06\\
 $[0.4,0.5]$ & 0.444 & 0.479 & $[0.2,0.3]$ & 0.244 & 0.377 & 0.828 & 0.706 & 0.678 & 3.70 $\pm$ 0.25 $\pm$ 0.21 & &1.69 $\pm$ 0.17 $\pm$ 0.06\\
 $[0.4,0.5]$ & 0.444 & 0.483 & $[0.3,0.4]$ & 0.344 & 0.439 & 0.752 & 0.702 & 0.673 & 4.40 $\pm$ 0.41 $\pm$ 0.30 & &2.04 $\pm$ 0.29 $\pm$ 0.11\\
 $[0.4,0.5]$ & 0.444 & 0.484 & $[0.4,0.5]$ & 0.444 & 0.483 & 0.699 & 0.699 & 0.669 & 4.87 $\pm$ 0.47 $\pm$ 0.31 & &1.96 $\pm$ 0.31 $\pm$ 0.11\\
 $[0.4,0.5]$ & 0.444 & 0.481 & $[0.5,0.7]$ & 0.578 & 0.516 & 0.647 & 0.693 & 0.664 & 6.20 $\pm$ 0.57 $\pm$ 0.46 & &3.36 $\pm$ 0.34 $\pm$ 0.20\\
 $[0.4,0.5]$ & 0.445 & 0.471 & $[0.7,0.9]$ & 0.774 & 0.441 & 0.564 & 0.681 & 0.659 & 12.62 $\pm$ 1.36 $\pm$ 0.82 & &4.84 $\pm$ 0.51 $\pm$ 0.24\\
 $[0.5,0.7]$ & 0.577 & 0.506 & $[0.15,0.2]$ & 0.174 & 0.315 & 1.155 & 0.697 & 0.677 & 4.81 $\pm$ 0.39 $\pm$ 0.33 & &1.62 $\pm$ 0.20 $\pm$ 0.10\\
 $[0.5,0.7]$ & 0.577 & 0.510 & $[0.2,0.3]$ & 0.244 & 0.379 & 1.042 & 0.698 & 0.675 & 5.55 $\pm$ 0.36 $\pm$ 0.34 & &2.07 $\pm$ 0.19 $\pm$ 0.10\\
 $[0.5,0.7]$ & 0.577 & 0.514 & $[0.3,0.4]$ & 0.344 & 0.438 & 0.921 & 0.696 & 0.671 & 6.09 $\pm$ 0.55 $\pm$ 0.37 & &2.85 $\pm$ 0.32 $\pm$ 0.19\\
 $[0.5,0.7]$ & 0.578 & 0.514 & $[0.4,0.5]$ & 0.444 & 0.481 & 0.840 & 0.693 & 0.667 & 7.82 $\pm$ 0.60 $\pm$ 0.39 & &3.17 $\pm$ 0.34 $\pm$ 0.19\\
 $[0.5,0.7]$ & 0.579 & 0.515 & $[0.5,0.7]$ & 0.579 & 0.515 & 0.769 & 0.688 & 0.662 & 10.97 $\pm$ 0.73 $\pm$ 0.62 & &4.57 $\pm$ 0.37 $\pm$ 0.25\\
 $[0.5,0.7]$ & 0.580 & 0.508 & $[0.7,0.9]$ & 0.775 & 0.440 & 0.653 & 0.675 & 0.656 & 19.80 $\pm$ 1.50 $\pm$ 1.57 & &7.66 $\pm$ 0.62 $\pm$ 0.31\\
 $[0.7,0.9]$ & 0.772 & 0.437 & $[0.15,0.2]$ & 0.174 & 0.319 & 1.515 & 0.683 & 0.672 & 7.28 $\pm$ 0.75 $\pm$ 0.55 & &2.37 $\pm$ 0.26 $\pm$ 0.11\\
 $[0.7,0.9]$ & 0.772 & 0.442 & $[0.2,0.3]$ & 0.244 & 0.379 & 1.314 & 0.685 & 0.672 & 8.67 $\pm$ 0.62 $\pm$ 0.56 & &2.99 $\pm$ 0.24 $\pm$ 0.11\\
 $[0.7,0.9]$ & 0.773 & 0.443 & $[0.3,0.4]$ & 0.344 & 0.433 & 1.111 & 0.684 & 0.668 & 9.52 $\pm$ 0.88 $\pm$ 0.61 & &4.00 $\pm$ 0.35 $\pm$ 0.20\\
 $[0.7,0.9]$ & 0.774 & 0.442 & $[0.4,0.5]$ & 0.445 & 0.469 & 0.978 & 0.682 & 0.665 & 11.92 $\pm$ 1.29 $\pm$ 0.69 & &5.91 $\pm$ 0.52 $\pm$ 0.24\\
 $[0.7,0.9]$ & 0.775 & 0.440 & $[0.5,0.7]$ & 0.580 & 0.506 & 0.869 & 0.676 & 0.659 & 20.73 $\pm$ 1.59 $\pm$ 1.41 & &7.26 $\pm$ 0.61 $\pm$ 0.30\\
 $[0.7,0.9]$ & 0.776 & 0.458 & $[0.7,0.9]$ & 0.776 & 0.461 & 0.735 & 0.664 & 0.651 & 39.31 $\pm$ 3.26 $\pm$ 5.06 & &11.10 $\pm$ 1.19 $\pm$ 0.39\\
\toprule 
\end{tabular}
\label{tab:totZul}
\end{center}
\end{table*}

\begin{table*}[!hbt]
\begin{center}
\caption{
Azimuthal asymmetries, in percent, obtained by fitting the UC double ratio in bins
\zbin. The errors are statistical and systematic.
The table also reports the average values of $z_{1,2}$, $p_{t1,t2,t0}$,  
 and $s^2_{i}/(1+c^2_{i})$, with $s=\sin\theta,\, c=\cos\theta, \, i=th,2$.
}
\begin{tabular}{ c c c c c c c c c c c c}
\toprule 
\multirow{2}{*}{$z_1$} & \multirow{2}{*}{$\langle z_1 \rangle$} &  $\langle p_{t1} \rangle$ & \multirow{2}{*}{$z_2$} & \multirow{2}{*}{$\langle z_2 \rangle$} & $\langle p_{t2} \rangle$ &  $\langle p_{t0} \rangle$ & \multirow{2}{*}{$ \left\langle\frac{s_{th}}{1+c_{th}}\right\rangle$} & \multirow{2}{*}{$\left\langle\frac{s_{2}}{1+c_{2}}\right\rangle$} &  \multirow{2}{*}{\aucTh (\%)} & & \multirow{2}{*}{ \auc  (\%)} \\
  & & \footnotesize{(GeV)}  & & & \footnotesize{(GeV)} & \footnotesize{(GeV)}    & & & & \\
 \toprule
 $[0.15,0.2]$ & 0.174 & 0.308 & $[0.15,0.2]$ & 0.174 & 0.307 & 0.372 & 0.716 & 0.687 & 1.01 $\pm$ 0.16 $\pm$ 0.05 & & 0.52 $\pm$ 0.11 $\pm$ 0.03\\
 $[0.15,0.2]$ & 0.174 & 0.310 & $[0.2,0.3]$ & 0.244 & 0.369 & 0.364 & 0.715 & 0.683 & 1.41 $\pm$ 0.15 $\pm$ 0.06 & & 0.66 $\pm$ 0.10 $\pm$ 0.03\\
 $[0.15,0.2]$ & 0.174 & 0.313 & $[0.3,0.4]$ & 0.344 & 0.431 & 0.356 & 0.711 & 0.676 & 1.37 $\pm$ 0.22 $\pm$ 0.09 & & 0.61 $\pm$ 0.14 $\pm$ 0.04\\
 $[0.15,0.2]$ & 0.174 & 0.315 & $[0.4,0.5]$ & 0.444 & 0.475 & 0.351 & 0.706 & 0.671 & 1.54 $\pm$ 0.23 $\pm$ 0.09 & & 0.65 $\pm$ 0.14 $\pm$ 0.04\\
 $[0.15,0.2]$ & 0.174 & 0.318 & $[0.5,0.7]$ & 0.577 & 0.506 & 0.346 & 0.697 & 0.663 & 2.19 $\pm$ 0.33 $\pm$ 0.16 & & 0.78 $\pm$ 0.17 $\pm$ 0.06\\
 $[0.15,0.2]$ & 0.174 & 0.322 & $[0.7,0.9]$ & 0.772 & 0.439 & 0.339 & 0.683 & 0.655 & 3.16 $\pm$ 0.61 $\pm$ 0.26 & & 1.05 $\pm$ 0.21 $\pm$ 0.06\\
 $[0.2,0.3]$ & 0.244 & 0.369 & $[0.15,0.2]$ & 0.174 & 0.309 & 0.511 & 0.715 & 0.685 & 1.38 $\pm$ 0.15 $\pm$ 0.06 & &0.57 $\pm$ 0.10 $\pm$ 0.03\\
 $[0.2,0.3]$ & 0.244 & 0.372 & $[0.2,0.3]$ & 0.244 & 0.372 & 0.490 & 0.713 & 0.682 & 1.67 $\pm$ 0.14 $\pm$ 0.07 & &0.70 $\pm$ 0.09 $\pm$ 0.03\\
 $[0.2,0.3]$ & 0.244 & 0.376 & $[0.3,0.4]$ & 0.344 & 0.435 & 0.467 & 0.710 & 0.676 & 1.81 $\pm$ 0.21 $\pm$ 0.10 & &0.82 $\pm$ 0.13 $\pm$ 0.04\\
 $[0.2,0.3]$ & 0.244 & 0.378 & $[0.4,0.5]$ & 0.444 & 0.479 & 0.453 & 0.706 & 0.672 & 1.94 $\pm$ 0.21 $\pm$ 0.10 & &0.77 $\pm$ 0.13 $\pm$ 0.04\\
 $[0.2,0.3]$ & 0.244 & 0.380 & $[0.5,0.7]$ & 0.577 & 0.511 & 0.438 & 0.698 & 0.665 & 2.65 $\pm$ 0.30 $\pm$ 0.16 & &1.00 $\pm$ 0.15 $\pm$ 0.06\\
 $[0.2,0.3]$ & 0.244 & 0.379 & $[0.7,0.9]$ & 0.772 & 0.442 & 0.412 & 0.685 & 0.658 & 3.86 $\pm$ 0.55 $\pm$ 0.26 & &1.25 $\pm$ 0.19 $\pm$ 0.06\\
 $[0.3,0.4]$ & 0.344 & 0.430 & $[0.15,0.2]$ & 0.174 & 0.311 & 0.707 & 0.710 & 0.682 & 1.24 $\pm$ 0.22 $\pm$ 0.08 & &0.58 $\pm$ 0.14 $\pm$ 0.04\\
 $[0.3,0.4]$ & 0.344 & 0.434 & $[0.2,0.3]$ & 0.244 & 0.375 & 0.661 & 0.710 & 0.680 & 1.51 $\pm$ 0.20 $\pm$ 0.09 & &0.63 $\pm$ 0.13 $\pm$ 0.04\\
 $[0.3,0.4]$ & 0.344 & 0.438 & $[0.3,0.4]$ & 0.344 & 0.438 & 0.613 & 0.706 & 0.674 & 2.11 $\pm$ 0.31 $\pm$ 0.15 & &0.95 $\pm$ 0.19 $\pm$ 0.07\\
 $[0.3,0.4]$ & 0.344 & 0.440 & $[0.4,0.5]$ & 0.444 & 0.482 & 0.580 & 0.702 & 0.671 & 1.84 $\pm$ 0.31 $\pm$ 0.15 & &0.80 $\pm$ 0.19 $\pm$ 0.07\\
 $[0.3,0.4]$ & 0.344 & 0.439 & $[0.5,0.7]$ & 0.577 & 0.515 & 0.548 & 0.696 & 0.665 & 2.50 $\pm$ 0.43 $\pm$ 0.24 & &1.11 $\pm$ 0.22 $\pm$ 0.11\\
 $[0.3,0.4]$ & 0.344 & 0.433 & $[0.7,0.9]$ & 0.772 & 0.442 & 0.493 & 0.684 & 0.659 & 3.84 $\pm$ 0.76 $\pm$ 0.35 & &1.66 $\pm$ 0.28 $\pm$ 0.12\\
 $[0.4,0.5]$ & 0.444 & 0.475 & $[0.15,0.2]$ & 0.174 & 0.313 & 0.901 & 0.706 & 0.680 & 1.44 $\pm$ 0.23 $\pm$ 0.09 & &0.47 $\pm$ 0.14 $\pm$ 0.04\\
 $[0.4,0.5]$ & 0.444 & 0.479 & $[0.2,0.3]$ & 0.244 & 0.377 & 0.828 & 0.706 & 0.678 & 1.63 $\pm$ 0.21 $\pm$ 0.09 & &0.74 $\pm$ 0.13 $\pm$ 0.04\\
 $[0.4,0.5]$ & 0.444 & 0.483 & $[0.3,0.4]$ & 0.344 & 0.439 & 0.752 & 0.702 & 0.673 & 1.87 $\pm$ 0.31 $\pm$ 0.15 & &0.84 $\pm$ 0.19 $\pm$ 0.07\\
 $[0.4,0.5]$ & 0.444 & 0.484 & $[0.4,0.5]$ & 0.444 & 0.483 & 0.699 & 0.699 & 0.669 & 1.99 $\pm$ 0.36 $\pm$ 0.15 & &0.78 $\pm$ 0.22 $\pm$ 0.07\\
 $[0.4,0.5]$ & 0.444 & 0.481 & $[0.5,0.7]$ & 0.578 & 0.516 & 0.647 & 0.693 & 0.664 & 2.59 $\pm$ 0.47 $\pm$ 0.24 & &1.32 $\pm$ 0.24 $\pm$ 0.11\\
 $[0.4,0.5]$ & 0.445 & 0.471 & $[0.7,0.9]$ & 0.774 & 0.441 & 0.564 & 0.681 & 0.659 & 4.80 $\pm$ 1.07 $\pm$ 0.37 & &1.77 $\pm$ 0.39 $\pm$ 0.12\\
 $[0.5,0.7]$ & 0.577 & 0.506 & $[0.15,0.2]$ & 0.174 & 0.315 & 1.155 & 0.697 & 0.677 & 2.22 $\pm$ 0.33 $\pm$ 0.17 & &0.72 $\pm$ 0.17 $\pm$ 0.06\\
 $[0.5,0.7]$ & 0.577 & 0.510 & $[0.2,0.3]$ & 0.244 & 0.379 & 1.042 & 0.698 & 0.675 & 2.49 $\pm$ 0.30 $\pm$ 0.17 & &0.90 $\pm$ 0.15 $\pm$ 0.06\\
 $[0.5,0.7]$ & 0.577 & 0.514 & $[0.3,0.4]$ & 0.344 & 0.438 & 0.921 & 0.696 & 0.671 & 2.64 $\pm$ 0.43 $\pm$ 0.23 & &1.13 $\pm$ 0.22 $\pm$ 0.10\\
 $[0.5,0.7]$ & 0.578 & 0.514 & $[0.4,0.5]$ & 0.444 & 0.481 & 0.840 & 0.693 & 0.667 & 3.25 $\pm$ 0.48 $\pm$ 0.23 & &1.23 $\pm$ 0.24 $\pm$ 0.11\\
 $[0.5,0.7]$ & 0.579 & 0.515 & $[0.5,0.7]$ & 0.579 & 0.515 & 0.769 & 0.688 & 0.662 & 4.56 $\pm$ 0.60 $\pm$ 0.37 & &1.74 $\pm$ 0.28 $\pm$ 0.16\\
 $[0.5,0.7]$ & 0.580 & 0.508 & $[0.7,0.9]$ & 0.775 & 0.440 & 0.653 & 0.675 & 0.656 & 7.69 $\pm$ 1.15 $\pm$ 0.62 & &2.70 $\pm$ 0.44 $\pm$ 0.17\\
 $[0.7,0.9]$ & 0.772 & 0.437 & $[0.15,0.2]$ & 0.174 & 0.319 & 1.515 & 0.683 & 0.672 & 3.18 $\pm$ 0.63 $\pm$ 0.27 & &1.00 $\pm$ 0.21 $\pm$ 0.06\\
 $[0.7,0.9]$ & 0.772 & 0.442 & $[0.2,0.3]$ & 0.244 & 0.379 & 1.314 & 0.685 & 0.672 & 3.79 $\pm$ 0.54 $\pm$ 0.27 & &1.22 $\pm$ 0.19 $\pm$ 0.07\\
 $[0.7,0.9]$ & 0.773 & 0.443 & $[0.3,0.4]$ & 0.344 & 0.433 & 1.111 & 0.684 & 0.668 & 3.93 $\pm$ 0.74 $\pm$ 0.32 & &1.53 $\pm$ 0.28 $\pm$ 0.11\\
 $[0.7,0.9]$ & 0.774 & 0.442 & $[0.4,0.5]$ & 0.445 & 0.469 & 0.978 & 0.682 & 0.665 & 4.72 $\pm$ 1.05 $\pm$ 0.34 & &2.16 $\pm$ 0.40 $\pm$ 0.12\\
 $[0.7,0.9]$ & 0.775 & 0.440 & $[0.5,0.7]$ & 0.580 & 0.506 & 0.869 & 0.676 & 0.659 & 7.72 $\pm$ 1.18 $\pm$ 0.61 & &2.56 $\pm$ 0.44 $\pm$ 0.17\\
 $[0.7,0.9]$ & 0.776 & 0.458 & $[0.7,0.9]$ & 0.776 & 0.461 & 0.735 & 0.664 & 0.651 & 13.50 $\pm$ 2.34 $\pm$ 1.50 & &3.52 $\pm$ 0.83 $\pm$ 0.18\\
\toprule
\end{tabular}
\label{tab:totZuc}
\end{center}
\end{table*}

\begin{table*}[!hbt]
\begin{center}
\caption{Azimuthal asymmetries obtained by fitting the UL 
and UC double ratios in bins of  $p_t$.
The upper (lower) table summarizes the results for RF12 (RF0).
The errors are statistical and systematic.
The table also reports the average values of $z_i$ and $p_{ti}$ and 
$\sin^2\theta/(1+\cos^2\theta)$ in the
corresponding $(p_{t1},p_{t2})$ or $p_{t0}$ bin.
} 
\begin{tabular}{ c c c c c c  c c c c}
\toprule
$p_{t1}$ & $\langle p_{t1} \rangle $    & \multirow{2}{*}{$\langle z_{1} \rangle$} &
$p_{t2}$ & $\langle p_{t2}\rangle$ & \multirow{2}{*}{$\langle z_{2} \rangle$} & \multirow{2}{*}{ $\left\langle\frac{\sin^2\theta_{th}}{1+\cos^2\theta_{th}}\right\rangle$ }
 & \multirow{2}{*}{ \aulTh }& & \multirow{2}{*}{ \aucTh} \\
 (\gev) & (\gev) & & (\gev) & (\gev) &   & & & \\
\toprule
$[0.,0.25]$ & 0.163 & 0.258 & $[0.,0.25]$ & 0.163 & 0.258 & 0.690 & 2.77 $\pm$ 0.70 $\pm$ 0.88 & & 1.26 $\pm$ 0.59 $\pm$ 0.43\\
 $[0.,0.25]$ & 0.163 & 0.260 & $[0.25,0.5]$ & 0.370 & 0.263 & 0.700 & 3.18 $\pm$ 0.36 $\pm$ 0.37 & & 1.44 $\pm$ 0.31 $\pm$ 0.18\\
 $[0.,0.25]$ & 0.161 & 0.261 & $[0.5,0.75]$ & 0.596 & 0.308 & 0.708 & 3.73 $\pm$ 0.52 $\pm$ 0.41 & & 1.73 $\pm$ 0.44 $\pm$ 0.21\\
 $[0.,0.25]$ & 0.161 & 0.263 & $[>0.75]$ & 0.895 & 0.412 & 0.708 & 6.17 $\pm$ 0.87 $\pm$ 0.73 & & 2.70 $\pm$ 0.71 $\pm$ 0.39\\
 $[0.25,0.5]$ & 0.370 & 0.263 & $[0.,0.25]$ & 0.163 & 0.260 & 0.700 & 4.28 $\pm$ 0.37 $\pm$ 0.53 & & 1.95 $\pm$ 0.31 $\pm$ 0.23\\
 $[0.25,0.5]$ & 0.367 & 0.270 & $[0.25,0.5]$ & 0.366 & 0.270 & 0.711 & 4.40 $\pm$ 0.18 $\pm$ 0.47 & & 2.01 $\pm$ 0.15 $\pm$ 0.22\\
 $[0.25,0.5]$ & 0.365 & 0.275 & $[0.5,0.75]$ & 0.596 & 0.322 & 0.720 & 3.90 $\pm$ 0.26 $\pm$ 0.41 & & 1.77 $\pm$ 0.22 $\pm$ 0.19\\
 $[0.25,0.5]$ & 0.363 & 0.278 & $[>0.75]$ & 0.890 & 0.424 & 0.721 & 6.10 $\pm$ 0.41 $\pm$ 0.65 & & 2.73 $\pm$ 0.34 $\pm$ 0.30\\
 $[0.5,0.75]$ & 0.596 & 0.308& $[0.,0.25]$ & 0.161 & 0.262 & 0.708 & 3.23 $\pm$ 0.51 $\pm$ 0.38 & & 1.51 $\pm$ 0.43 $\pm$ 0.19\\
 $[0.5,0.75]$ & 0.596 & 0.321& $[0.25,0.5]$ & 0.365 & 0.275 & 0.720 & 4.05 $\pm$ 0.27 $\pm$ 0.43 & & 1.83 $\pm$ 0.22 $\pm$ 0.19\\
 $[0.5,0.75]$ & 0.595 & 0.324& $[0.5,0.75]$ & 0.595 & 0.326 & 0.731 & 4.71 $\pm$ 0.53 $\pm$ 0.50 & & 2.09 $\pm$ 0.35 $\pm$ 0.24\\
 $[0.5,0.75]$ & 0.595 & 0.330& $[>0.75]$ & 0.885 & 0.423 & 0.735 & 6.04 $\pm$ 0.66 $\pm$ 0.69 & & 2.63 $\pm$ 0.51 $\pm$ 0.35\\
 $[>0.75]$ & 0.895 & 0.412 & $[0.,0.25]$ & 0.161 & 0.264 & 0.709 & 5.29 $\pm$ 0.84 $\pm$ 0.74 & & 2.39 $\pm$ 0.70 $\pm$ 0.37\\
 $[>0.75]$ & 0.890 & 0.423 & $[0.25,0.5]$ & 0.363 & 0.279 & 0.721 & 5.27 $\pm$ 0.41 $\pm$ 0.55 & & 2.40 $\pm$ 0.34 $\pm$ 0.26\\
 $[>0.75]$ & 0.885 & 0.422 & $[0.5,0.75]$ & 0.595 & 0.331 & 0.735 & 5.91 $\pm$ 0.67 $\pm$ 0.63 & & 2.50 $\pm$ 0.51 $\pm$ 0.31\\
 $[>0.75]$ & 0.881 & 0.425 & $[>0.75]$ & 0.880 & 0.426 & 0.743 & 6.62 $\pm$ 1.14 $\pm$ 0.80 & & 2.93 $\pm$ 0.86 $\pm$ 0.46\\
\toprule 
$p_{t0}$ &$\langle p_{t0}\rangle$ & \multirow{2}{*}{$\langle z_1 \rangle$}  & & &
\multirow{2}{*}{ $\langle z_2 \rangle$} & \multirow{2}{*}{ $\left\langle\frac{\sin^2\theta_{2}}{1+\cos^2\theta_{2}}\right\rangle$}
  & \multirow{2}{*}{ \aul} & & \multirow{2}{*}{ \auc} \\
  (\gev) & (\gev) & &  &  &   & & & \\
\toprule 
$[0.,0.125]$ & 0.083 & 0.230   & & & 0.300 & 0.685 & -0.20 $\pm$ 0.28 $\pm$ 0.08 & & -0.09 $\pm$ 0.23 $\pm$ 0.06\\
 $[0.125,0.25]$ & 0.194 & 0.231   & & & 0.299 & 0.683 & 0.34 $\pm$ 0.17 $\pm$ 0.06 & &  0.15 $\pm$ 0.14 $\pm$ 0.04\\
 $[0.25,0.375]$ & 0.315 & 0.233   & & & 0.295 & 0.680 & 1.15 $\pm$ 0.14 $\pm$ 0.11 & & 0.52 $\pm$ 0.12 $\pm$ 0.06\\
 $[0.375,0.5]$ & 0.438 & 0.239   & & & 0.289 & 0.678 & 1.67 $\pm$ 0.13 $\pm$ 0.11 & & 0.76 $\pm$ 0.11 $\pm$ 0.06\\
 $[0.5,0.625]$ & 0.558 & 0.258   & & & 0.281 & 0.677 & 2.24 $\pm$ 0.15 $\pm$ 0.14 & & 1.01 $\pm$ 0.12 $\pm$ 0.07\\
 $[0.625,0.75]$ & 0.683 & 0.302   & & & 0.276 & 0.677 & 2.02 $\pm$ 0.18 $\pm$ 0.14 & & 0.91 $\pm$ 0.14 $\pm$ 0.07\\
 $[0.75,0.9]$ & 0.818 & 0.349   & & & 0.270 & 0.677 & 2.54 $\pm$ 0.21 $\pm$ 0.17 & & 1.13 $\pm$ 0.16 $\pm$ 0.09\\
 $[0.9,1.1]$ & 0.989 & 0.406   & & & 0.262 & 0.677 & 2.20 $\pm$ 0.21 $\pm$ 0.17 & & 0.96 $\pm$ 0.17 $\pm$ 0.09\\
 $[1.1,1.5]$ & 1.258 & 0.488   & & & 0.252 & 0.678 & 2.12 $\pm$ 0.20 $\pm$ 0.17 & & 0.92 $\pm$ 0.16 $\pm$ 0.09\\
\toprule 
\end{tabular}
\label{tab:totPt}
\end{center}
\end{table*}

\begin{table*}[ht]
\begin{center}
\caption{Collins asymmetries in percent obtained by fitting the UL
  double ratios in bins of  $(z_1,z_2,p_{t1},p_{t2})$, in RF12. 
Each 4-dimensional bin is identified by two pairs of digits.
The pair on the first raw identify the  $p_{t1}$ and $p_{t2}$
intervals,   with ``0'', ``1'', and ``2'', corresponding to the three
bins  $p_t<0.25\,\gevc$, $0.25<p_t<0.5\,\gevc$, and $0.5\,\gevc<p_t$, respectively.
The pair on the first column identify the $z_1$ and $z_2$ intervals,
 with ``0'', ``1'', ``2'', and ``3'', referring to $0.15<z<0.2$, 
$0.2<z<0.3$, $0.3<z<0.5$, and $0.5<z<0.9$, respectively.
The error shown is the sum in quadrature of the statistical and systematic uncertainties.}

\footnotesize{
\begin{tabular}{ c || c| c| c| c| c| c| c| c| c }
\toprule
\multicolumn{10}{c}{ \multirow{2}{*} {\aulTh\ $(10^{-2})$}} \\
\multicolumn{10}{c}{ } \\
\toprule
 \backslashbox{($z_1$,$z_2$) \kern-1em}{\kern-1em ($p_{t1}$,$p_{t2}$)} &  $(0,0)$ &  $(0,1)$ &  $(0,2)$ &  $(1,0)$ &  $(1,1)$ &  $(1,2)$ &  $(2,0)$ &  $(2,1)$ &  $(2,2)$  \\ \toprule
 \multirow{2}{*}{$(0,0)$}  & \multirow{2}{*}{-0.01 $\pm$ 1.33} & \multirow{2}{*}{1.63 $\pm$ 0.83} & \multirow{2}{*}{1.37 $\pm$ 2.03} & \multirow{2}{*}{2.68 $\pm$ 0.87} &  \multirow{2}{*}{3.41 $\pm$ 0.65} & \multirow{2}{*}{0.25 $\pm$ 1.28} & \multirow{2}{*}{3.17 $\pm$ 2.01} & \multirow{2}{*}{2.16 $\pm$ 1.24} & \multirow{2}{*}{-5.29 $\pm$ 2.97} \\ 
&  & &  & & &   &   &  &    \\ \hline
\multirow{2}{*}{$(0,1)$} & \multirow{2}{*}{3.72 $\pm$ 1.57} & \multirow{2}{*}{1.80 $\pm$ 0.77} & \multirow{2}{*}{2.48 $\pm$ 1.09} & \multirow{2}{*}{3.14 $\pm$ 0.95} &  \multirow{2}{*}{4.50 $\pm$ 0.56} & \multirow{2}{*}{3.48 $\pm$ 0.73} & \multirow{2}{*}{2.24 $\pm$ 2.17} & \multirow{2}{*}{2.09 $\pm$ 1.11} & \multirow{2}{*}{2.16 $\pm$ 1.64} \\
&  &  &  & & &   &   &  &    \\ \hline
\multirow{2}{*}{$(0,2)$} & \multirow{2}{*}{3.52 $\pm$ 2.11} & \multirow{2}{*}{2.28 $\pm$ 0.96} & \multirow{2}{*}{3.06 $\pm$ 0.89} & \multirow{2}{*}{3.22 $\pm$ 1.41} &  \multirow{2}{*}{3.71 $\pm$ 0.64} & \multirow{2}{*}{3.62 $\pm$ 0.60} & \multirow{2}{*}{-0.60 $\pm$ 3.33} & \multirow{2}{*}{0.02 $\pm$ 1.24} & \multirow{2}{*}{2.74 $\pm$ 1.46} \\
&  &  &  & & &   &   &  &    \\ \hline
\multirow{2}{*}{$(0,3)$} & \multirow{2}{*}{8.66 $\pm$ 4.26} & \multirow{2}{*}{3.71 $\pm$ 1.53} & \multirow{2}{*}{4.70 $\pm$ 1.02} & \multirow{2}{*}{5.66 $\pm$ 2.74} &  \multirow{2}{*}{5.86 $\pm$ 1.09} & \multirow{2}{*}{6.61 $\pm$ 0.70} & \multirow{2}{*}{2.29 $\pm$ 6.04} & \multirow{2}{*}{1.20 $\pm$ 1.64} & \multirow{2}{*}{3.41 $\pm$ 1.74} \\
&  &  &  & & &   &   &  &    \\ \hline 
\multirow{2}{*}{$(1,0)$} & \multirow{2}{*}{2.59 $\pm$ 1.53} & \multirow{2}{*}{1.07 $\pm$ 0.93} & \multirow{2}{*}{5.11 $\pm$ 2.32} & \multirow{2}{*}{3.36 $\pm$ 0.79} &  \multirow{2}{*}{2.82 $\pm$ 0.50} & \multirow{2}{*}{3.68 $\pm$ 1.21} & \multirow{2}{*}{1.60 $\pm$ 1.10} & \multirow{2}{*}{2.08 $\pm$ 0.78} & \multirow{2}{*}{3.28 $\pm$ 1.65} \\
&  &  &  & & &   &   &  &    \\ \hline
\multirow{2}{*}{$(1,1)$} & \multirow{2}{*}{1.83 $\pm$ 1.61} & \multirow{2}{*}{3.78 $\pm$ 0.90} & \multirow{2}{*}{4.66 $\pm$ 1.27} & \multirow{2}{*}{4.78 $\pm$ 0.92} &  \multirow{2}{*}{4.19 $\pm$ 0.47} & \multirow{2}{*}{3.38 $\pm$ 0.66} & \multirow{2}{*}{5.34 $\pm$ 1.28} & \multirow{2}{*}{4.10 $\pm$ 0.68} & \multirow{2}{*}{3.25 $\pm$ 1.01} \\
&  &  &  & & &   &   &  &    \\ \hline
\multirow{2}{*}{$(1,2)$} & \multirow{2}{*}{4.77 $\pm$ 2.35} & \multirow{2}{*}{4.88 $\pm$ 1.10} & \multirow{2}{*}{4.57 $\pm$ 1.05} & \multirow{2}{*}{2.56 $\pm$ 1.32} &  \multirow{2}{*}{5.91 $\pm$ 0.68} & \multirow{2}{*}{4.36 $\pm$ 0.56} & \multirow{2}{*}{5.20 $\pm$ 1.94} & \multirow{2}{*}{2.13 $\pm$ 0.71} & \multirow{2}{*}{4.31 $\pm$ 0.88} \\
&  &  &  & & &   &   &  &    \\ \hline
\multirow{2}{*}{$(1,3)$} & \multirow{2}{*}{2.35 $\pm$ 4.01} & \multirow{2}{*}{3.86 $\pm$ 1.37} & \multirow{2}{*}{6.28 $\pm$ 1.10} & \multirow{2}{*}{8.50 $\pm$ 3.00} &  \multirow{2}{*}{5.63 $\pm$ 1.00} & \multirow{2}{*}{7.94 $\pm$ 0.64} & \multirow{2}{*}{-2.73 $\pm$ 3.45} & \multirow{2}{*}{4.16 $\pm$ 1.04} & \multirow{2}{*}{5.30 $\pm$ 0.95} \\
&  &  &  & & &   &   &  &    \\ \hline 
\multirow{2}{*}{$(2,0)$}  & \multirow{2}{*}{0.65 $\pm$ 2.22} & \multirow{2}{*}{3.20 $\pm$ 1.44} & \multirow{2}{*}{1.10 $\pm$ 3.37} & \multirow{2}{*}{2.41 $\pm$ 0.96} &  \multirow{2}{*}{3.27 $\pm$ 0.63} & \multirow{2}{*}{2.54 $\pm$ 1.47} & \multirow{2}{*}{2.95 $\pm$ 0.91} & \multirow{2}{*}{4.85 $\pm$ 0.74} & \multirow{2}{*}{2.17 $\pm$ 1.47} \\
&  &  &  & & &   &   &  &    \\ \hline
\multirow{2}{*}{$(2,1)$}  & \multirow{2}{*}{1.12 $\pm$ 2.32} & \multirow{2}{*}{1.93 $\pm$ 1.36} & \multirow{2}{*}{0.27 $\pm$ 1.87} & \multirow{2}{*}{4.88 $\pm$ 1.14} &  \multirow{2}{*}{4.36 $\pm$ 0.62} & \multirow{2}{*}{2.09 $\pm$ 0.83} & \multirow{2}{*}{3.33 $\pm$ 1.01} & \multirow{2}{*}{4.61 $\pm$ 0.65} & \multirow{2}{*}{3.80 $\pm$ 0.86} \\
&  &  &  & & &   &   &  &    \\ \hline
\multirow{2}{*}{$(2,2)$}  & \multirow{2}{*}{5.59 $\pm$ 3.15} & \multirow{2}{*}{3.68 $\pm$ 1.58} & \multirow{2}{*}{5.28 $\pm$ 1.61} & \multirow{2}{*}{5.36 $\pm$ 1.80} &  \multirow{2}{*}{4.58 $\pm$ 0.73} & \multirow{2}{*}{5.08 $\pm$ 0.75} & \multirow{2}{*}{4.96 $\pm$ 1.63} & \multirow{2}{*}{4.70 $\pm$ 0.73} & \multirow{2}{*}{3.82 $\pm$ 0.71} \\
&  &  &  & & &   &   &  &    \\ \hline
\multirow{2}{*}{$(2,3)$} & \multirow{2}{*}{-1.59 $\pm$ 3.93} & \multirow{2}{*}{7.04 $\pm$ 2.27} & \multirow{2}{*}{5.91 $\pm$ 1.57} & \multirow{2}{*}{6.24 $\pm$ 2.95} &  \multirow{2}{*}{7.50 $\pm$ 1.27} & \multirow{2}{*}{8.13 $\pm$ 0.84} & \multirow{2}{*}{-3.36 $\pm$ 2.99} & \multirow{2}{*}{4.62 $\pm$ 0.98} & \multirow{2}{*}{7.39 $\pm$ 0.85}\\ 
&  &  &  & & &   &   &  &    \\ \hline 
\multirow{2}{*}{$(3,0)$} & \multirow{2}{*}{3.96 $\pm$ 3.94} & \multirow{2}{*}{3.36 $\pm$ 2.94} & \multirow{2}{*}{0.88 $\pm$ 6.24} & \multirow{2}{*}{3.16 $\pm$ 1.54} &  \multirow{2}{*}{6.64 $\pm$ 1.13} & \multirow{2}{*}{3.51 $\pm$ 2.39} & \multirow{2}{*}{4.34 $\pm$ 1.00} & \multirow{2}{*}{9.24 $\pm$ 1.12} & \multirow{2}{*}{5.59 $\pm$ 1.79} \\
&  &  &  & & &   &   &  &    \\ \hline 
\multirow{2}{*}{$(3,1)$} & \multirow{2}{*}{4.04 $\pm$ 3.97} & \multirow{2}{*}{7.95 $\pm$ 2.86} & \multirow{2}{*}{1.95 $\pm$ 3.21} & \multirow{2}{*}{4.98 $\pm$ 1.51} &  \multirow{2}{*}{5.78 $\pm$ 1.00} & \multirow{2}{*}{4.34 $\pm$ 1.35} & \multirow{2}{*}{6.08 $\pm$ 1.10} & \multirow{2}{*}{9.56 $\pm$ 0.92} & \multirow{2}{*}{5.23 $\pm$ 0.96} \\
&  &  &  & & &   &   &  &    \\ \hline 
\multirow{2}{*}{$(3,2)$} & \multirow{2}{*}{-0.34 $\pm$ 4.49} & \multirow{2}{*}{8.76 $\pm$ 3.27} & \multirow{2}{*}{8.97 $\pm$ 3.43} & \multirow{2}{*}{5.48 $\pm$ 2.23} &  \multirow{2}{*}{6.19 $\pm$ 1.14} & \multirow{2}{*}{7.04 $\pm$ 1.21} & \multirow{2}{*}{3.14 $\pm$ 1.52} & \multirow{2}{*}{9.75 $\pm$ 1.00} & \multirow{2}{*}{8.12 $\pm$ 0.82} \\
&  &  &  & & &   &   &  &    \\ \hline 
\multirow{2}{*}{$(3,3)$} & \multirow{2}{*}{16.79 $\pm$ 8.84} & \multirow{2}{*}{8.99 $\pm$ 4.10} & \multirow{2}{*}{7.06 $\pm$ 2.84} & \multirow{2}{*}{17.79 $\pm$ 4.12} &  \multirow{2}{*}{12.23 $\pm$ 2.12} & \multirow{2}{*}{12.61 $\pm$ 1.36} & \multirow{2}{*}{12.24 $\pm$ 3.34} & \multirow{2}{*}{10.68 $\pm$ 1.31} & \multirow{2}{*}{16.76 $\pm$ 1.14}\\ 
&  &  &  & & &   &   &  &    \\ \hline 
\toprule
\end{tabular}
}
\label{tab:zpt}
\end{center}
\end{table*}

\begin{table*}[ht]
\begin{center}
\caption{Collins asymmetries in percent obtained by fitting the UC
  double ratios in bins of  $(z_1,z_2,p_{t1},p_{t2})$, in RF12. 
Each 4-dimensional bin is identified by two pairs of digits.
The pair on the first raw identify the  $p_{t1}$ and $p_{t2}$
intervals,   with ``0'', ``1'', and ``2'', corresponding to the three
bins  $p_t<0.25\,\gevc$, $0.25<p_t<0.5\,\gevc$, and $0.5\,\gevc<p_t$, respectively.
The pair on the first column identify the $z_1$ and $z_2$ intervals,
 with ``0'', ``1'', ``2'', and ``3'', referring to $0.15<z<0.2$, 
$0.2<z<0.3$, $0.3<z<0.5$, and $0.5<z<0.9$, respectively.
The error shown is the sum in quadrature of the statistical and systematic uncertainties.}
\footnotesize{
\begin{tabular}{ c || c| c| c| c| c| c| c| c| c }
\toprule
\multicolumn{10}{c}{ \multirow{2}{*} {\aucTh\ $(10^{-2})$}} \\
\multicolumn{10}{c}{ } \\
\toprule
 \backslashbox{($z_1$,$z_2$) \kern-0.7em}{ \kern-0.7em $\,$ ($p_{t1}$,$p_{t2}$)} &  $(0, 0)$ &  $(0, 1)$ &  $(0, 2)$ &  $(1, 0)$ &  $(1, 1)$ &  $(1, 2)$ &  $(2, 0)$ &  $(2, 1)$ &  $(2, 2)$  \\ \toprule
\toprule
 \multirow{2}{*}{$(0,0)$}  & \multirow{2}{*}{-0.07 $\pm$ 1.14} & \multirow{2}{*}{0.76 $\pm$ 0.71} & \multirow{2}{*}{0.62 $\pm$ 1.71} & \multirow{2}{*}{1.28 $\pm$ 0.72} &  \multirow{2}{*}{1.64 $\pm$ 0.50} & \multirow{2}{*}{0.27 $\pm$ 1.08} & \multirow{2}{*}{1.54 $\pm$ 1.71} & \multirow{2}{*}{1.06 $\pm$ 1.05} & \multirow{2}{*}{-2.12 $\pm$ 2.26}\\
&  &  &  & & &   &   &  &    \\ \hline  
 \multirow{2}{*}{$(0,1)$}  & \multirow{2}{*}{1.70 $\pm$ 1.30} & \multirow{2}{*}{0.88 $\pm$ 0.65} & \multirow{2}{*}{1.15 $\pm$ 0.90} & \multirow{2}{*}{1.46 $\pm$ 0.79} &  \multirow{2}{*}{2.10 $\pm$ 0.44} & \multirow{2}{*}{1.63 $\pm$ 0.61} & \multirow{2}{*}{0.89 $\pm$ 1.84} & \multirow{2}{*}{1.05 $\pm$ 0.94} & \multirow{2}{*}{0.98 $\pm$ 1.34}\\
&  &  &  & & &   &   &  &    \\ \hline
 \multirow{2}{*}{$(0,2)$}  & \multirow{2}{*}{1.97 $\pm$ 1.84} & \multirow{2}{*}{1.14 $\pm$ 0.81} & \multirow{2}{*}{1.43 $\pm$ 0.74} & \multirow{2}{*}{1.50 $\pm$ 1.21} &  \multirow{2}{*}{1.73 $\pm$ 0.52} & \multirow{2}{*}{1.65 $\pm$ 0.49} & \multirow{2}{*}{0.53 $\pm$ 2.76} & \multirow{2}{*}{0.17 $\pm$ 1.04} & \multirow{2}{*}{1.28 $\pm$ 1.20} \\
&  &  &  & & &   &   &  &    \\ \hline
 \multirow{2}{*}{$(0,3)$}  & \multirow{2}{*}{3.04 $\pm$ 3.44} & \multirow{2}{*}{1.68 $\pm$ 1.28} & \multirow{2}{*}{2.18 $\pm$ 0.84} & \multirow{2}{*}{2.66 $\pm$ 2.35} &  \multirow{2}{*}{2.69 $\pm$ 0.91} & \multirow{2}{*}{3.04 $\pm$ 0.57} & \multirow{2}{*}{0.29 $\pm$ 5.15} & \multirow{2}{*}{1.13 $\pm$ 1.42} & \multirow{2}{*}{1.73 $\pm$ 1.46} \\
&  &  &  & & &   &   &  &    \\ \hline
 \multirow{2}{*}{$(1,0)$} & \multirow{2}{*}{1.19 $\pm$ 1.29} & \multirow{2}{*}{0.49 $\pm$ 0.80} & \multirow{2}{*}{2.34 $\pm$ 1.95} & \multirow{2}{*}{1.62 $\pm$ 0.66} &  \multirow{2}{*}{1.28 $\pm$ 0.42} & \multirow{2}{*}{1.67 $\pm$ 1.01} & \multirow{2}{*}{0.73 $\pm$ 0.92} & \multirow{2}{*}{0.97 $\pm$ 0.65} & \multirow{2}{*}{1.65 $\pm$ 1.35}\\
&  &  &  & & &   &   &  &    \\ \hline
 \multirow{2}{*}{$(1,1)$} & \multirow{2}{*}{0.86 $\pm$ 1.37} & \multirow{2}{*}{1.72 $\pm$ 0.75} & \multirow{2}{*}{2.20 $\pm$ 1.03} & \multirow{2}{*}{2.19 $\pm$ 0.76} &  \multirow{2}{*}{1.93 $\pm$ 0.39} & \multirow{2}{*}{1.55 $\pm$ 0.54} & \multirow{2}{*}{2.45 $\pm$ 1.03} & \multirow{2}{*}{1.88 $\pm$ 0.55} & \multirow{2}{*}{1.59 $\pm$ 0.81} \\
&  &  &  & & &   &   &  &    \\ \hline 
 \multirow{2}{*}{$(1,2)$} &  \multirow{2}{*}{2.14 $\pm$ 1.93} & \multirow{2}{*}{2.25 $\pm$ 0.90} & \multirow{2}{*}{2.03 $\pm$ 0.85} & \multirow{2}{*}{1.06 $\pm$ 1.12} &  \multirow{2}{*}{2.64 $\pm$ 0.50} & \multirow{2}{*}{1.96 $\pm$ 0.44} & \multirow{2}{*}{2.37 $\pm$ 1.55} & \multirow{2}{*}{0.98 $\pm$ 0.57} & \multirow{2}{*}{1.94 $\pm$ 0.67} \\
&  &  &  & & &   &   &  &    \\ \hline
 \multirow{2}{*}{$(1,3)$} &  \multirow{2}{*}{0.17 $\pm$ 3.42} & \multirow{2}{*}{1.97 $\pm$ 1.29} & \multirow{2}{*}{2.61 $\pm$ 0.91} & \multirow{2}{*}{3.68 $\pm$ 2.28} &  \multirow{2}{*}{2.48 $\pm$ 0.81} & \multirow{2}{*}{3.49 $\pm$ 0.50} & \multirow{2}{*}{-1.77 $\pm$ 3.05} & \multirow{2}{*}{2.03 $\pm$ 0.84} & \multirow{2}{*}{2.59 $\pm$ 0.77} \\
&  &  &  & & &   &   &  &    \\ \hline
 \multirow{2}{*}{$(2,0)$} & \multirow{2}{*}{0.34 $\pm$ 1.88} & \multirow{2}{*}{1.43 $\pm$ 1.21} & \multirow{2}{*}{0.89 $\pm$ 2.77} & \multirow{2}{*}{1.16 $\pm$ 0.80} &  \multirow{2}{*}{1.50 $\pm$ 0.52} & \multirow{2}{*}{1.14 $\pm$ 1.24} & \multirow{2}{*}{1.32 $\pm$ 0.75} & \multirow{2}{*}{2.24 $\pm$ 0.59} & \multirow{2}{*}{1.33 $\pm$ 1.19} \\
&  &  &  & & &   &   &  &    \\ \hline 
 \multirow{2}{*}{$(2,1)$} & \multirow{2}{*}{0.67 $\pm$ 1.95} & \multirow{2}{*}{0.89 $\pm$ 1.13} & \multirow{2}{*}{0.24 $\pm$ 1.53} & \multirow{2}{*}{2.19 $\pm$ 0.90} &  \multirow{2}{*}{1.93 $\pm$ 0.48} & \multirow{2}{*}{0.99 $\pm$ 0.67} & \multirow{2}{*}{1.58 $\pm$ 0.84} & \multirow{2}{*}{2.09 $\pm$ 0.50} & \multirow{2}{*}{1.76 $\pm$ 0.68} \\
&  &  &  & & &   &   &  &    \\ \hline 
 \multirow{2}{*}{$(2,2)$} & \multirow{2}{*}{2.71 $\pm$ 2.58} & \multirow{2}{*}{1.65 $\pm$ 1.27} & \multirow{2}{*}{2.43 $\pm$ 1.27} & \multirow{2}{*}{2.41 $\pm$ 1.34} &  \multirow{2}{*}{2.01 $\pm$ 0.58} & \multirow{2}{*}{2.26 $\pm$ 0.55} & \multirow{2}{*}{2.19 $\pm$ 1.27} & \multirow{2}{*}{2.04 $\pm$ 0.55} & \multirow{2}{*}{1.70 $\pm$ 0.52} \\
&  &  &  & & &   &   &  &    \\ \hline 
 \multirow{2}{*}{$(2,3)$} & \multirow{2}{*}{-2.41 $\pm$ 3.99} & \multirow{2}{*}{2.59 $\pm$ 1.81} & \multirow{2}{*}{2.43 $\pm$ 1.29} & \multirow{2}{*}{2.67 $\pm$ 2.44} &  \multirow{2}{*}{2.94 $\pm$ 0.93} & \multirow{2}{*}{3.35 $\pm$ 0.61} & \multirow{2}{*}{-1.43 $\pm$ 2.47} & \multirow{2}{*}{1.87 $\pm$ 0.75} & \multirow{2}{*}{2.98 $\pm$ 0.61} \\
&  &  &  & & &   &   &  &    \\ \hline
 \multirow{2}{*}{$(3,0)$}  & \multirow{2}{*}{1.52 $\pm$ 3.38} & \multirow{2}{*}{1.98 $\pm$ 2.48} & \multirow{2}{*}{2.37 $\pm$ 5.49} & \multirow{2}{*}{1.49 $\pm$ 1.25} &  \multirow{2}{*}{2.91 $\pm$ 0.91} & \multirow{2}{*}{1.75 $\pm$ 2.10} & \multirow{2}{*}{1.99 $\pm$ 0.83} & \multirow{2}{*}{4.21 $\pm$ 0.84} & \multirow{2}{*}{2.58 $\pm$ 1.50} \\
&  &  &  & & &   &   &  &    \\ \hline 
 \multirow{2}{*}{$(3,1)$}  &  \multirow{2}{*}{2.43 $\pm$ 3.35} & \multirow{2}{*}{3.70 $\pm$ 2.28} & \multirow{2}{*}{0.83 $\pm$ 2.87} & \multirow{2}{*}{2.42 $\pm$ 1.28} &  \multirow{2}{*}{2.49 $\pm$ 0.81} & \multirow{2}{*}{2.20 $\pm$ 1.12} & \multirow{2}{*}{2.72 $\pm$ 0.89} & \multirow{2}{*}{4.38 $\pm$ 0.71} & \multirow{2}{*}{2.43 $\pm$ 0.77} \\
&  &  &  & & &   &   &  &    \\ \hline
 \multirow{2}{*}{$(3,2)$}  &  \multirow{2}{*}{-1.10 $\pm$ 4.19} & \multirow{2}{*}{4.32 $\pm$ 2.50} & \multirow{2}{*}{3.91 $\pm$ 2.71} & \multirow{2}{*}{2.29 $\pm$ 1.79} &  \multirow{2}{*}{2.57 $\pm$ 0.92} & \multirow{2}{*}{2.92 $\pm$ 0.95} & \multirow{2}{*}{1.22 $\pm$ 1.29} & \multirow{2}{*}{4.19 $\pm$ 0.78} & \multirow{2}{*}{3.40 $\pm$ 0.62} \\
&  &  &  & & &   &   &  &    \\ \hline
 \multirow{2}{*}{$(3,3)$}  & \multirow{2}{*}{5.82 $\pm$ 7.57} & \multirow{2}{*}{3.43 $\pm$ 3.40} & \multirow{2}{*}{2.64 $\pm$ 2.57} & \multirow{2}{*}{8.57 $\pm$ 3.43} &  \multirow{2}{*}{4.47 $\pm$ 1.48} & \multirow{2}{*}{5.05 $\pm$ 1.06} & \multirow{2}{*}{5.35 $\pm$ 2.78} & \multirow{2}{*}{4.57 $\pm$ 1.05} & \multirow{2}{*}{6.21 $\pm$ 0.78} \\
&  &  &  & & &   &   &  &    \\ \hline
\toprule
\end{tabular}
}
\label{tab:zptUC}
\end{center}
\end{table*}

\begin{table*}[ht]
\begin{center}
\caption{ Mean values of $z_1$, $z_2$ (first line),  $p_{t1}$, $p_{t2}$ (second line),
and $\sin^2\theta_{th}/(1+\cos^2\theta_{th})$ (third line) for each $z-p_t$ bin.
Each 4-dimensional bin is identified by two pairs of digits.
The pair on the second raw identify the  $p_{t1}$ and $p_{t2}$
intervals,   with ``0'', ``1'', and ``2'', corresponding to the three
bins  $p_t<0.25\,\gevc$, $0.25<p_t<0.5\,\gevc$, and $0.5\gevc\,<p_t$, respectively.
The pair on the first column identify the $z_1$ and $z_2$ intervals,
 with ``0'', ``1'', ``2'', and ``3'', referring to $0.15<z<0.2$, 
$0.2<z<0.3$, $0.3<z<0.5$, and $0.5<z<0.9$, respectively.
The error shown is the sum in quadrature of the statistical and systematic uncertainties.}
\footnotesize{
\begin{tabular}{ c || c| c| c| c| c| c| c| c| c }
\toprule
\multirow{3}{*} {$z-p_t$ bins} & \multicolumn{9}{c}{ $\langle z_1 \rangle$ $\langle z_1 \rangle$} \\
& \multicolumn{9}{c}{ $\langle p_{t1} \rangle$ $\langle p_{t2} \rangle$} \\
& \multicolumn{9}{c}{ $\left\langle\frac{\sin^2\theta_{th}}{1+\cos^2\theta_{th}}\right\rangle$} \\
\toprule
  \backslashbox{($z_1$,$z_2$) \kern-0.7em}{ \kern-0.7em $\,$ ($p_{t1}$,$p_{t2}$)} &  $(0, 0)$ &  $(0, 1)$ &  $(0, 2)$ &  $(1, 0)$ &  $(1, 1)$ &  $(1, 2)$ &  $(2, 0)$ &  $(2, 1)$ &  $(2, 2)$  \\ \toprule
   \multirow{3}{*}{$(0,0)$} & 0.172  0.172 & 0.173  0.173 & 0.173  0.179 & 0.173  0.173 & 0.174  0.174 & 0.173  0.178 & 0.179  0.173 & 0.178  0.173 & 0.177  0.178 \\ 
  & 0.164  0.164 & 0.162  0.364 & 0.155  0.560 & 0.364  0.161 & 0.355  0.355 & 0.360  0.564 & 0.561  0.155 & 0.565  0.360 & 0.567  0.565 \\ 
  & 0.694 & 0.709 & 0.723 & 0.709  & 0.727 & 0.742 & 0.722 & 0.742  & 0.753 \\ \toprule
    \multirow{3}{*}{$(0,1)$} & 0.172  0.241 & 0.172  0.242 & 0.173  0.249 & 0.173  0.242 & 0.174  0.244 & 0.174  0.250 & 0.179  0.242 & 0.178  0.243 & 0.178  0.248 \\ 
  & 0.164  0.163 & 0.164  0.371 & 0.159  0.622 & 0.366  0.162 & 0.357  0.365 & 0.355  0.621 & 0.559  0.157 & 0.564  0.365 & 0.566  0.627 \\ 
  & 0.692 & 0.703 & 0.718 & 0.706  & 0.719 & 0.739 & 0.718 & 0.735  & 0.753 \\ \toprule
    \multirow{3}{*}{$(0,2)$} & 0.172  0.371 & 0.172  0.372 & 0.173  0.380 & 0.173  0.372 & 0.173  0.374 & 0.174  0.384 & 0.179  0.373 & 0.179  0.375 & 0.178  0.381 \\ 
  & 0.164  0.163 & 0.164  0.375 & 0.163  0.695 & 0.367  0.163 & 0.363  0.373 & 0.355  0.675 & 0.558  0.160 & 0.560  0.368 & 0.564  0.680 \\ 
  & 0.689 & 0.696 & 0.709 & 0.700  & 0.709 & 0.725 & 0.711 & 0.723  & 0.745 \\ \toprule
    \multirow{3}{*}{$(0,3)$} & 0.173  0.626 & 0.173  0.612 & 0.173  0.602 & 0.173  0.627 & 0.173  0.613 & 0.173  0.605 & 0.179  0.629 & 0.179  0.614 & 0.179  0.604 \\ 
  & 0.165  0.162 & 0.165  0.375 & 0.164  0.752 & 0.367  0.162 & 0.366  0.374 & 0.359  0.728 & 0.557  0.160 & 0.558  0.370 & 0.560  0.709 \\ 
  & 0.685 & 0.688 & 0.686 & 0.694  & 0.697 & 0.698 & 0.705 & 0.708  & 0.717 \\ \toprule
    \multirow{3}{*}{$(1,0)$} & 0.242  0.172 & 0.242  0.173 & 0.242  0.179 & 0.242  0.172 & 0.244  0.174 & 0.243  0.178 & 0.249  0.173 & 0.249  0.174 & 0.248  0.178 \\ 
  & 0.163  0.164 & 0.162  0.366 & 0.157  0.558 & 0.371  0.164 & 0.365  0.357 & 0.365  0.562 & 0.621  0.159 & 0.621  0.355 & 0.628  0.564 \\ 
  & 0.692 & 0.705 & 0.718 & 0.703  & 0.719 & 0.735 & 0.718 & 0.739  & 0.753 \\ \toprule
    \multirow{3}{*}{$(1,1)$} & 0.242  0.242 & 0.242  0.242 & 0.242  0.248 & 0.242  0.242 & 0.243  0.243 & 0.244  0.250 & 0.248  0.242 & 0.250  0.244 & 0.249  0.249 \\ 
  & 0.163  0.163 & 0.163  0.371 & 0.161  0.624 & 0.371  0.163 & 0.369  0.369 & 0.362  0.617 & 0.624  0.161 & 0.617  0.362 & 0.624  0.623 \\ 
  & 0.691 & 0.701 & 0.715 & 0.701  & 0.713 & 0.730 & 0.715 & 0.730  & 0.751 \\ \toprule
    \multirow{3}{*}{$(1,2)$} & 0.242  0.372 & 0.242  0.373 & 0.242  0.380 & 0.242  0.372 & 0.242  0.373 & 0.243  0.382 & 0.248  0.372 & 0.249  0.375 & 0.250  0.384 \\ 
  & 0.164  0.163 & 0.164  0.375 & 0.163  0.699 & 0.371  0.163 & 0.371  0.374 & 0.367  0.682 & 0.627  0.162 & 0.619  0.370 & 0.617  0.671 \\ 
  & 0.688 & 0.694 & 0.706 & 0.696  & 0.704 & 0.719 & 0.708 & 0.718  & 0.738 \\ \toprule
    \multirow{3}{*}{$(1,3)$} & 0.242  0.627 & 0.242  0.614 & 0.242  0.603 & 0.243  0.628 & 0.243  0.614 & 0.243  0.604 & 0.248  0.628 & 0.248  0.612 & 0.250  0.605 \\ 
  & 0.164  0.162 & 0.164  0.375 & 0.163  0.754 & 0.371  0.162 & 0.371  0.375 & 0.369  0.743 & 0.628  0.161 & 0.622  0.373 & 0.612  0.712 \\ 
  & 0.684 & 0.687 & 0.687 & 0.690  & 0.695 & 0.696 & 0.701 & 0.705  & 0.712 \\ \toprule
    \multirow{3}{*}{$(2,0)$} & 0.372  0.172 & 0.372  0.173 & 0.373  0.179 & 0.372  0.172 & 0.374  0.173 & 0.375  0.179 & 0.380  0.173 & 0.384  0.174 & 0.381  0.178 \\ 
  & 0.163  0.164 & 0.163  0.367 & 0.160  0.557 & 0.375  0.164 & 0.372  0.363 & 0.368  0.559 & 0.695  0.162 & 0.674  0.354 & 0.681  0.563 \\ 
  & 0.688 & 0.700 & 0.711 & 0.696  & 0.709 & 0.724 & 0.709 & 0.725  & 0.745 \\ \toprule
    \multirow{3}{*}{$(2,1)$} & 0.372  0.242 & 0.372  0.242 & 0.372  0.248 & 0.373  0.242 & 0.373  0.242 & 0.375  0.249 & 0.380  0.242 & 0.382  0.243 & 0.384  0.250 \\ 
  & 0.163  0.164 & 0.163  0.371 & 0.162  0.626 & 0.375  0.163 & 0.374  0.371 & 0.370  0.618 & 0.698  0.163 & 0.682  0.367 & 0.671  0.616 \\ 
  & 0.688 & 0.696 & 0.709 & 0.694  & 0.704 & 0.718 & 0.706 & 0.718  & 0.739 \\ \toprule
    \multirow{3}{*}{$(2,2)$} & 0.372  0.372 & 0.372  0.373 & 0.372  0.380 & 0.373  0.372 & 0.373  0.373 & 0.373  0.381 & 0.380  0.372 & 0.381  0.373 & 0.383  0.383 \\ 
  & 0.163  0.163 & 0.163  0.375 & 0.163  0.700 & 0.375  0.163 & 0.374  0.374 & 0.373  0.692 & 0.700  0.163 & 0.692  0.373 & 0.674  0.674 \\ 
  & 0.684 & 0.690 & 0.700 & 0.690  & 0.697 & 0.710 & 0.700 & 0.710  & 0.727 \\ \toprule
    \multirow{3}{*}{$(2,3)$} & 0.374  0.631 & 0.374  0.620 & 0.373  0.606 & 0.374  0.631 & 0.375  0.618 & 0.374  0.605 & 0.381  0.630 & 0.382  0.614 & 0.383  0.604 \\ 
  & 0.164  0.163 & 0.163  0.375 & 0.163  0.755 & 0.374  0.162 & 0.374  0.375 & 0.374  0.751 & 0.704  0.161 & 0.696  0.374 & 0.681  0.731 \\ 
  & 0.678 & 0.683 & 0.683 & 0.684  & 0.689 & 0.692 & 0.692 & 0.699  & 0.708 \\ \toprule
    \multirow{3}{*}{$(3,0)$} & 0.626  0.173 & 0.627  0.173 & 0.629  0.179 & 0.612  0.173 & 0.613  0.173 & 0.614  0.180 & 0.602  0.173 & 0.605  0.174 & 0.604  0.179 \\ 
  & 0.162  0.165 & 0.162  0.367 & 0.160  0.555 & 0.375  0.165 & 0.374  0.366 & 0.370  0.556 & 0.752  0.164 & 0.727  0.359 & 0.710  0.559 \\ 
  & 0.685 & 0.694 & 0.704 & 0.688  & 0.697 & 0.708 & 0.687 & 0.698  & 0.718 \\ \toprule
    \multirow{3}{*}{$(3,1)$} & 0.627  0.242 & 0.628  0.242 & 0.628  0.248 & 0.614  0.242 & 0.614  0.243 & 0.612  0.248 & 0.603  0.242 & 0.604  0.243 & 0.605  0.250 \\ 
  & 0.162  0.164 & 0.162  0.371 & 0.161  0.628 & 0.375  0.164 & 0.375  0.371 & 0.373  0.622 & 0.754  0.163 & 0.743  0.369 & 0.711  0.611 \\ 
  & 0.684 & 0.690 & 0.701 & 0.687  & 0.694 & 0.705 & 0.687 & 0.696  & 0.712 \\ \toprule
    \multirow{3}{*}{$(3,2)$} & 0.631  0.374 & 0.631  0.374 & 0.630  0.381 & 0.619  0.374 & 0.618  0.375 & 0.614  0.382 & 0.606  0.373 & 0.605  0.374 & 0.604  0.383 \\ 
  & 0.162  0.164 & 0.162  0.374 & 0.161  0.704 & 0.375  0.164 & 0.375  0.374 & 0.374  0.696 & 0.755  0.163 & 0.750  0.374 & 0.732  0.681 \\ 
  & 0.679 & 0.685 & 0.692 & 0.683  & 0.689 & 0.699 & 0.683 & 0.692  & 0.709 \\ \toprule
    \multirow{3}{*}{$(3,3)$} & 0.637  0.637 & 0.634  0.624 & 0.637  0.610 & 0.624  0.634 & 0.624  0.624 & 0.620  0.610 & 0.610  0.637 & 0.609  0.620 & 0.607  0.608 \\ 
  & 0.163  0.163 & 0.163  0.374 & 0.161  0.775 & 0.374  0.162 & 0.374  0.374 & 0.374  0.754 & 0.775  0.161 & 0.754  0.374 & 0.741  0.741 \\ 
  & 0.667 & 0.675 & 0.675 & 0.675  & 0.681 & 0.684 & 0.675 & 0.684  & 0.699 \\ 
\toprule
\end{tabular}
}
\label{tab:zptmean}
\end{center}
\end{table*}

\begin{table*}[!hbt]
\begin{center}
\caption{Azimuthal asymmetries, in percent, obtained by fitting the UL 
and UC double ratios in bins of  $\sin^2\theta/(1+\cos^2\theta)$, 
where $\theta = \theta_{th}$ for RF12, and $\theta = \theta_2$ for RF0.
The upper (lower) table summarizes the results for RF12 (RF0).
The errors are statistical and systematic, respectively.
The table also reports the average values of $z_i$, $p_{ti}$ and 
$\sin^2\theta/(1+\cos^2\theta)$ in the
corresponding reference frames.
} 
\begin{tabular}{ c c c c c c  c c c}
\toprule
\multirow{2}{*}{$  \frac{\sin^2\theta_{th}}{1+\cos^2\theta_{th}} $ } & \multirow{2}{*}{$\langle z_{1} \rangle$}   & \multirow{2}{*}{$\langle z_{2} \rangle$} 
 & $\langle p_{t1}\rangle$ & $\langle p_{t2}\rangle$ & \multirow{2}{*}{ $\left\langle\frac{\sin^2\theta_{th}}{1+\cos^2\theta_{th}}\right\rangle$ }
 & \multirow{2}{*}{ \aulTh (\%)}& & \multirow{2}{*}{ \aucTh (\%)}  \\
 & &  & (\gev) & (\gev) & &  & & \\
\toprule
$[0.25-0.30]$ & 0.281 & 0.282 & 0.344 & 0.344 & 0.276  & 2.38 $\pm$ 0.61 $\pm$ 0.34 & &1.19 $\pm$ 0.55 $\pm$ 0.20\\
$[0.30-0.35]$ & 0.281 & 0.282 & 0.342 & 0.342 & 0.326  & 0.90 $\pm$ 0.55 $\pm$ 0.35 & &0.47 $\pm$ 0.48 $\pm$ 0.20\\
$[0.35-0.40]$ & 0.280 & 0.281 & 0.343 & 0.343 & 0.376  & 1.87 $\pm$ 0.49 $\pm$ 0.36 & &0.89 $\pm$ 0.42 $\pm$ 0.20\\
$[0.40-0.45]$ & 0.280 & 0.279 & 0.345 & 0.345 & 0.425  & 1.38 $\pm$ 0.46 $\pm$ 0.36 & &0.64 $\pm$ 0.38 $\pm$ 0.20\\
$[0.45-0.50]$ & 0.277 & 0.278 & 0.348 & 0.350 & 0.475  & 3.52 $\pm$ 0.44 $\pm$ 0.40 & &1.59 $\pm$ 0.37 $\pm$ 0.21\\
$[0.50-0.55]$ & 0.276 & 0.277 & 0.350 & 0.353 & 0.525  & 2.21 $\pm$ 0.41 $\pm$ 0.36 & &1.00 $\pm$ 0.34 $\pm$ 0.19\\
$[0.55-0.60]$ & 0.275 & 0.276 & 0.353 & 0.355 & 0.575  & 3.01 $\pm$ 0.40 $\pm$ 0.38 & &1.36 $\pm$ 0.33 $\pm$ 0.20\\
$[0.60-0.65]$ & 0.275 & 0.275 & 0.355 & 0.355 & 0.625  & 3.91 $\pm$ 0.38 $\pm$ 0.39 & &1.76 $\pm$ 0.32 $\pm$ 0.20\\
$[0.65-0.70]$ & 0.274 & 0.275 & 0.356 & 0.356 & 0.675  & 3.82 $\pm$ 0.38 $\pm$ 0.39 & &1.72 $\pm$ 0.31 $\pm$ 0.20\\
$[0.70-0.75]$ & 0.274 & 0.274 & 0.358 & 0.358 & 0.725  & 3.88 $\pm$ 0.36 $\pm$ 0.38 & &1.74 $\pm$ 0.30 $\pm$ 0.20\\
$[0.75-0.80]$ & 0.274 & 0.274 & 0.359 & 0.359 & 0.775  & 4.14 $\pm$ 0.35 $\pm$ 0.39 & &1.87 $\pm$ 0.29 $\pm$ 0.20\\
 $[0.80-0.85]$& 0.274 & 0.274 & 0.360 & 0.359 & 0.825  & 5.08 $\pm$ 0.33 $\pm$ 0.40 & &2.29 $\pm$ 0.27 $\pm$ 0.21\\
 $[0.85-0.90]$& 0.273 & 0.274 & 0.360 & 0.360 & 0.876  & 5.38 $\pm$ 0.31 $\pm$ 0.41 & &2.42 $\pm$ 0.26 $\pm$ 0.21\\
$[0.90-0.95]$ & 0.273 & 0.273 & 0.360 & 0.360 & 0.926  & 4.51 $\pm$ 0.26 $\pm$ 0.35 & &2.57 $\pm$ 0.27 $\pm$ 0.25\\
$[0.95-1]$ & 0.273 & 0.273 & 0.361 & 0.361 & 0.982  & 4.93 $\pm$ 0.17 $\pm$ 0.37 & &2.82 $\pm$ 0.18 $\pm$ 0.26\\ 
\toprule\toprule
\multirow{2}{*}{$  \frac{\sin^2\theta_{2}}{1+\cos^2\theta_{2}} $ } & \multirow{2}{*}{$\langle z_{1} \rangle$}   & \multirow{2}{*}{$\langle z_{2} \rangle$} 
 & $\langle p_{t0}\rangle$ & \multirow{2}{*}{-} & \multirow{2}{*}{ $\left\langle\frac{\sin^2\theta_{2}}{1+\cos^2\theta_{2}}\right\rangle$ }
 & \multirow{2}{*}{ \aul (\%)}& & \multirow{2}{*}{ \auc (\%)}  \\
 & &  & (\gev) &  & &  & & \\
\toprule
$[0.25-0.30]$ & 0.281 & 0.282 & 0.498 &  & 0.276  & 1.87 $\pm$ 0.27 $\pm$ 0.27 & & 0.84 $\pm$ 0.22 $\pm$ 0.13\\
$[0.30-0.35]$ & 0.281 & 0.282 & 0.492 &  & 0.325  & 1.41 $\pm$ 0.25 $\pm$ 0.27 & & 0.63 $\pm$ 0.21 $\pm$ 0.13\\
$[0.35-0.40]$ & 0.280 & 0.281 & 0.489 &  & 0.375  & 1.85 $\pm$ 0.25 $\pm$ 0.27 & & 0.84 $\pm$ 0.21 $\pm$ 0.13\\
$[0.40-0.45]$ & 0.280 & 0.279 & 0.489 &  & 0.425  & 1.40 $\pm$ 0.25 $\pm$ 0.27 & & 0.63 $\pm$ 0.21 $\pm$ 0.13\\
$[0.45-0.50]$ & 0.277 & 0.278 & 0.490 &  & 0.475  & 1.63 $\pm$ 0.25 $\pm$ 0.27 & & 0.73 $\pm$ 0.21 $\pm$ 0.13\\
$[0.50-0.55]$ & 0.276 & 0.277 & 0.491 &  & 0.525  & 1.86 $\pm$ 0.25 $\pm$ 0.27 & & 0.83 $\pm$ 0.21 $\pm$ 0.13\\
$[0.55-0.60]$ & 0.275 & 0.276 & 0.493 &  & 0.575  & 2.37 $\pm$ 0.25 $\pm$ 0.27 & & 1.06 $\pm$ 0.21 $\pm$ 0.13\\
$[0.60-0.65]$ & 0.275 & 0.275 & 0.495 &  & 0.625  & 1.74 $\pm$ 0.25 $\pm$ 0.27 & & 0.78 $\pm$ 0.21 $\pm$ 0.13\\
$[0.65-0.70]$ & 0.274 & 0.275 & 0.498 &  & 0.675  & 1.49 $\pm$ 0.24 $\pm$ 0.27 & & 0.67 $\pm$ 0.20 $\pm$ 0.13\\
$[0.70-0.75]$ & 0.274 & 0.274 & 0.500 &  & 0.725  & 2.30 $\pm$ 0.24 $\pm$ 0.27 & & 1.04 $\pm$ 0.20 $\pm$ 0.13\\
$[0.75-0.80]$ & 0.274 & 0.274 & 0.502 &  & 0.775  & 2.13 $\pm$ 0.23 $\pm$ 0.27 & & 0.96 $\pm$ 0.19 $\pm$ 0.13\\
 $[0.80-0.85]$& 0.274 & 0.274 & 0.505 &  & 0.825  & 2.03 $\pm$ 0.22 $\pm$ 0.27 & & 0.91 $\pm$ 0.18 $\pm$ 0.13\\
 $[0.85-0.90]$& 0.273 & 0.274 & 0.507 &  & 0.876  & 2.27 $\pm$ 0.21 $\pm$ 0.27 & & 1.02 $\pm$ 0.17 $\pm$ 0.13\\
$[0.90-0.95]$ & 0.273 & 0.273 & 0.509 &  & 0.926  & 2.21 $\pm$ 0.19 $\pm$ 0.27 & & 1.00 $\pm$ 0.15 $\pm$ 0.13\\
$[0.95-1]$ & 0.273 & 0.273 & 0.510 &  & 0.983  & 2.14 $\pm$ 0.12 $\pm$ 0.27 & & 0.97 $\pm$ 0.10 $\pm$ 0.13\\
\toprule 
\end{tabular}
\label{tab:thetaasy}
\end{center}
\end{table*}

\begin{table*}[!htbp]
\begin{center}
\caption{ Results of the linear fits to the Collins asymmetries as functions of 
  $\sin^2\theta/(1+\cos^2\theta)$, where $\theta = \theta_{th}$ for
  RF12, and $\theta = \theta_2$ for RF0. 
}
\small{
\begin{tabular}{ |c || c c c c c c c |}
\toprule 
 & \multirow{2}{*}{\aulTh} &&  \multirow{2}{*}{\aucTh} &&  \multirow{2}{*}{\aul} && \multirow{2}{*}{\auc}\\
&&&&&&&\\ \toprule
$p_0$ & $-0.001\pm0.005$ && $-0.003\pm0.003$ && $0.014\pm0.003$ && $0.006\pm0.002$\\
$p_1$ & $0.055\pm0.007$ && $0.031\pm0.005$ && $0.009\pm0.004$ && $0.004\pm0.003$\\
$\chi^2/ndf$ & 1.1 && 0.4 && 0.5 & &0.2 \\
\toprule 
$p_0$ fixed& $ 0 $ && $ 0 $ && $ 0 $ && $ 0 $\\
$p_1$ & $0.053\pm0.002$ && $0.027\pm0.001$ && $0.028\pm0.002$ && $0.012\pm0.001$\\
$\chi^2/ndf$ & 1.0 && 0.4 && 2.0 & & 1.0 \\
\toprule
\end{tabular}
} 
\label{tab:thetaTh}
\end{center}
\end{table*}

\bibliography{paper}

\end{document}

%% file: authors_may2013.tex
%
\author{J.~P.~Lees}
\author{V.~Poireau}
\author{V.~Tisserand}
\affiliation{Laboratoire d'Annecy-le-Vieux de Physique des Particules (LAPP), Universit\'e de Savoie, CNRS/IN2P3,  F-74941 Annecy-Le-Vieux, France}
\author{E.~Grauges}
\affiliation{Universitat de Barcelona, Facultat de Fisica, Departament ECM, E-08028 Barcelona, Spain }
\author{A.~Palano$^{ab}$ }
\affiliation{INFN Sezione di Bari$^{a}$; Dipartimento di Fisica, Universit\`a di Bari$^{b}$, I-70126 Bari, Italy }
\author{G.~Eigen}
\author{B.~Stugu}
\affiliation{University of Bergen, Institute of Physics, N-5007 Bergen, Norway }
\author{D.~N.~Brown}
\author{L.~T.~Kerth}
\author{Yu.~G.~Kolomensky}
\author{M.~J.~Lee}
\author{G.~Lynch}
\affiliation{Lawrence Berkeley National Laboratory and University of California, Berkeley, California 94720, USA }
% ...
\collaboration{The \babar\ Collaboration}
\noaffiliation